\documentclass[twocolumn]{aastex62}
\pdfoutput=1 %for arXiv submission

\usepackage{natbib}
\usepackage{epsfig}

\begin{document}

\DeclareGraphicsExtensions{.pdf,.gif,.jpg,.eps,.ps}

\title{Warm Molecular Hydrogen in Nearby, Luminous Infrared Galaxies}
 
\author{Andreea O.  Petric}
\affil{Institute for Astronomy, 2680 Woodlawn Drive, Honolulu,HI, 96822,USA}
\affil{Canada-France-Hawaii Telescope, 65-1238 Mamalahoa Highway, Kamuela, HI, 96743, USA}

\author{Lee Armus}
\affil{Infrared Processing \& Analysis Center, MS 100-22, California Institute of Technology, Pasadena, CA, 91125, USA} 

\author{Nicolas Flagey}
\affil{Canada-France-Hawaii Telescope, 65-1238 Mamalahoa Highway, Kamuela, HI, 96743, USA}

\author{Pierre Guillard}
\affil{Institut d'Astrophysique de Paris, 96 bis boulevard Arago, 75014 Paris, France}

\author{Justin Howell}
\affil{Infrared Processing \& Analysis Center, MS 100-22, California Institute of Technology, Pasadena, CA, 91125, USA} 

\author{Hanae Inami}
\affil{Universite de Lyon, 92 Rue Pasteur, 69007, Lyon, France}

\author{Vassillis Charmandaris}
\affil{Department of Physics, University of Crete, GR-71003 Heraklion, Greece}
\affil{IAASARS, National Observatory of Athens, GR-15236, Penteli, Greece}

\author{Aaron Evans}
\affil{Department of Astronomy, University of Virginia, Charlottesville, VA, 22903, USA}
\affil{National Radio Astronomy Observatory, 520 Edgemont Road, Charlottesville, VA, 22903, USA}

\author{Sabrina Stierwalt}
\affil{Department of Astronomy, University of Virginia, Charlottesville, VA, 22903, USA}
\affil{National Radio Astronomy Observatory, 520 Edgemont Road, Charlottesville, VA, 22903, USA}

\author{Tanio Diaz-Santos}
\affil{Infrared Processing \& Analysis Center, MS 100-22, California Institute of Technology, Pasadena, CA, 91125, USA} 
\affil{Nucleo de Astronomia de la Facultad de Ingenieria, Universitad Diego Portales, Av. Ejercito Libertador 441, Santiago, Chile}

\author{Nanyao Lu}
\affil{National Astronomical Observatories, Chinese Academy of Sciences (CAS), Beijing 100012, China; nanyao.lu@gmail.com}
\affil{China-Chile Joint Center for Astronomy, Camino El Observatorio 1515, Las Condes, Santiago, Chile}

\author{Henrik Spoon}
\affil{Department of Astronomy, Cornell University, 616A Space Science Building, Ithaca, NY, 14853, USA} 

\author{Joe Mazzarella}
\affil{Infrared Processing \& Analysis Center, MS 100-22, California Institute of Technology, Pasadena, CA, 91125, USA} 

\author{Phil Appleton}
\affil{Infrared Processing \& Analysis Center, MS 100-22, California Institute of Technology, Pasadena, CA, 91125, USA} 

\author{Ben Chan}
\affil{Infrared Processing \& Analysis Center, MS 100-22, California Institute of Technology, Pasadena, CA, 91125, USA} 

\author{Jason Chu}
\affil{Institute for Astronomy, 2680 Woodlawn Drive, Honolulu,HI, 96822,USA}

\author{Derek Hand}
\affil{Department of Physics and Astronomy, University of Hawaii, 200 W. Kawili St., Hilo, HI 96720}

\author{George Privon}
\affil{Instituto de Astrofisica, Facultad de Fisica, Pontificia Universitad Catolica de Chile, Casilla 306, Santiago 22, Chile}

\author{David Sanders}
\affil{Institute for Astronomy, 2680 Woodlawn Drive, Honolulu,HI, 96822,USA}

\author{Jason Surace}
\affil{Infrared Processing \& Analysis Center, MS 100-22, California Institute of Technology, Pasadena, CA, 91125, USA} 

\author{Kevin Xu}
\affil{Infrared Processing \& Analysis Center, MS 100-22, California Institute of Technology, Pasadena, CA, 91125, USA} 

\author{Yinghe Zhao}
\affil{Infrared Processing \& Analysis Center, MS 100-22, California Institute of Technology, Pasadena, CA, 91125, USA} 
\affil{Yunnan Observatories, CAS, Kumming 65011, China}
\affil{Key laboratory for the Structure and Evolution of Celestial Objects, CAS, Kumming 65011, China}
\affil{Purple Mountain Observatory, CAS, Nanjing, 210008, China}

\begin{abstract}
  Mid-infrared molecular hydrogen (H$_2$) emission is a powerful cooling agent in galaxy mergers and in radio galaxies; it is a potential key tracer of gas evolution and energy dissipation associated with mergers, star formation, and accretion onto supermassive black holes. We detect mid-IR H$_2$ line emission in at least one rotational transition in 91\% of the 214 Luminous Infrared Galaxies (LIRGs) observed with Spitzer as part of the Great Observatories All-sky LIRG Survey (GOALS). We use H$_2$ excitation diagrams to estimate the range of masses and temperatures of warm molecular gas in these galaxies. We find that LIRGs in which the IR emission originates mostly from the Active Galactic Nuclei (AGN) have about 100K higher H$_2$ mass-averaged excitation temperatures than LIRGs in which the IR emission originates mostly from star formation.
  Between 10 and 15\% of LIRGs have H$_2$ emission lines that are sufficiently broad to be resolved or partially resolved by the high resolution modules of Spitzer's Infrared Spectrograph (IRS). Those sources tend to be mergers and contain AGN. This suggests that a significant fraction of the H$_2$ line emission is powered by AGN activity through X-rays, cosmic rays, and turbulence.
  We find a statistically significant correlation between the kinetic energy in the H$_2$ gas and the H$_2$ to IR luminosity ratio. The sources with the largest warm gas kinetic energies are mergers. We speculate that mergers increase the production of bulk in-flows leading to observable broad H$_2$ profiles and possibly denser environments.
\end{abstract}

\section{Introduction}
Molecular hydrogen (H$_2$) is the material from which stars form and black holes grow. In turn, young massive stars and active galactic nuclei (AGN) transfer energy to the molecular hydrogen and change its physical conditions.  In this paper we use several approaches to estimate if and how the molecular gas changes on kiloparsec scales in response to changes in the gravitational potential due to galaxy mergers and in response to AGN emission. 

In interacting galaxies, large gas-flows move low-metallicity gas from the outer regions of the galaxy toward the center \citep[e.g.][]{kew2010}. 
The time-scales for this process seem to be on the order of 1 Gyr: \citet{rupke2010} find that in the approximate time between first and second passage in a merger, the central metallicity becomes diluted by low-metallicity gas flowing in from the metal-poor outskirts of the merging galaxies. Kinematic signatures in the warm molecular gas may indicate bulk flows on similar timescales. 

The amount of interstellar medium (ISM) available for star formation determines a galaxy's evolution. Because star formation is enhanced in interacting galaxies \citep[e.g.][]{pat2013}, mergers may consume their gas supply at a higher rate than non-mergers \citep[e.g][]{mih1996, spring2000, hay2011}. However, observational studies of gas in post-mergers find them to have more neutral hydrogen than non-mergers \citep[e.g][]{elis2015, lar2016}.  

While there are large variations of gas properties in interacting galaxies \citep[e.g][]{haan2011,fern2014}, mergers may have more cold gas than non-mergers because of inflows from cooled ionized halo gas \citep{elis2015, bc1993}. If such inflows include a warm molecular phase it could take the form of asymmetric emission profiles or red-shifted components. One of our goals in this work is to complement studies of atomic hydrogen in mergers, and trace the fate of the warm molecular gas component in the ISM of nearby star-forming galaxies.

Our investigation focuses on warm molecular gas as traced by H$_2$ rotational emission lines (Table 1) in a sample of nearby, Luminous Infrared Galaxies (LIRGs). LIRGs are galaxies with L$(8 -- 1000\mu m) = \rm{L}_{IR} \ge 10^{11} \rm{L}_{\odot}$ a subset of which have L$_{IR} \ge 10^{12} \rm{L}_{\odot}$ and are called Ultraluminous Infrared Galaxies \citep[ULIRGs:][]{sam1996}. Because LIRGs can have a wide range of optical classifications and because they span the full range of galaxy interactions from non-merging spirals to late stage mergers they are well suited for the study of how AGN and mergers impact the ISM \citep{armus09, petric2011, lar2016, psy2016, stir2013, priv2015}. Furthermore, LIRGs bridge the luminosity gap between nearby star-forming galaxies and quasars and so they may provide the link between the extreme objects we see at high redshift and typical nearby systems.

The LIRGs we study here are part of the Great Observatories All-sky LIRG Survey (GOALS) which targets a representative sample of 202 systems in the local Universe ($z~\leq $ 0.088) selected from the IRAS Revised Bright Galaxy Sample \citep{sanders03}. An outline of the GOALS project and a multi-wavelength analysis of the LIRG VV 340 are given in \citet{armus09}. Results from the MIR spectroscopy of the GOALS sample were presented in \citet{eva2008}, \citet{inami2010}, \citet{ds2010}, \citet{ds2011}, \citet{ds2014}, \citet{petric2011}, \citet{mazz2012}, \citet{modi2012}, \citet{stir2013}, and \citet{stir2014}.

The connections between AGN, merger stage, and the state of the gas are complex. In a detailed, high-spectral resolution, Spitzer IRS, study of the GOALS LIRGS, \citet{ina2013} found no correlation between fine-structure line ratios and the merger stage but found that emission lines from more highly ionized ions have broader line widths, e.g. Ne~{\sc v}]  emission lines are broader than [Ne~{\sc iii}] lines, which in turn are broader than  [Ne~{\sc ii}] lines. They also find five LIRGs whose shifted [Ne~{\sc iii}],  [Ne~{\sc v}] lines suggest the presence of fast moving highly ionized gas that may be part of galactic bulk-flows. In this paper, we compare the line widths of the warm molecular gas with the contribution of the AGN to the mid-infrared (MIR) luminosity to test if the thermal energy in the warm gas is contributed by the AGN. 

In most LIRGs and ULIRGs, star formation dominates the heating of H$_2$ \citep{stir2014, hill2014, hig2006}.  However, a fraction of LIRGs and even a larger fraction of ULIRGs have more MIR H$_2$ emission than what could be expected if the H$_2$ emission originates in photo-dissociation regions. \citet{stir2014} use low resolution IRS data to study the H$_2$  and dust properties of LIRGs and find that most nearby LIRGs have higher ratios of $L$(H$_2)/L({\rm{PAHs}})$ than normal star-forming galaxies. \citet{stir2014} show that this ratio increases with H$_2$ luminosity and that in around 10\% of LIRGs, the H$_2$ emission may be excited by shocks either from powerful starbursts or AGN. ULIRGs show on average three times more emission in the rotational transitions of molecular hydrogen than expected based on their star formation rates \citep{hill2014}. \citet{hill2014} also found a weak positive correlation between H$_2$ emission and the length of the tidal tails and a strong correlation between H$_2$  and [Fe~{\sc{ii}}] suggesting that the excess H$_2$ is produced by shocks. 

Studies of warm molecular H$_2$ kinematics may help disentangle the impact that gravitational interactions and AGN have on the interstellar medium. In this paper, we extract kinematic information from resolved H$_2$ line profiles to test if mergers lead to bulk gas motions. We also estimate if and how the masses, temperatures and excitation conditions of H$_2$ change with merger stage and with the AGN contribution to the LIRG's IR luminosity.
The paper is organized as follows. In section 2 we describe the IRS observations and reduction methods. Section 3 includes a presentation of the H$_{2}$ flux measurements, a description of the method used to estimate the total warm H$_2$ masses and temperatures as well as basic statistics of these quantities as a function of merger stage, IR luminosity, and AGN contribution to the IR emission. In section 4 we discuss our findings and in section 5 we summarize our conclusions.

\section{Sample and Data}
The GOALS sample properties and selection are described in detail in \citet{armus09}. For this investigation we use the spectra of 248 individual nuclei in 202 LIRG systems, observed in the high-resolution IRS modules (Short-High, and Long-High) and complementary low-resolution (IRS Short-Low, Long-Low) spectra for 234 sources. The widths of the SL, SH, LL, LH slits (3.6\arcsec , 4.7\arcsec, 10.7\arcsec, 11.1\arcsec) correspond to 1.5, 2.0, 4.5, and 4.6 kpc respectively at a distance of 88 Mpc -- the median galaxy distance in our sample. The distances to the GOALS's galaxies are between 17.5 and 387 Mpc. We obtained the IRS spectra in our own observing program (PI: Armus, PID 30323) for 158 LIRG systems, with the IRS spectra for the remaining 44 LIRG systems taken from the Spitzer archive.
In all data from PID 30323, the IRS Staring Mode was employed, using ``cluster target'' observations for those sources with well separated ($\geq 10$ arcsec) companions. Among the 202 LIRGs studied, secondary nuclei were targeted only when the flux ratio of primary to secondary nucleus (as measured in the Spitzer MIPS 24 $\mu $m data) is less than or equal to five, in order to capture the spectra of the nuclei actively participating in the far-infrared emission of the system.  

A more detailed description of how the spectra, used for the analysis presented in this paper, were reduced is given in three previous papers: \citet{petric2011, ina2013, stir2013}. The spectra were extracted with the SPICE\footnote{\url{http://irsa.ipac.caltech.edu/data/SPITZER/docs/datanalysys/tools/spice/}} software, assuming that the flux in the slit originates from a point source. The profile for the extraction is set automatically to match the PSF at different wavelengths. The PSF was determined by the IRS team from standard calibrators.
Twenty eight systems were observed in spectral mapping mode. These data were assembled and cleaned to remove noisy pixels. Nuclear spectra were then extracted with CUBISM \citep{smith07} using extraction regions of sizes equal to those of extraction regions for point sources in the spectra taken in staring mode.
A more detailed description of the spectra including the position of the IRS slits are given in \citet{stir2013}, \citet{ina2013}, and the GOALS delivery documents\footnote{\url{https://irsa.ipac.caltech.edu/data/GOALS/overview.html}}.

The GOALS sources were classified in 5 merger stages: (0) no obvious sign of a disturbance either in the IRAC or HST morphologies, or published evidence that the gas is not in dynamical equilibrium (i.e., undisturbed circular orbits); (1) early stage, where the galaxies are within 1 arcmin of each other, but little or no morphological disturbance can be observed; (2) the galaxies exhibit bridges and tidal tails but they do not have a common envelope, and each optical disk is relatively intact; (3) the optical disks are completely destroyed but two nuclei can be distinguished; (4) the two interacting nuclei are merged but structure in the disk indicates that the source has gone through a merger. The merger classifications for LIRGs are published in \citet{stir2013}, see also Fig. 10 in \citet{petric2011}.  A subset of 65 LIRGs was re-evaluated by \citet{lar2016}.  We acknowledge the pitfalls of morphological classifications, and refer the reader to the more precise techniques requiring high sensitivity and high spatial and velocity resolution\citep{privon2013}. 

Here we seek to be consistent with previous merger-class investigations of the GOALS sample of LIRGs presented in  \citet{lar2016,stir2013,haan2011a} and we compress the merger stages into three categories: non mergers ($nm$) which are targets without obvious signs of morphological disturbances; early-mergers ($em$) which are systems in which the interacting galaxies are within 1 arcmin of each other but show little or no morphological disturbance; and mergers ($m$) which are all the other stages of gravitational interactions.

\section{Results}
In this section we will present: our measurements of emission line fluxes, and line-widths, how we identify resolved emission lines, how we estimate warm molecular H$_2$ masses, temperatures, ortho to para ratios, and H$_2$ excitation conditions.

\subsection{Fluxes}
To measure line fluxes and line-widths from our SH, LH and SL spectra we fit Gaussian functions to the atomic and molecular gas emission line profiles. We inspect all the fits to ensure that spurious hot pixels were excluded. 
We did not use PAHFIT \citep{smith2007} because it does not account for potential ice absorption.
We also refer the reader to \citet{stir2014} who present H$_2$ fluxes estimated from simultaneous fits of the dust and gas features and continuum in the low resolution data after scaling the SL spectra to match the LL data. Here we chose to measure the H$_2$ emission line fluxes the same way from the low and high resolution spectra because the S0 line is not easily detected in the low resolution spectra, yet its measurement is important for determining warm molecular gas mass and temperatures. 

For each line flux measurement we average the fluxes estimated from the two IRS nods; to assess the flux errors we combine the uncertainties from the Gaussian fits in quadrature. We calculate upper limits from line free regions near the line of interest by estimating the total emission for a hypothetical line with: a width equal to the spectral resolution at that wavelength and a three $\sigma$ intensity peak. 

\begin{figure*}[h!] 
  \includegraphics[width=0.32\linewidth]{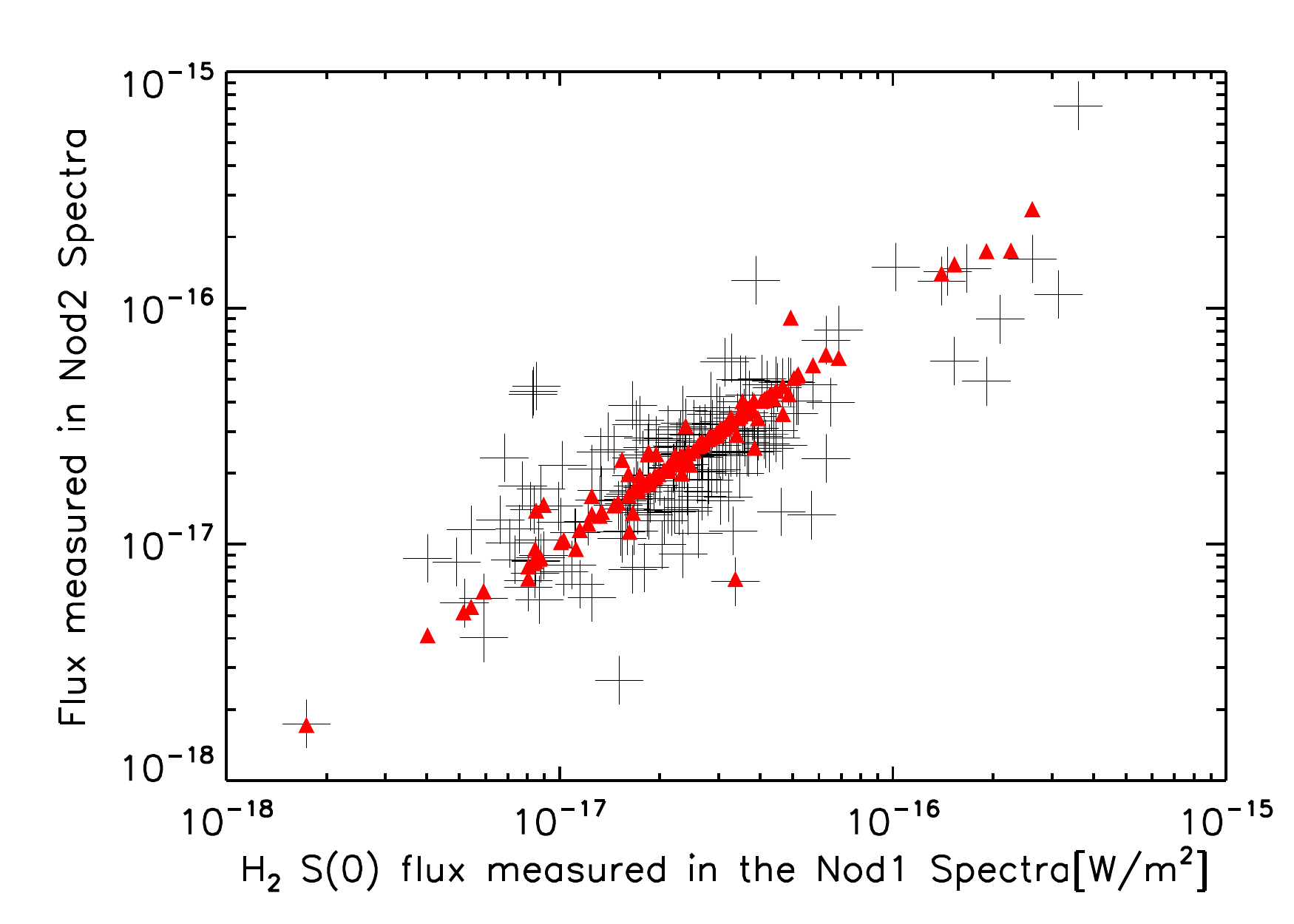}
  \includegraphics[width=0.32\linewidth]{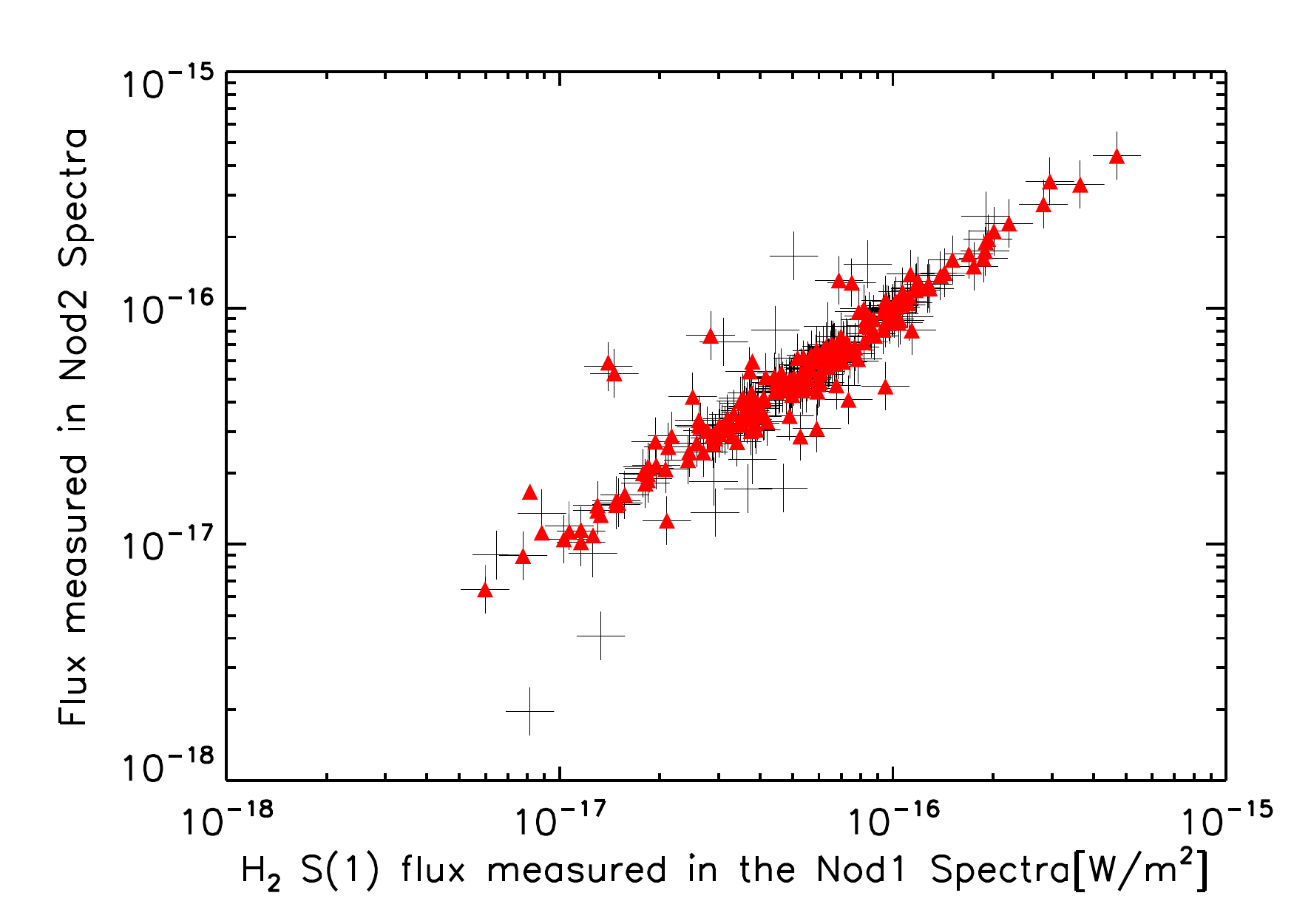}
  \includegraphics[width=0.32\linewidth]{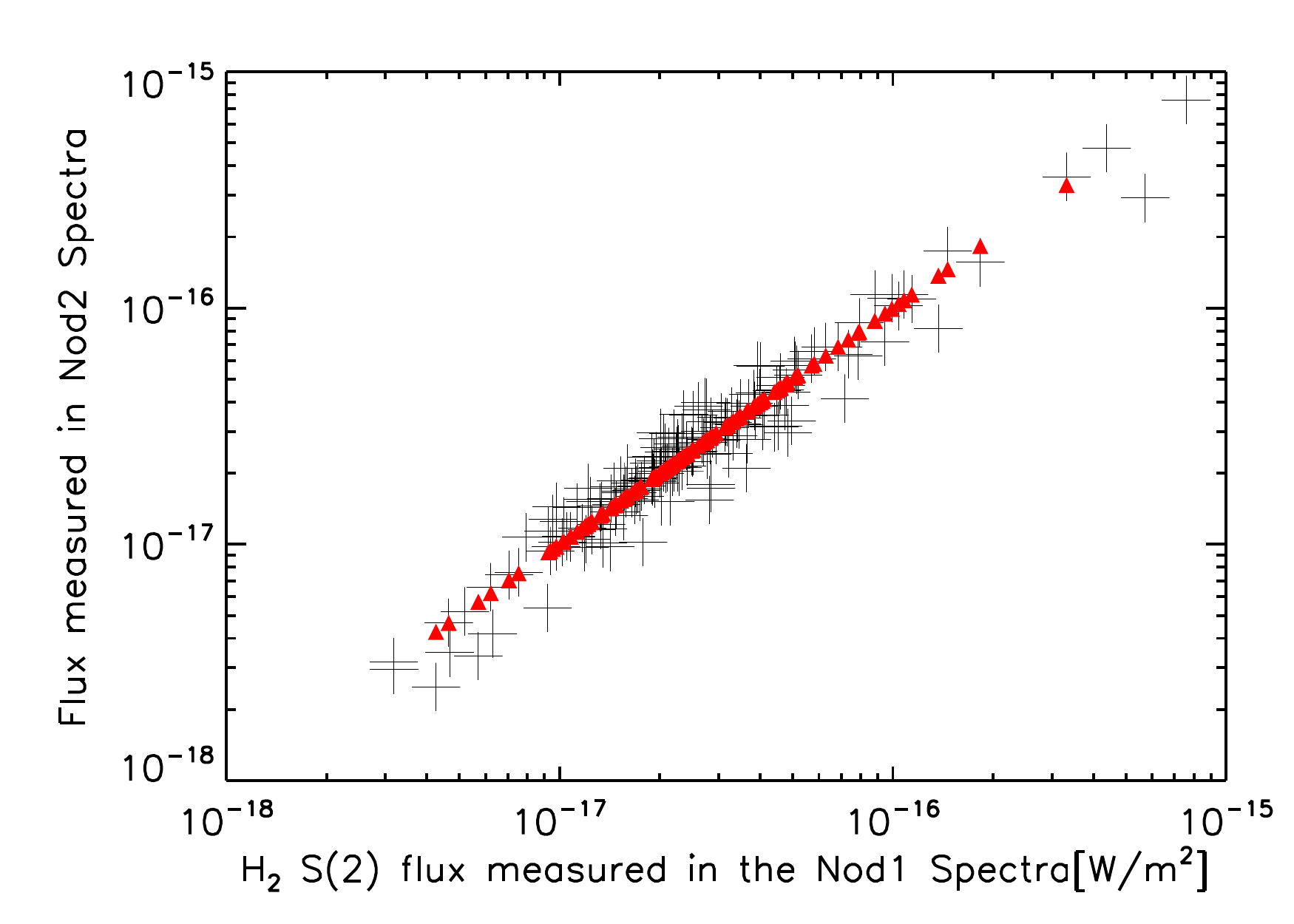}\\
  \includegraphics[width=0.32\linewidth]{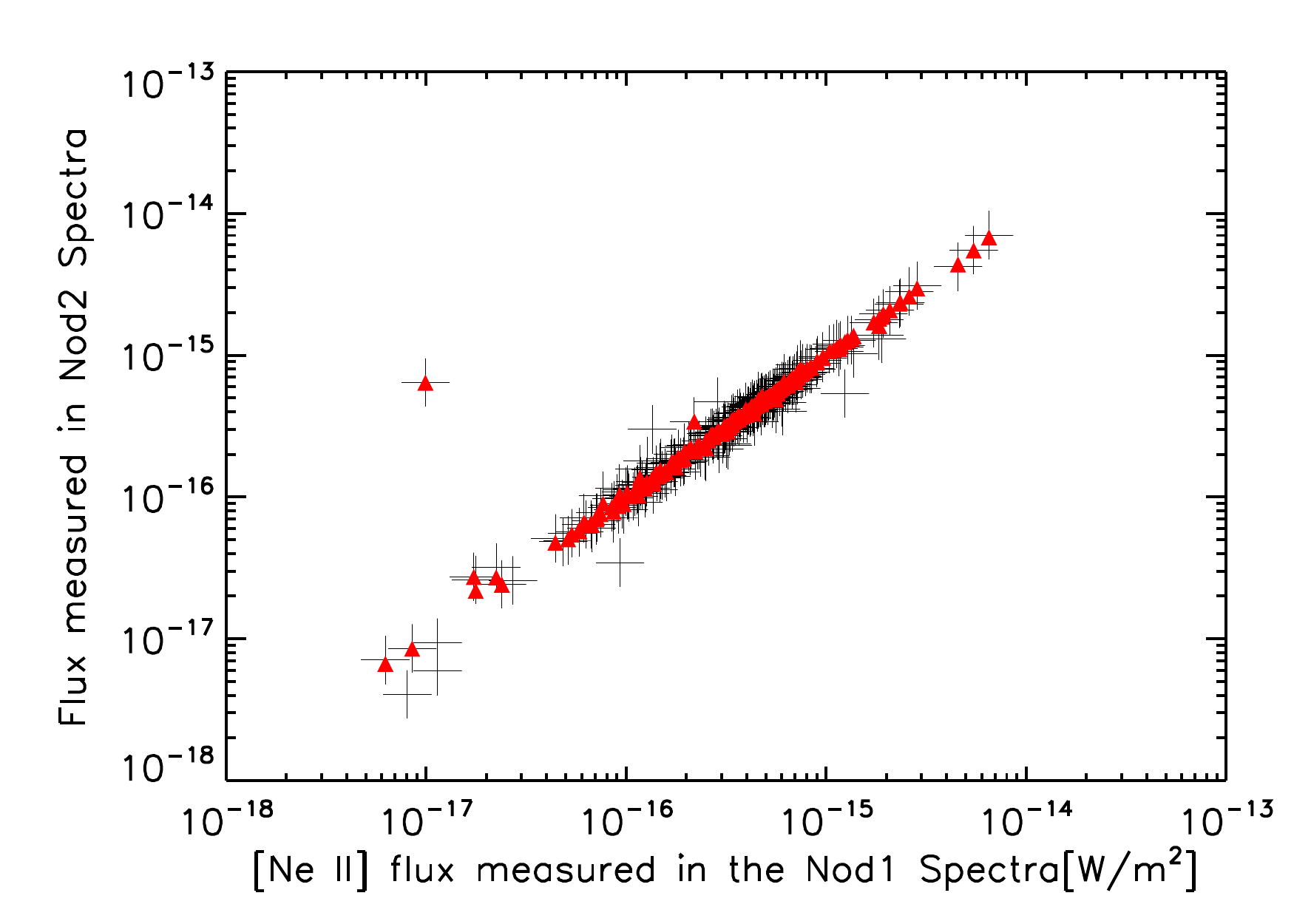}
  \includegraphics[width=0.32\linewidth]{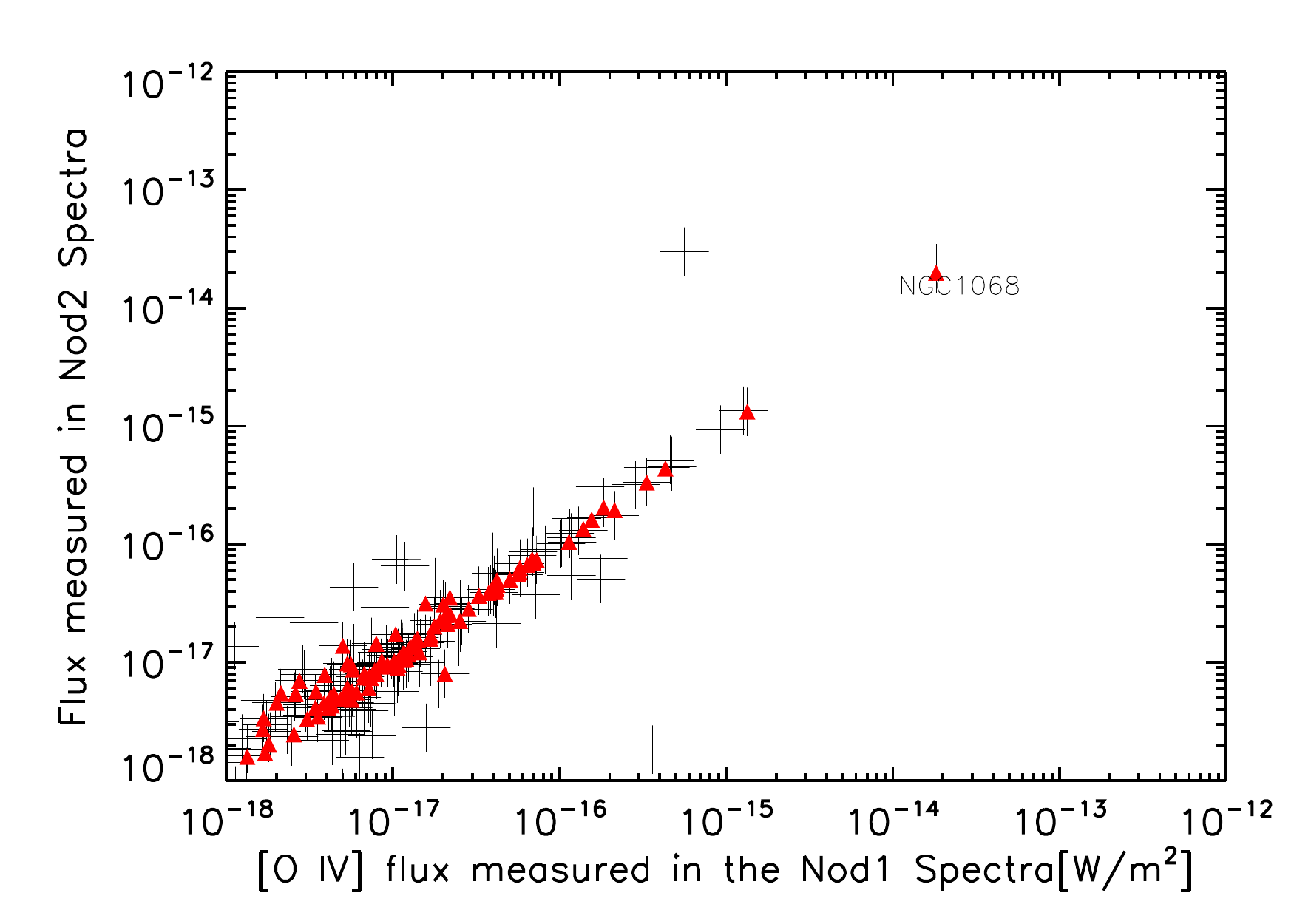}
  \includegraphics[width=0.32\linewidth]{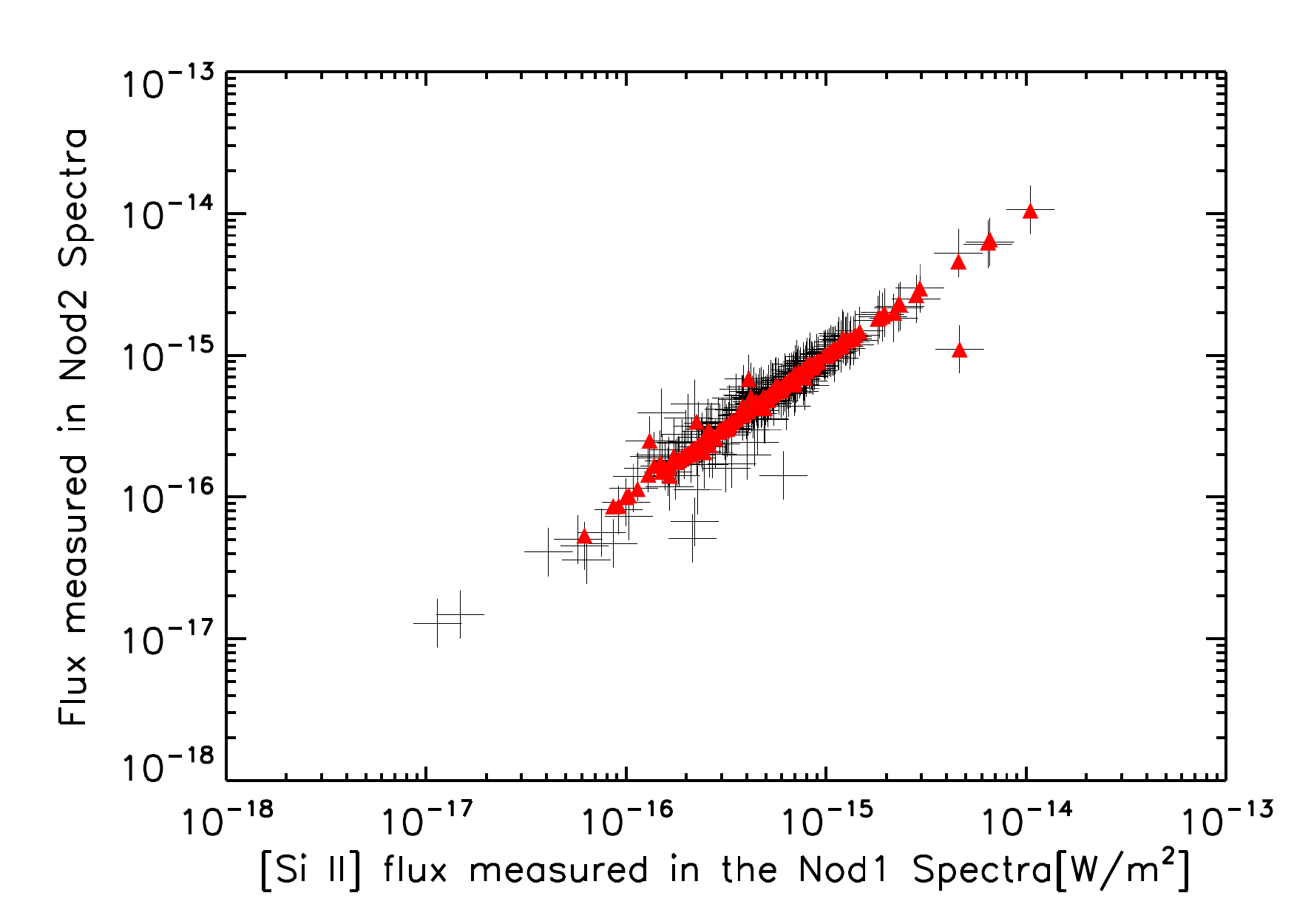}
\caption{ H2 (S0), H2 (S1), H2 (S2), [Ne~{\sc{ii}}],  [O~{\sc{iv}}] and [Si~{\sc{ii}}]  fluxes measured from IRS spectra modules SH and LH.  As described in the text each fit was inspected by hand and the continuum adjusted for bad pixels in the two nods associated with each SH or LH observation.  These plots show in black the measured Nod 1 vs Nod 2 fluxes and in red Nod 1 fluxes versus the final flux for each of the detected lines.\label{bfluxes}} 
\end{figure*}

Figure \ref{bfluxes} shows the H$_2$ S(0), S(1), S(2), S(3), [Fe II], [Ne II], [O IV] and [Si II] fluxes measured from IRS spectra.
These plots show Nod 1 vs Nod 2 fluxes and Nod 1 fluxes versus the final flux for each of the detected lines.
This figure illustrates that, for most of the sources, our estimates from each of the IRS node positions match well. For lines where the flux from the two nods differed by more than $3~\times ~\sqrt{ (\sigma F1 ^2 + \sigma F2 ^2) }$, where $\sigma F1 $ and $\sigma F 2$ represent the errors on the flux measured in Nod 1 and Nod 2 respectively, we choose results from the fit with the best overall signal to noise ration (SNR). In addition, we re-inspect visually the spectra to ensure the fit we use is better than the one we discard and to understand the reason for the difference between the two nods.

\begin{figure*}[h!]
  \includegraphics[width=0.49\linewidth]{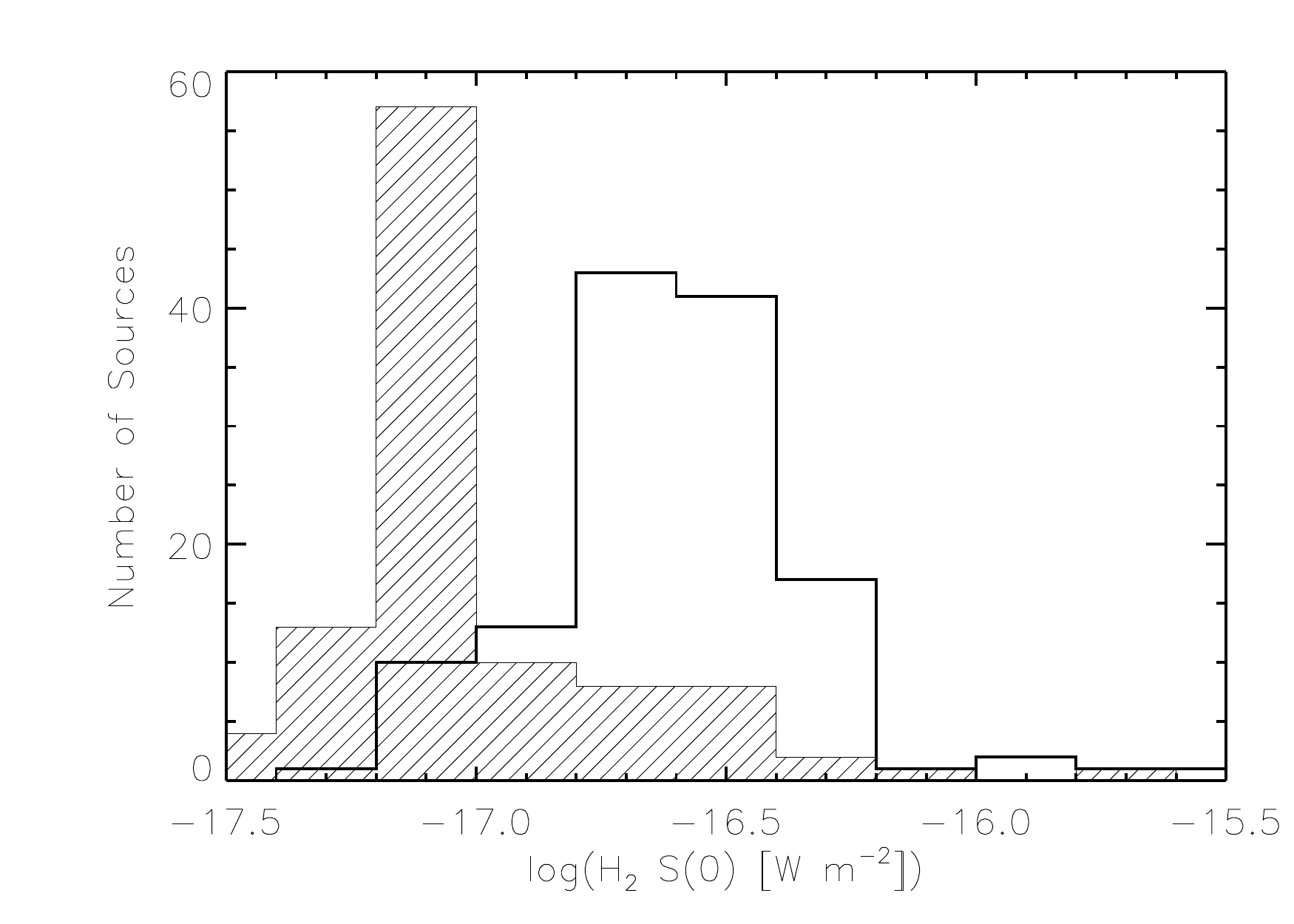}
  \includegraphics[width=0.49\linewidth]{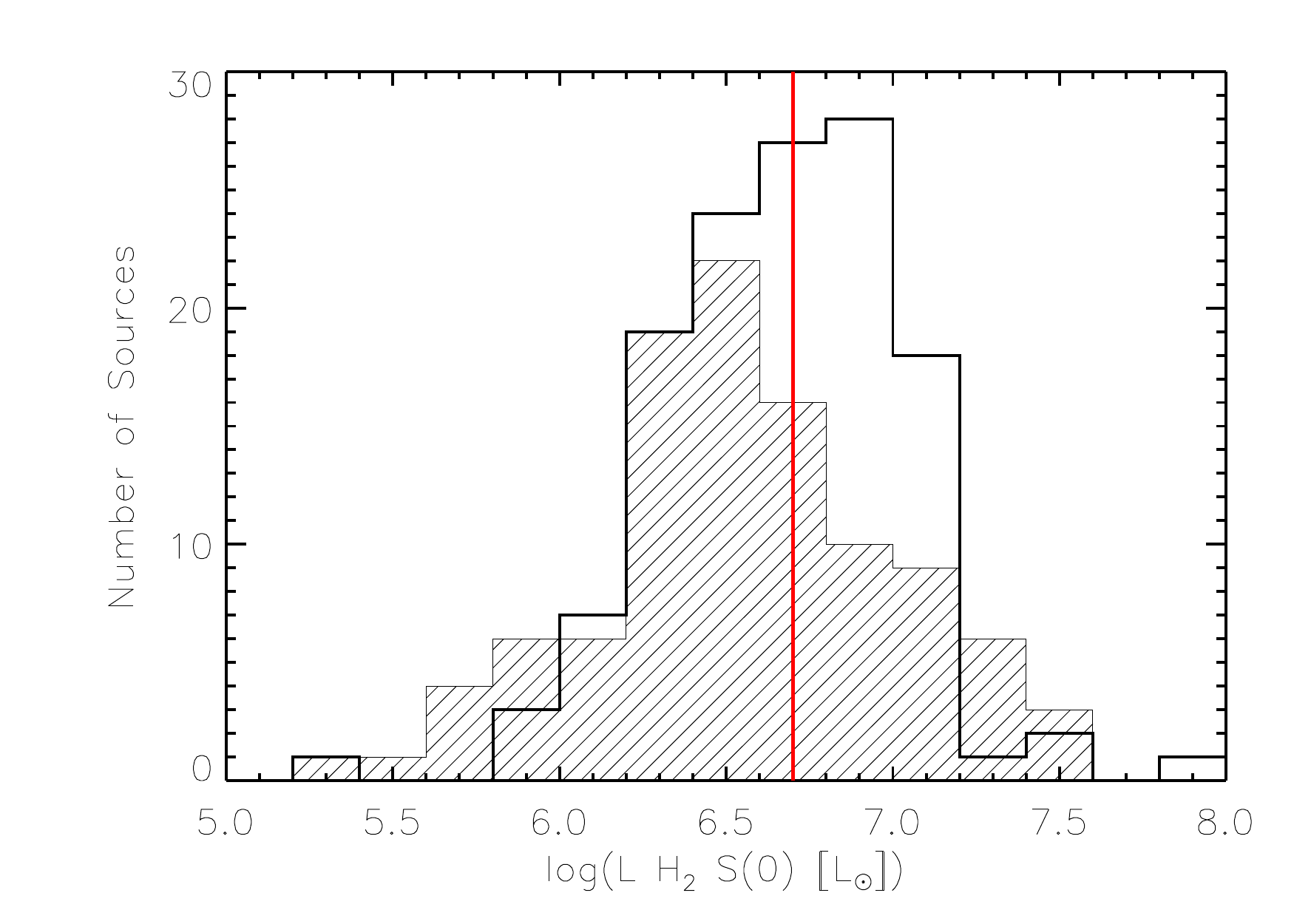}\\
  \includegraphics[width=0.49\linewidth]{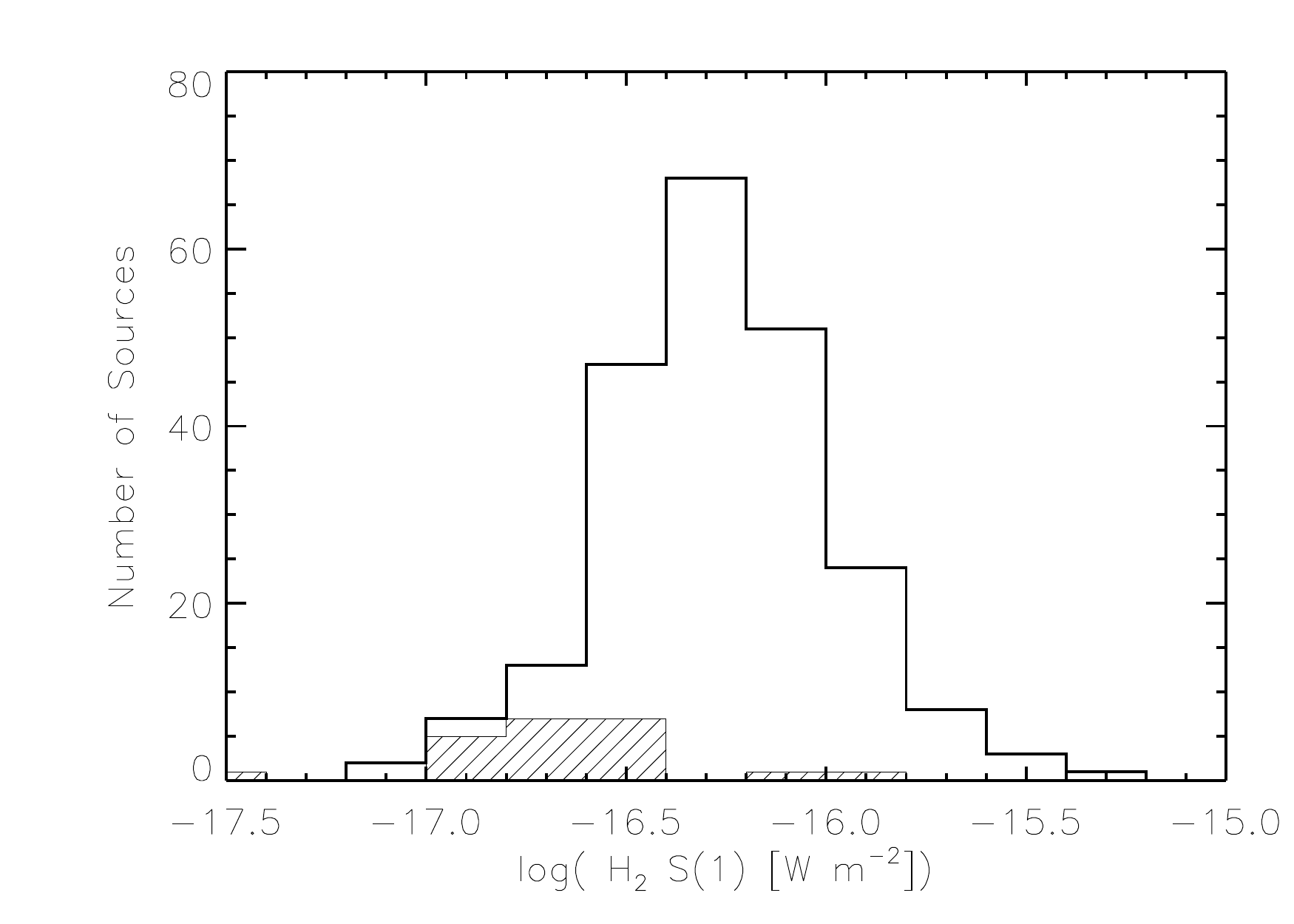}
  \includegraphics[width=0.49\linewidth]{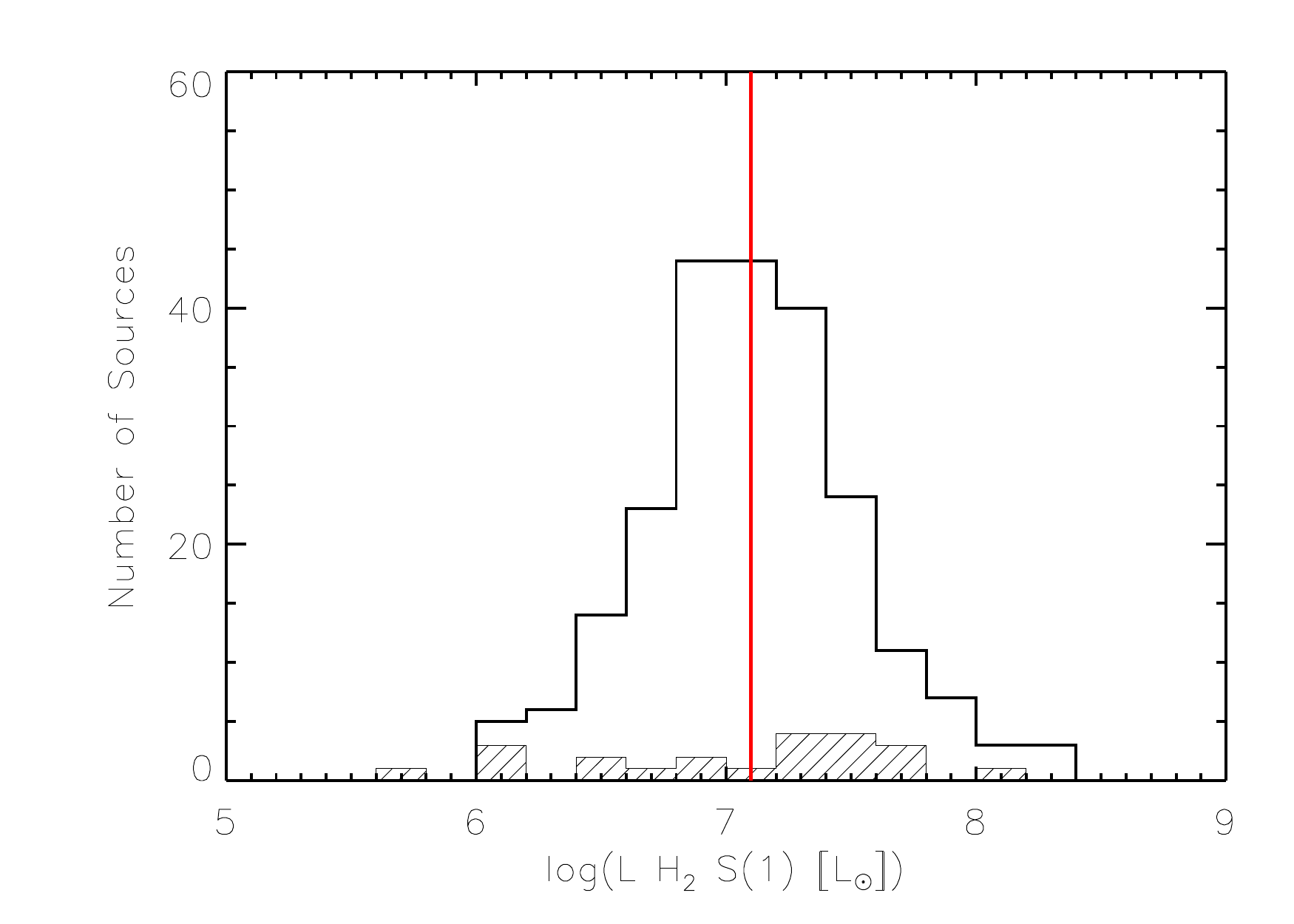}\\
  \includegraphics[width=0.49\linewidth]{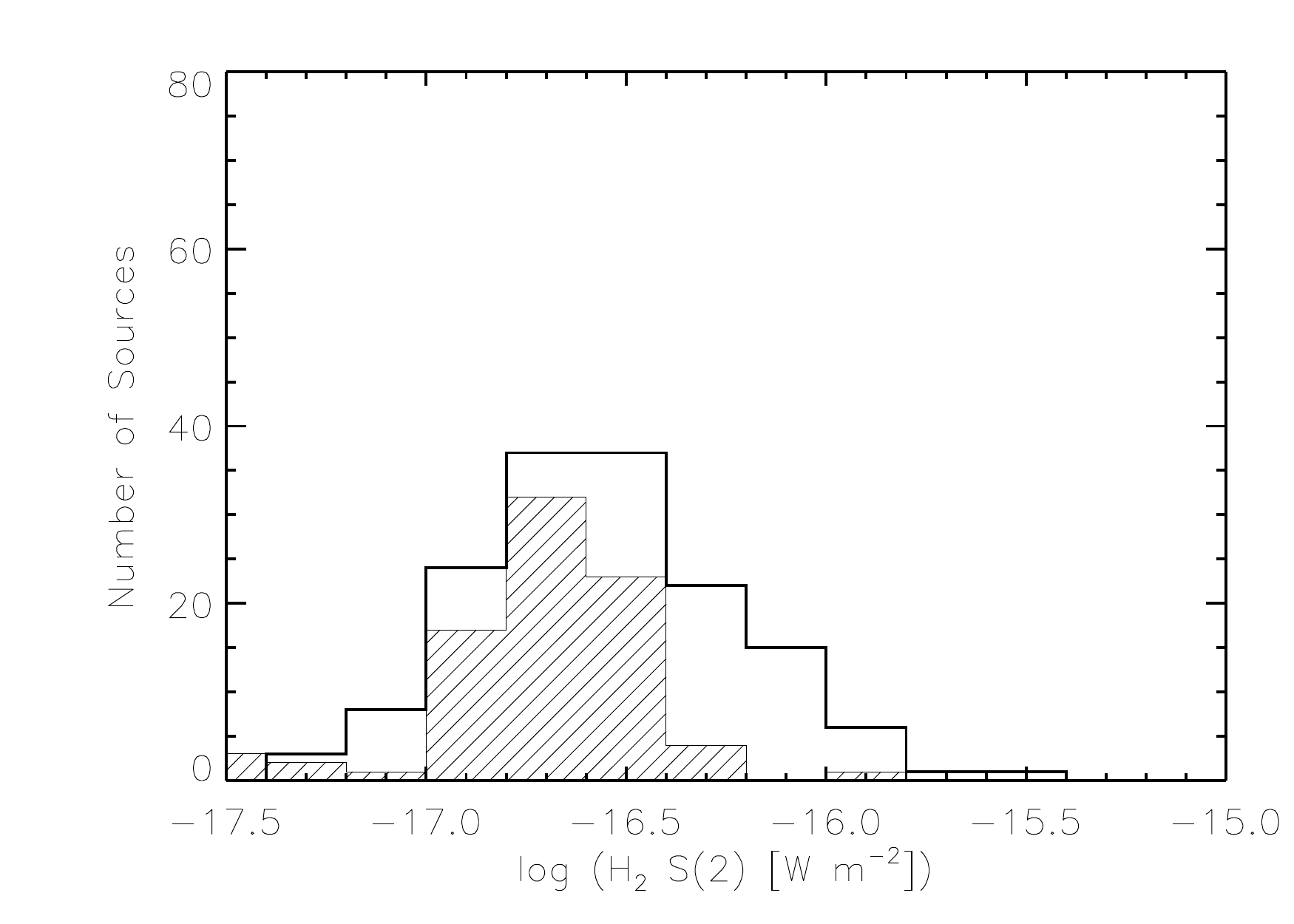}
  \includegraphics[width=0.49\linewidth]{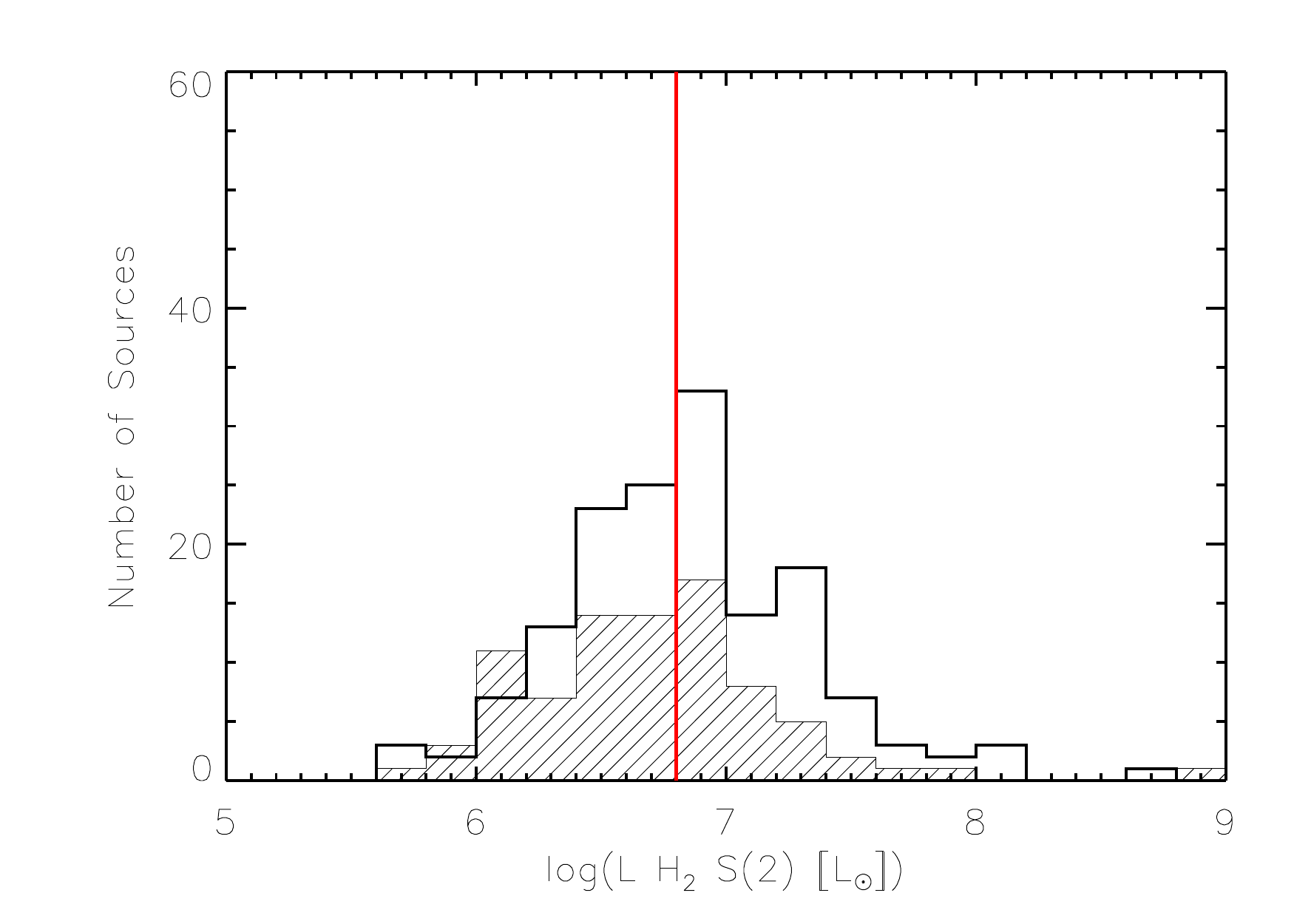}
  \caption{ Histograms of measured fluxes (left) and luminosities (right) for the H$_2$ molecular gas corresponding to the following rotational transitions, S(0) S(1), and S(2) The fluxes are given in log[ W/m$^{-2}$]  and the luminosities in log of solar units. Solid histogram present our detections. Dashed histograms show the upper limits, and the red vertical lines indicate the median luminosities. \label{FHISTO}}
\end{figure*}

Table 2 gives the H$_2$ rotational emission line fluxes and uncertainties, measured from the IRS high resolution spectra. Figure \ref{FHISTO} shows histograms of the line fluxes  and luminosities corresponding to the rotational transitions, S(0), S(1), and S(2). Table 3 gives detection statistics for the H$_2$ emission lines: percentages of detected sources, minimum, mean, median, maximum, and standard deviations of the detected fluxes and luminosities.  In particular, we find that the median H$_2$ S(0), S(1), and S(2) luminosities are $10^{6.7} , 10^{7.1}, {\rm{~and~}} 10^{6.8}  ~\rm{L}_{\odot} $  respectively. Appendix 1 describes how we combined fluxes from different modules. 

\subsection{Resolved Lines}
The IRS high resolution modules are described in the official instrumental handbook\footnote{(Version 5.0, last modified December 12, 2012)} as cross-dispersed echelle spectrographs that provide a resolving power $R~=~\lambda/\Delta \lambda ~ \sim 600$. The velocity resolution of the high resolution modules is nominally $c/R~=~494$ km/s between 4 and 18 $\mu$m and 503 km/s between 25 and  34.2 $\mu$m. Previous investigations of high-resolution line profiles used measurements of the widths of standard IRS calibration targets (P Cygni, HD 190429, HD 174638) to assess that the uncertainty in the instrumental resolution is 59 km/s in SH and 63 km/s in LH \citep[e.g][]{das2008,gui2012}.

To take advantage of the large number of high resolution, high signal-to-noise spectra of the GOALS targets, here we take a complementary approach by looking at the distribution of line widths of the H$_{2}$ S(1) and S(0) lines in our sample of LIRGs. To determine the expected distribution of widths for a sample of unresolved source we use two independent methods and compare our results. We include several derived line-width estimates that are smaller than the instrumental resolution, those are sources with poor sampling, and/or low signal to noise. We keep them here because they indicate the error budget in our overall measurements.

\begin{figure*}[h!] 
  \includegraphics[width=0.49\linewidth]{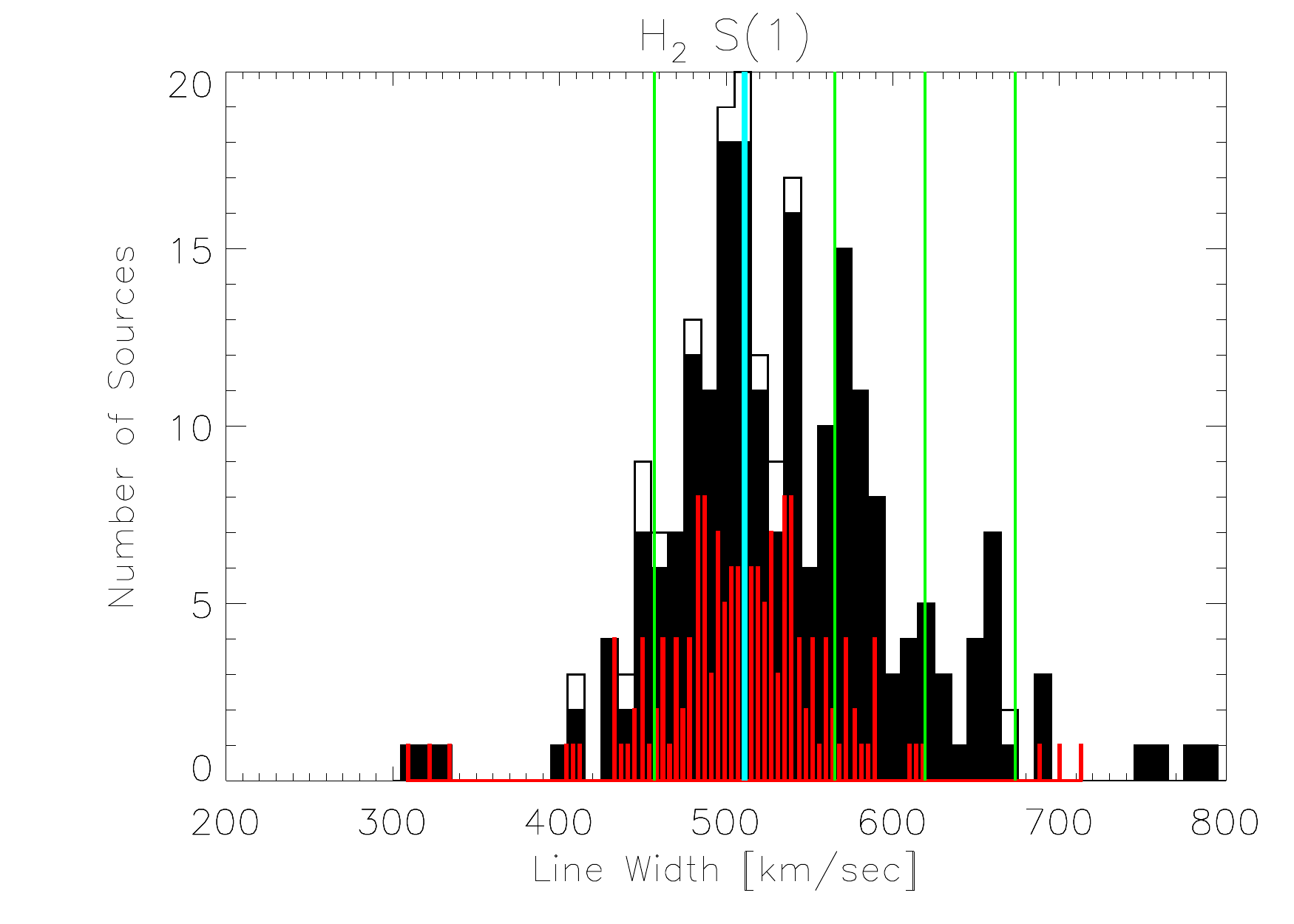}
  \includegraphics[width=0.49\linewidth]{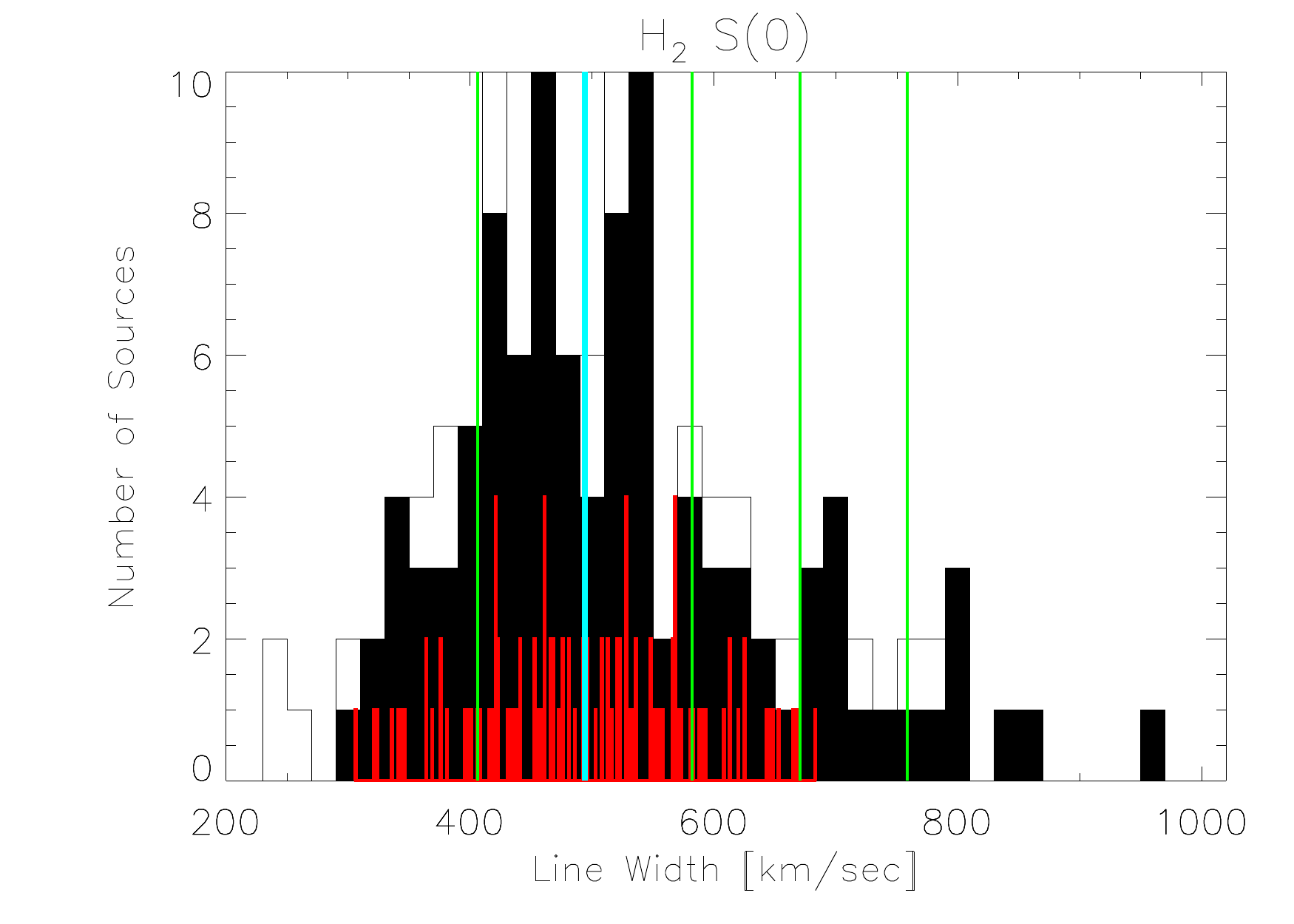}
  \caption{Histograms for H$_2$ S(1) (left) and H$_2$ S(0) (right) $FWHM$ measured in the high resolution IRS modules (using bins of 10 km/s). In red we show our model for the distribution of possible measurements if none of the sources were resolved. (See text for details.) The standard deviation of this distribution, shown in red, is 56 km/s for the $H_2$ S(1) transitions. The mode and $\sigma _{inst}$ of the S(0) distribution of FWHMs are 515 km/s and 103 km/s. The mode is marked in cyan, and the mode + 1, 2, and 3 $\sigma _{inst}$ and mode - 1$\times$ $\sigma _{inst}$ are labeled in green. The black solid histograms show only those widths with errors $\sigma _{\rm{FWHM}} \leq \rm{FWHM}/3$. \label{HistoResLines}}
\end{figure*}

{\bf{Method 1:}}  We assume that, in the absence of resolved sources, the distribution of H$_2$ line widths should be symmetric.  Figure \ref{HistoResLines} shows the histogram of the H$_2$ S(1) and S(0) lines with 10 km/s bins.  The distributions of derived line-widths show a tail of sources with larger line-widths, presumably representing the resolved and marginally resolved objects.  We estimate the standard deviation and mode from measurements with a SNR of three or better.

We find that the H$_{2}$ S(1) distribution has a mode of 511 km/s (RES1 thereafter), and a weighted mean of 540 km/s. Using the mode and the measurements of lines lower than the mode, we obtain a standard deviation of 56 km/s for the unresolved sources that we claim is purely of instrumental origin ($\sigma _{inst}$ ~= ~ 56 km/s). The value for this standard deviation is almost identical to the uncertainty quoted by \citet{das2008}.
We look at three classes of sources: clearly resolved, marginally resolved and not resolved. 

Writing:  $\rm{FWHM} - \sigma _{\rm{FWHM}} ~\geq {RES1} + n*\sigma _{inst} $, where $\rm{FWHM}$ is the measured Full Width at Half Maximum, we find three sources with $n~=~3$ and six with $n~=~2$. We refer to these sources as being clearly resolved.  Eighteen nuclei have marginally resolved  H$_2$ S(1) emission with $n~=~1$. Those eighteen marginally resolved sources would have been considered clearly resolved in earlier studies \citep[e.g][]{das2008,gui2012}. We present the measured properties of all these 27 sources with $\rm{FWHM} - \sigma _{\rm{FWHM}} \geq 567$ kms/sec in Table \ref{RTAB}. Note that these sources are not resolved at a 1$\sigma$ statistical significance level but higher, because we look at both the distribution of possible widths and the error associated with each measurement. 

\begin{figure*}[h!] 
  \includegraphics[width=0.99\linewidth]{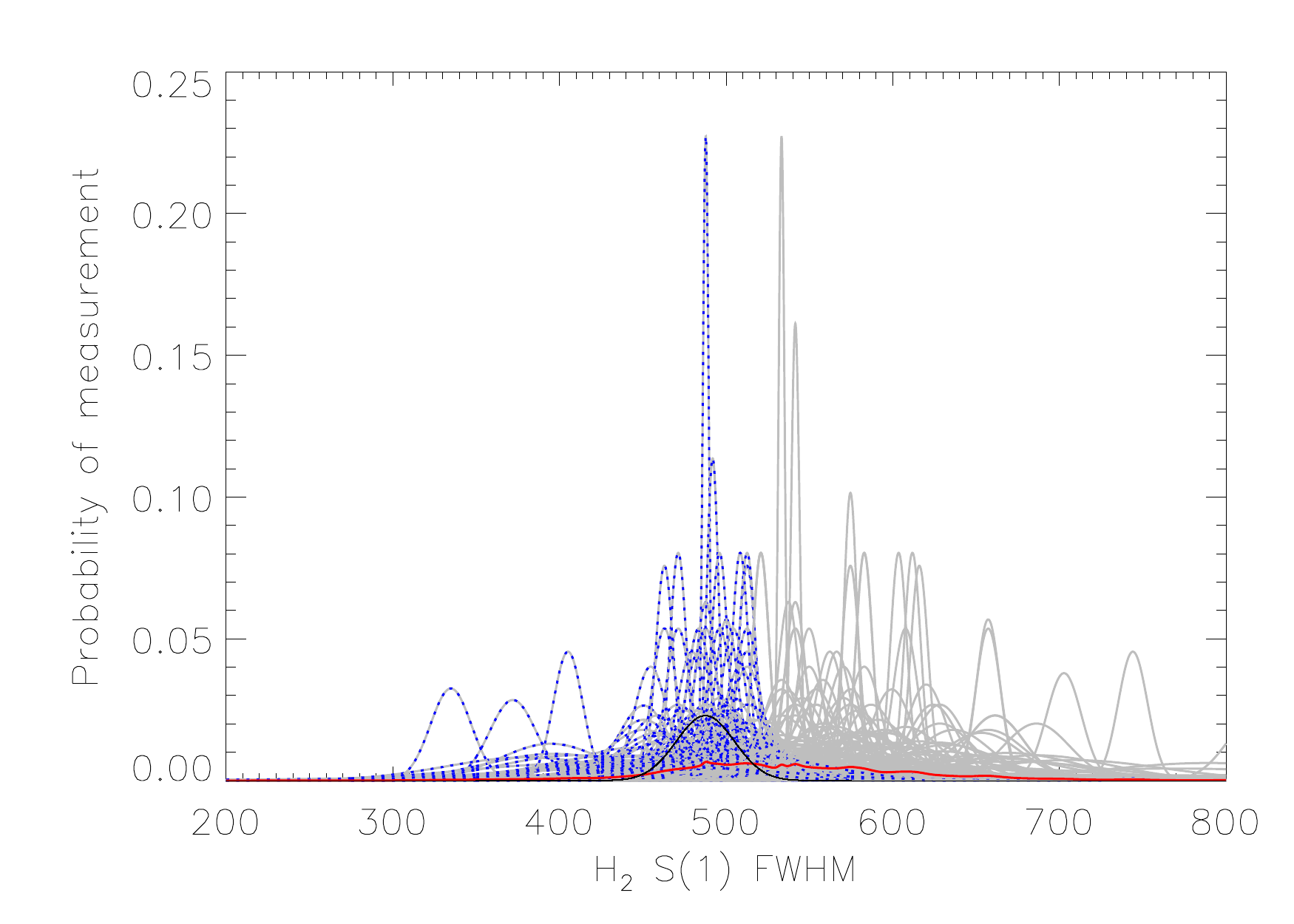}
  \caption{The distribution of derived H$_2$ line $FWHM$ (obtained from fitting each line with a Gaussian) each represented as a (grey) gaussian centered on the measured H$_2$ line-width and a width equal to the estimated error on the measured H$_2$ line-width; the peak of each gaussian is set to normalize its area to 1. Adding up all the individual gaussian distributions gives gives the most probable line-width measurement for the LIRGs in our sample and is represented as the red solid-line curve. The solid-line black Gaussian gives the probability distribution of possible measurements of H$_2$ line-width for an unresolved source and was obtained by adding up all the probability distribution for the 94 unresolved sources shown as blue, dotted Gaussian curves. We used this probability distribution and the error on each measurement to compute the likelihood a source is unresolved. The derived estimates that appear lower than the instrumental resolution are sources with poor sampling, and/or low signal to noise. We keep them here because they indicate the error budget in our measurements. \label{NewSigmaDist}}
\end{figure*}

{\bf{Method 2:}} In the second approach we derive the distribution of possible measurements for an unresolved source by looking at both the measured H$_2$ line-widths (FWHM) and the errors associated with those individual measurements. Figure {\ref{NewSigmaDist}} shows the distribution of derived H$_2$ S(1) FWHM, each represented as a gaussian centered on the measured H$_2$ line-width and a width equal to the estimated error on the measurement. Each gaussian is normalized to an area of 1. Adding up all the individual gaussian distributions gives the most probable line-width measurement.
Adding up all the individual gaussian distribution of sources with measured H$_2$ FWHM smaller than this value gives the probability distribution of possible measurements of H$_2$ line-width for an unresolved source. Using this probability distribution and the error on each measurement for the sources we marked as marginally resolved or resolved from the previous technique, we compute the likelihood it is actually unresolved. We find that the sources we marked as marginally resolved and resolved have a probability between $~2\times 10^{-6}$ (for NGC 6240) and  $0.1$  (for VV250) to be unresolved with a median value of 0.01. 

The S(0) line width distribution was more difficult to characterize because of larger error bars on the estimated line-widths. We found different modes when we binned the line-widths distribution using bin sizes of 10, 15, 20 and 25 km/sec. The modes we find range between 421 and 545 km/sec, with an average value of 494 km/sec. The weighted mean of the un-binned distribution of line-widths is also 494 km/sec. We thus used the weighted mean of 494 km/sec as the most likely value for a measurement of an unresolved source. The associated dispersion is 62 km/sec. We find eight galaxies that have marginally resolved S(0) lines: NGC 0828, ESO 255-IG007, ESO 507-G070, IRAS 13052-5711, NGC 5257, CGCG 142-034, NGC 6240 and MCG +04-18-002. The measured S(1) FWHM of these source range between 603 and 781 km/s, with median of  653 km/s, while their S(0) FWHMs range between 611 and 835 km/s with a median of 785 km/s. 

\subsection{Excitation Diagrams, Masses and Temperatures of Molecular H$_{2}$}

\begin{figure*}
  \includegraphics[width=0.49\linewidth]{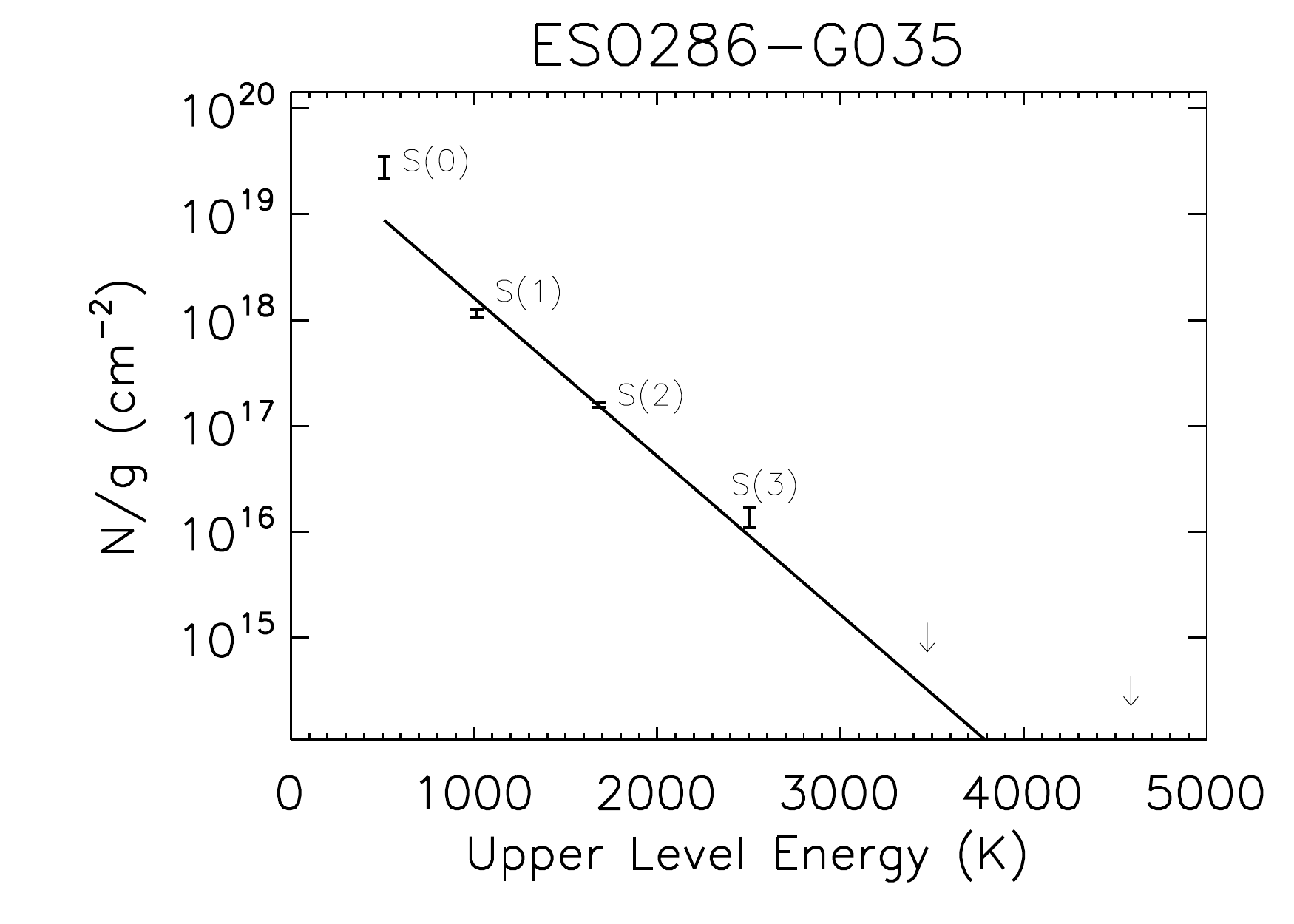}
  \includegraphics[width=0.49\linewidth]{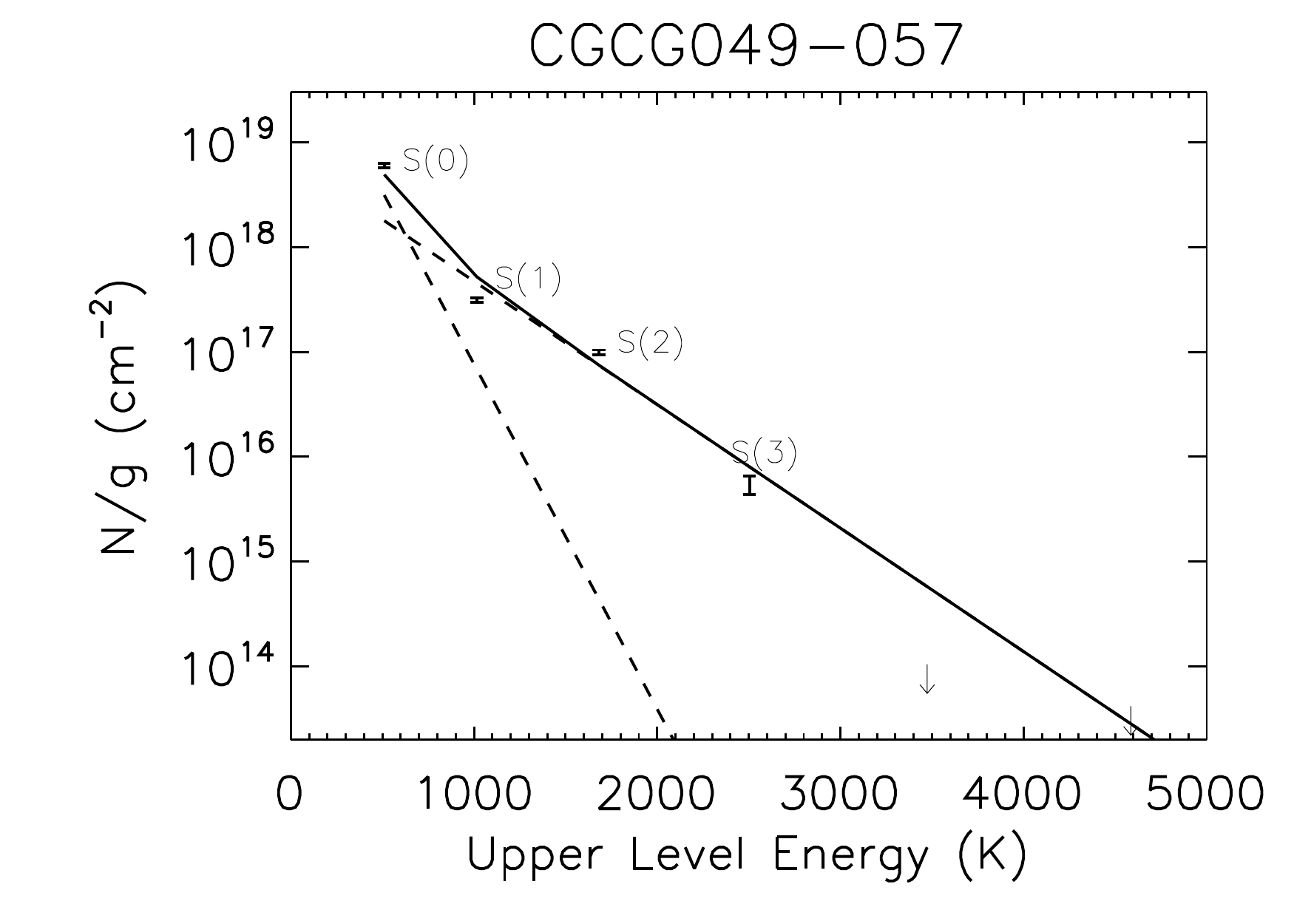}
  \caption{Excitation diagrams. The electronic version and appendix C contain all the excitation diagrams. Here we only show 2 examples of sources well fit by 1 and 2 temperature components. \label{ExDiag}}
\end{figure*}

Figure \ref{ExDiag} presents the excitation diagrams for the sources where we detected at least two of the rotational transitions in the IRS SH, LH, or SL spectra. An excitation diagram is a plot of the column density in the upper level of each transition (N$_{u}$), normalized by its statistical weight (g$_{u}$) as a function of the temperature T$_{u}$ associated with the upper level energy E$_{u}$. We inspect visually all the excitation diagrams to determine if more than one temperature component is needed to model the data. For most sources, two temperature fits are not well constrained, i.e. the masses and temperatures we derive are not the results from a fit, instead they are estimates of four unknown parameters from four emission line fluxes; therefore we cannot provide comparisons of $\chi^2$ as a function of the number of temperature components. We are cognizant of the limitations of this approach but it allows us to qualitatively and consistently compare with other samples of galaxies analyzed in a similar fashion. There are no obvious systemic errors in this method that would erroneously lead to trends between the warm molecular gas properties and the target's morphologies (mergers versus non-mergers) or AGN contribution to the IR emission from their host galaxy. 

Errors on the estimated warm H$_2$ masses and temperatures come from: (1) measurement errors and (2) the assumption that we can describe the data with a simple distribution of one or two temperature components. When we add the H$_2$ emission line flux uncertainties in quadrature we find mass estimates errors on the order of 10-15\%. We assess the second source of error by using three methods to estimate warm H$_2$ masses and temperatures and comparing the results: (1) we use only the S(1) and S(3) lines, as it was done in \citet{hill2014}, (2) we fit the excitation diagrams with detected lines only, and (3) for excitation diagrams where S(0) is not detected, we use the S(0) upper limits as if they were detections and derive an upper limit on the total mass. We find that adding the S(0) line to the excitation diagram fit has the largest impact: the estimated masses become 1.5 larger when we include S(0) measurements. Because LL data often lack the sensitivity to detect the S(0) line, estimates of warm molecular gas masses based on lower resolution IRS data may be systematically lower than the true values \citep{rous2007, hill2014}. 

To compute warm molecular gas masses and temperatures we use the same method as \citet{ogle2010}, Higdon et al (2006), and Roussel et al. (2007). We assume that the gas is in local thermodynamic equilibrium (LTE).  The relation between the observed transitions of the H$_{2}$ rotational levels and the total column density $N_{TOT}$ are given by:

\begin{equation}
  N_{u}~=~g_{u}~N_{TOT}~exp[-E_{u}/(kT)]/Z(T),
\end{equation}

\noindent where $Z(T)$ is the partition function for the J$^{\rm{th}}$ state given by 

\begin{equation}
  Z(T) ~=~\sum _{J} exp(-E_{J}/kT.)
\end{equation}

\begin{equation}  
  T~=~\frac{E_{u2}~-~E_{u1}}{k\times ln(N_{u1}/N_{u2} \times g_{u2}/g_{u1})}
\end{equation}

\noindent where $N_{u}$ is the column density in the upper level of each transition, $g_{u}$ its statistical weight, and $E_{u}$ its energy. The E$_{u}/k$ associated with S(0), S(1), S(2) and S(3) as well as the statistical weights for the lines are given in Table 1. The column density N$_{u}$ is related to the measured flux $F$ emitted in a transition:

\begin{equation}
  N_{u} = \frac{F}{h\nu A}\times \frac{4\pi}{\Omega}
\end{equation} 

\noindent where $A$ is the Einstein coefficient giving the probability for spontaneous emission, $h\nu$ is the transition energy and $\Omega$ is the beam solid angle. The statistical weight is a function of the rotational number $J$, and the spin number $I$, given by:

\begin{equation}
  g_{u} ~=~ (2I~+~1)(2J~+1~).
\end{equation}

The mass can be determined from $N_{TOT}$ with the source size derived from the size of the spectral extraction region: $4.7"~ \times ~4.7"$ at the distance of each source (e.g. 2 kpc for a source at a redshift of 0.02) . 

We estimate that the molecular H$_{2}$ gas has mass-averaged effective temperatures between 92 and 650 K, and the sums of the individual mass components are between $10^6$ and $10^9 ~\rm{M} _{\odot}$. The derived values are estimates because the gas may have a distribution of temperatures and may originate from regions with different physical properties within the few kpc region probed by our spectra.  However, such estimates are useful because they provide comparisons with similar analysis done on normal galaxies and ULIRGs. This analysis provides a practical way to characterize the true underlying gas temperatures.  With the exception of two sources for which we find the coldest temperatures at 92 K and 97 K, the coldest components are above 100 K.
The LIRGs IRAS 19542+1110 and  ESO 339-G011 are the only two sources with total warm molecular gas masses greater than 1$\times 10^9 M_{\odot}$. 

\begin{figure*}[h!] 
  \includegraphics[width=0.47\linewidth]{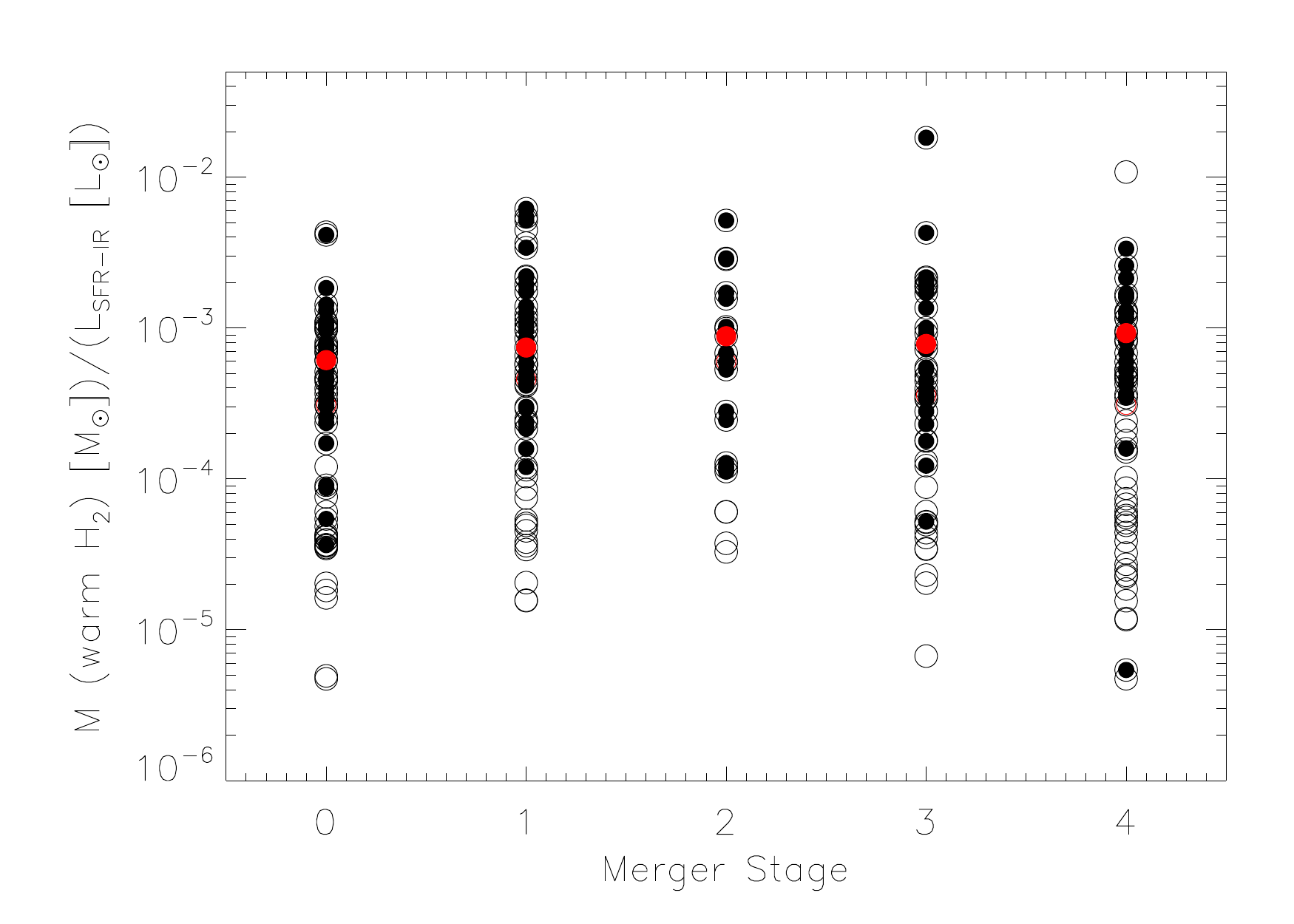}
  \includegraphics[width=0.49\linewidth]{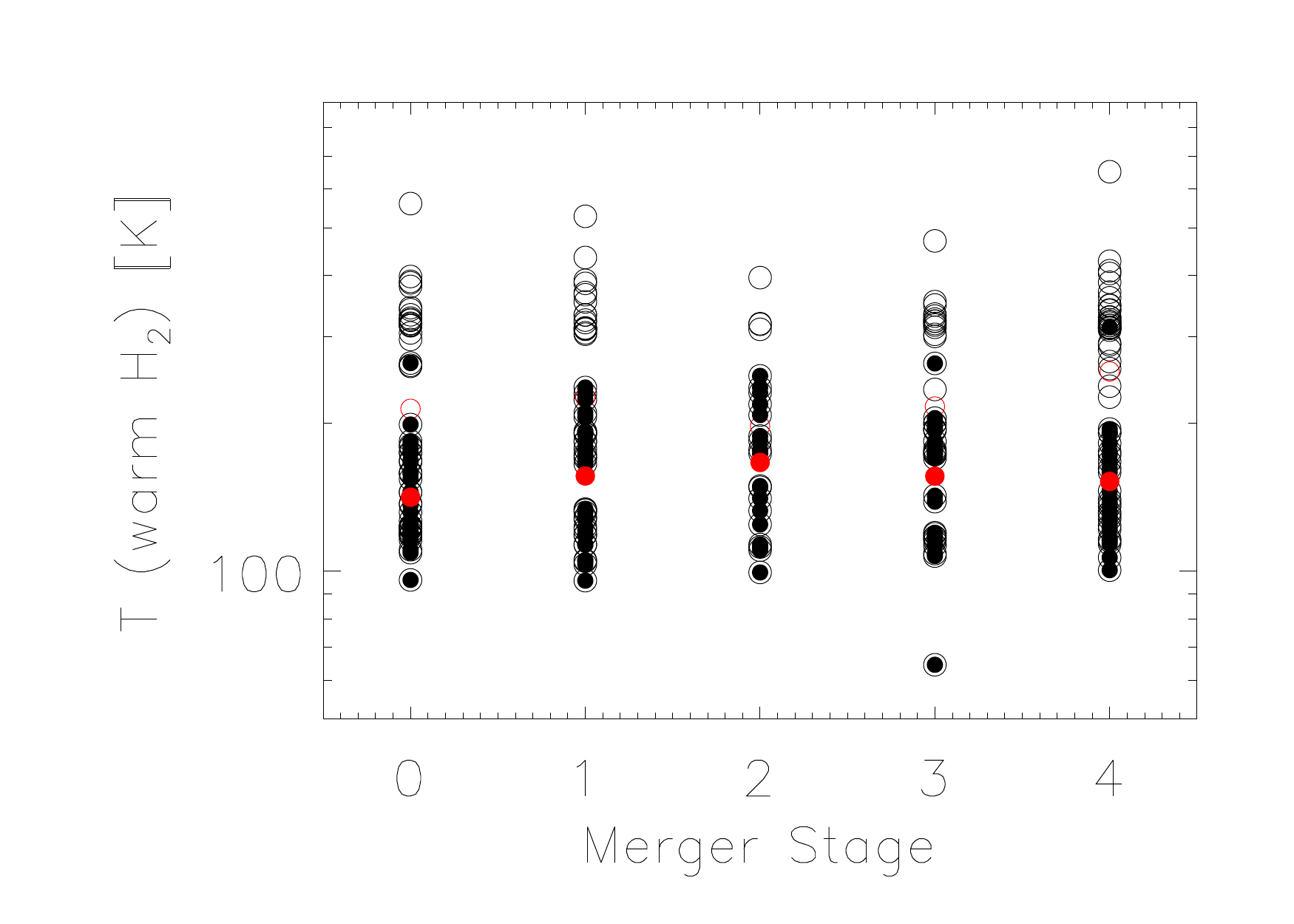}
  \caption{H$_2$ masses normalized by the L$_{IR}$ (left) and temperatures (right)  versus merger stage. Filled circles use S(0) detections, empty circles use S(0) upper limits. Red circles give the average value for each merger state. }
  \label{MATvsMerg}
\end{figure*}

The estimated masses and temperatures of the warm molecular gas emitting in the MIR are presented in Table \ref{FitHMT}. The median temperature for sources that are well fit by one temperature component is $\sim$300 K. The median of the total masses is $\log \rm{M}_{\odot} = 7.2 \pm 0.5 $.
Figure \ref{MATvsMerg} shows the distribution of total gas masses normalized by  L$_{IR}$ and temperatures (mass - averaged temperatures for sources where we required at least two temperatures to fit the observed rotational lines) as a function of merger stage.

Out of the 214 nuclei for which we determine excitation diagrams, 103 require two temperature components. The average masses of warm molecular gas for objects in the non-interacting (n), early-mergers (em), and late stage mergers (m) are 8.3, 8.9, and 12.5 $~\times ~10^7~\rm{M}_{\odot}$ respectively.
The average temperatures of warm molecular gas we derive for sources in each of these respective interaction stages are 242, 243, and 277 K respectively.

\subsection{Ortho to Para Ratios}
To understand the impact that AGN and gravitational interactions have on the warm molecular gas we also look at the relative strengths of emission from states with odd total angular momentum to emission from states with even angular momentum (ortho to para ratios: OPR, appendix B). 
The H$_{2}$ molecule consists of two covalently bound hydrogen atoms. Because its center of mass is the same as that of its electrical charges it does not have a permanent dipole moment. Therefore H$_{2}$ cannot transition from its ortho state (i.e. odd total angular momentum number J) to its para state (i.e even J). The value of the OPR is related to the history of the molecular cloud. If the gas is in local thermodynamic equilibrium (LTE), a Boltzman distribution describes the populations at each energy level, and the OPR is a known function of temperature. At typical 300 K temperatures, the OPR is 3. Measuring OPRs that are lower than 3 may suggest that the gas is not in LTE or that the gas is thermalized at a temperature lower than 300 K \citep[see][for a similar discussion about H$_2$O OPR]{fla2013}. The thermalization time for H$_{2}$ is of the order of 5000 years so the OPR traces the temperature of the matter with which the H$_2$ has last thermalized beyond that time. Other interpretations for measuring OPRs that are different than 3 are related to extinction and/or the presences of multiple components along the line of sight \citep[][and ref. within]{fla2013,rous2007}. Following \citet{rous2007} we first determined which sources have non-LTE OPRs using the apparent excitation temperatures derived from the S(0), S(1), S(2) and S(3) lines. More details about how the OPRs were estimated are provided in Appendix 1 and Figure \ref{OPPlots} shows several illustrative examples.

We find 30 LIRGs with OPR values that appear incompatible with LTE conditions. Among these sources are Arp~220, NGC~3690, and NGC~0992. To test if we observe OPRs incompatible with LTE because of dust we assume that the amount of extinction is proportional to the silicate strength measured by \citet{stir2013} using the method described in \citet{spoon2007}. The silicate strength at 9.7 $\mu$m is defined as: $log(f_{9.7 \mu m}/C_{9.7 \mu m})$ ~where~ $f_{9.7 \mu m}$ is the flux measured at 9.7$\mu$m and $C_{9.7\mu m}$ is the continuum flux in the absence of the absorption feature \citep[see:][]{stir2013,spoon2007}. The silicates strength of Arp~220 is one of the highest in the GOALS sample and NGC~3690 has a silicate strength of -1.65 $\pm$ 0.02 which is relatively high for LIRGs though not one of the highest \citep{stir2013}. Extinction might be the reason why we observe an OPR incompatible with LTE in those two LIRGs. NGC~0992, however, does not appear to be heavily obscured: \citet{stir2013} report a silicate strength of 0.05 $\pm$ 0.04 for this source. It thus seems that at least in the case of NGC~0992, either the silicate absorption does not originate in the same region as the warm H$_{2}$ or, as is the case for several normal galaxies, the gas is not yet thermalized \citep{rous2007}. This could mean (1) that we are observing this galaxy at a peculiar moment in its evolution, which would be surprising given the low critical densities of H$_{2}$ rotational transitions or (2) that given the large size of our beam we are recovering emission from regions with heterogenous physical conditions, with gas at different temperatures. 

\begin{figure*}
  \includegraphics[width=0.44\linewidth]{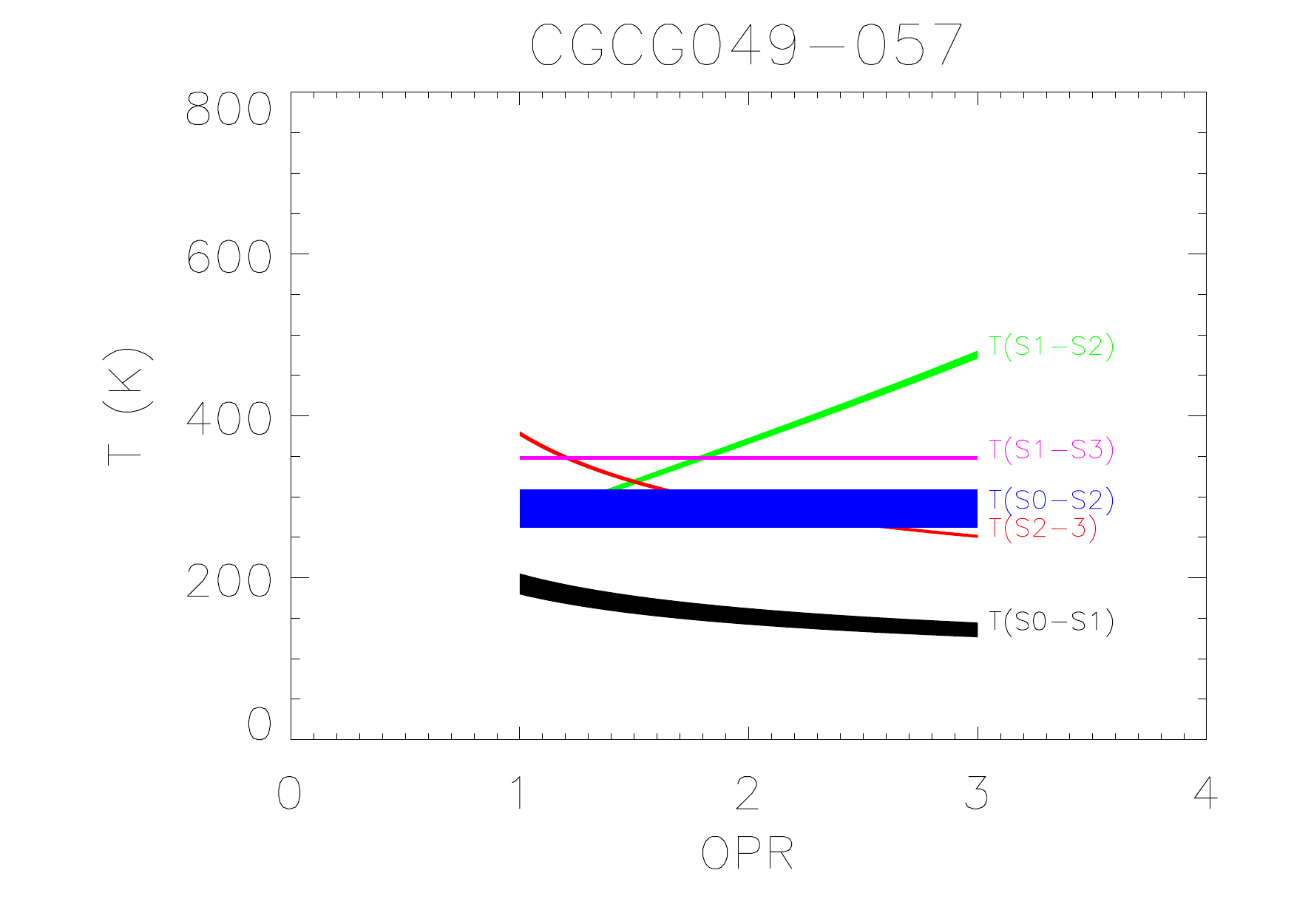}
  \includegraphics[width=0.44\linewidth]{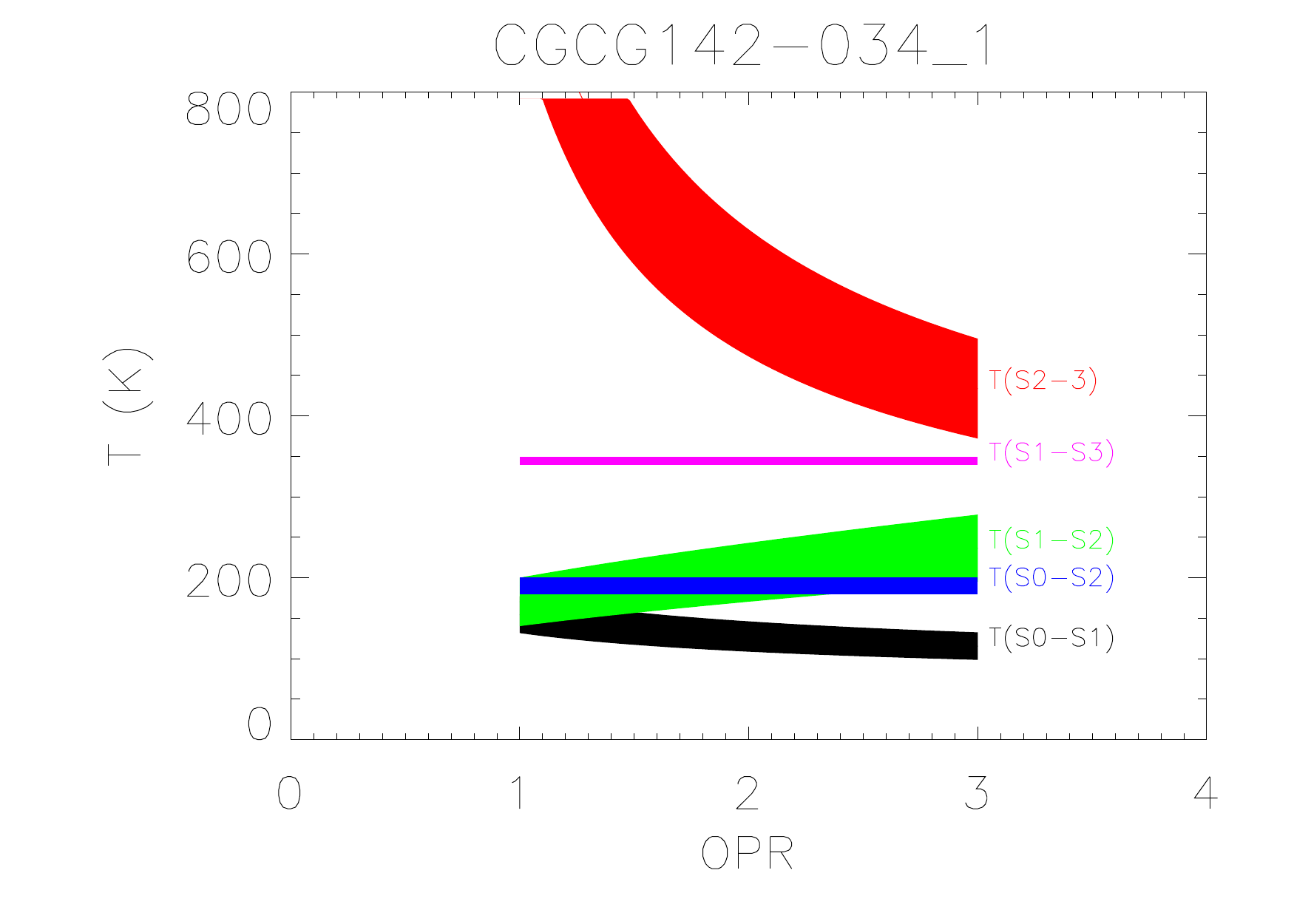}\\
  \includegraphics[width=0.44\linewidth]{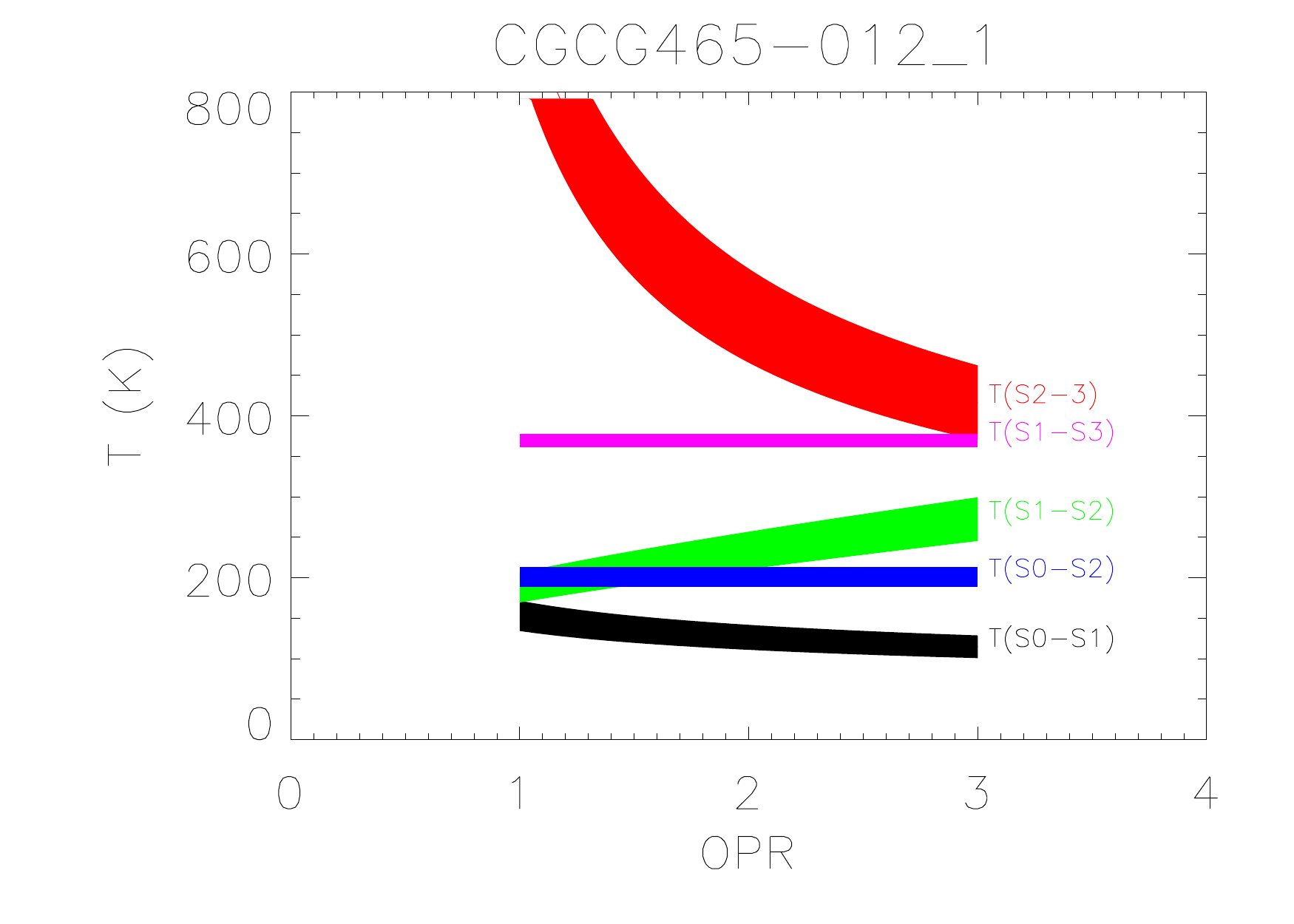}
  \includegraphics[width=0.44\linewidth]{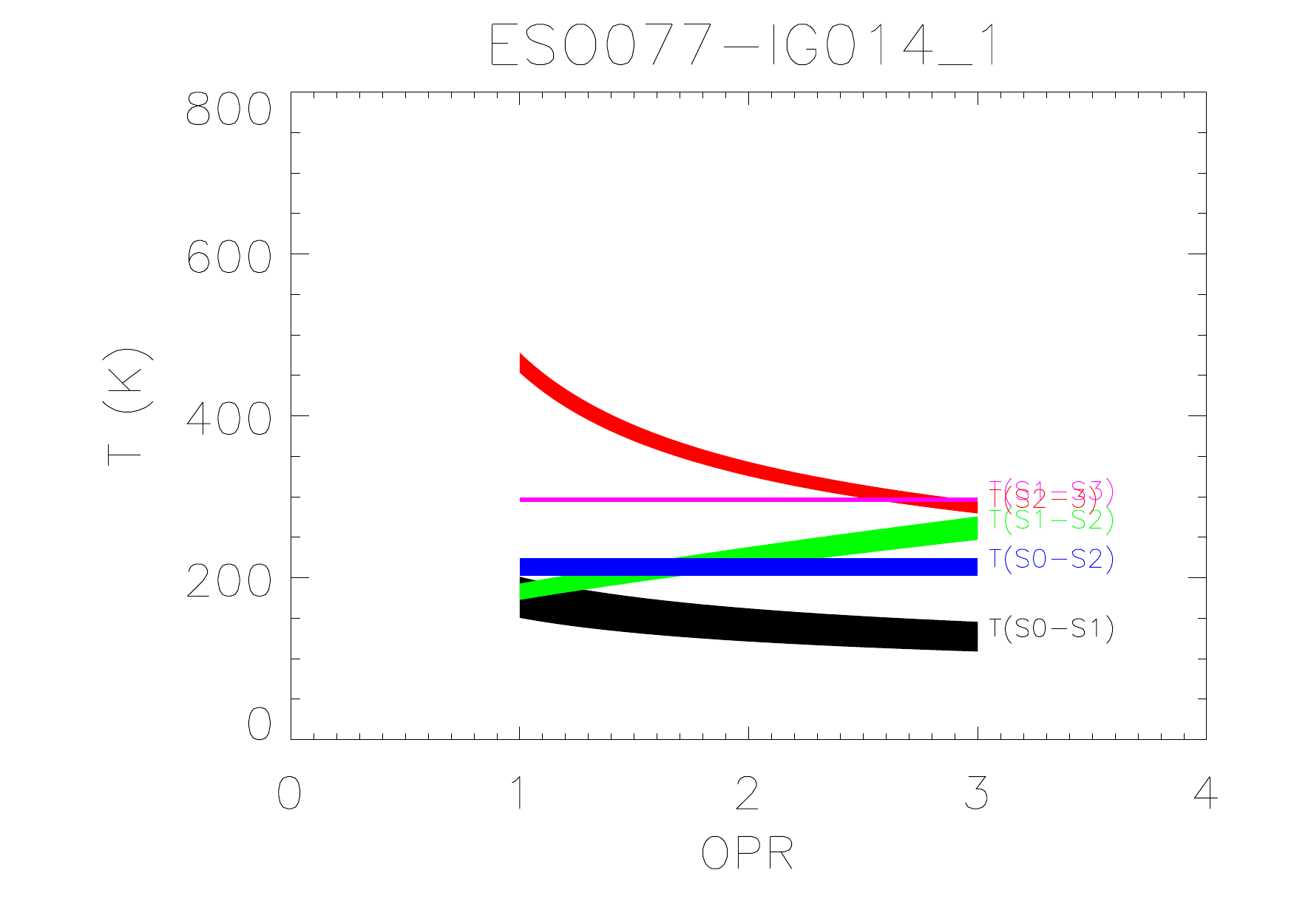}\\
  \includegraphics[width=0.44\linewidth]{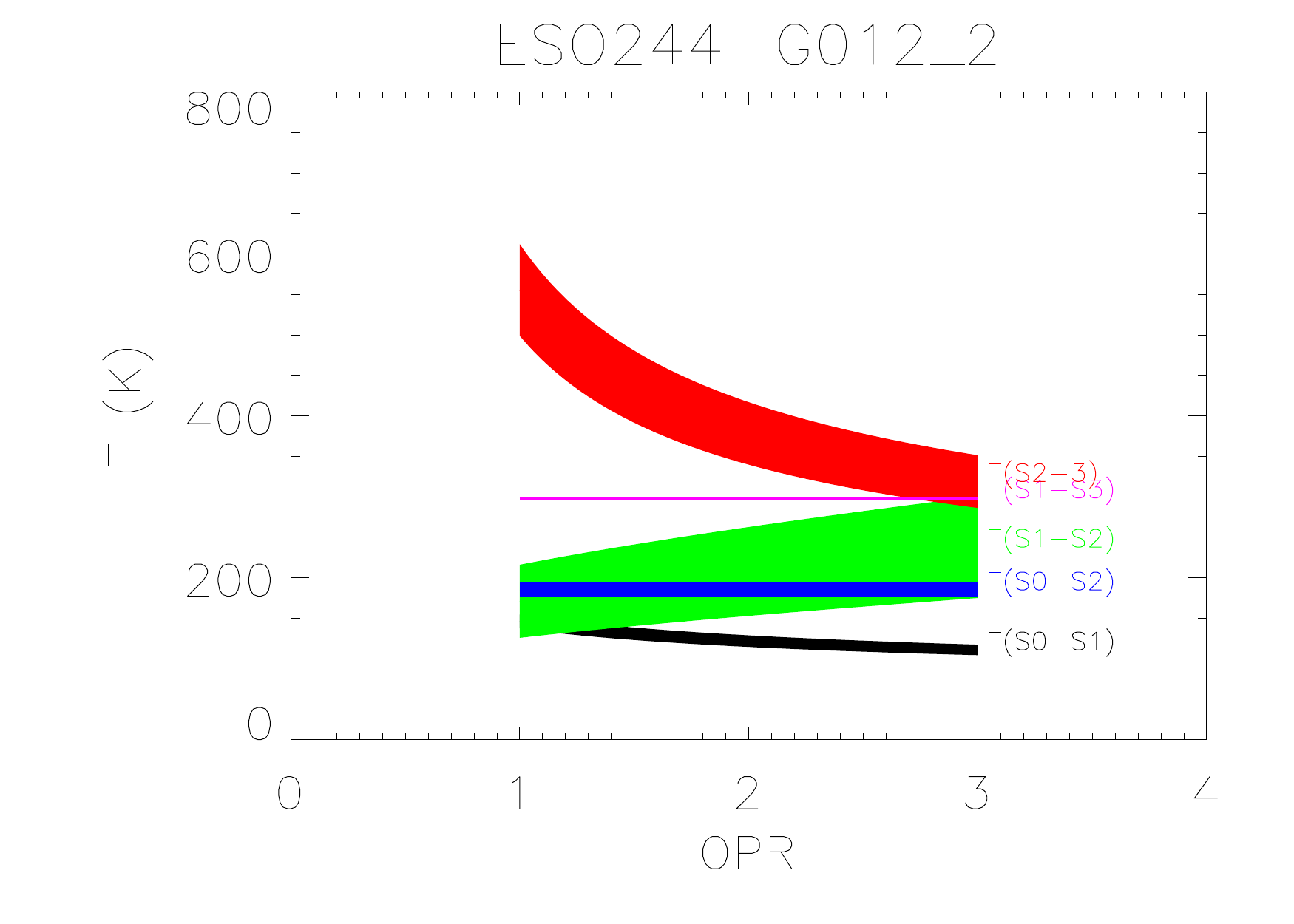}
  \includegraphics[width=0.44\linewidth]{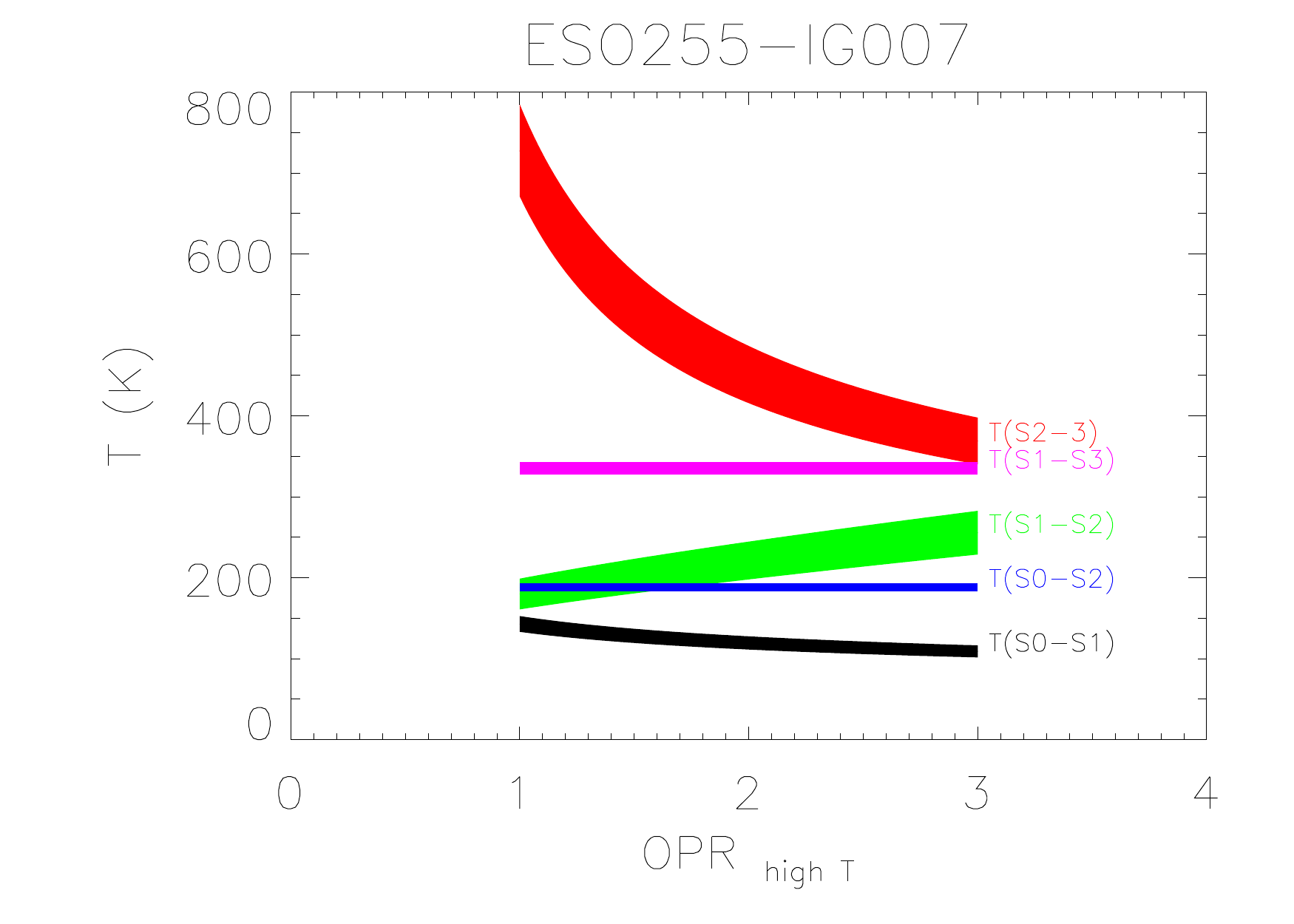}\\
  \includegraphics[width=0.44\linewidth]{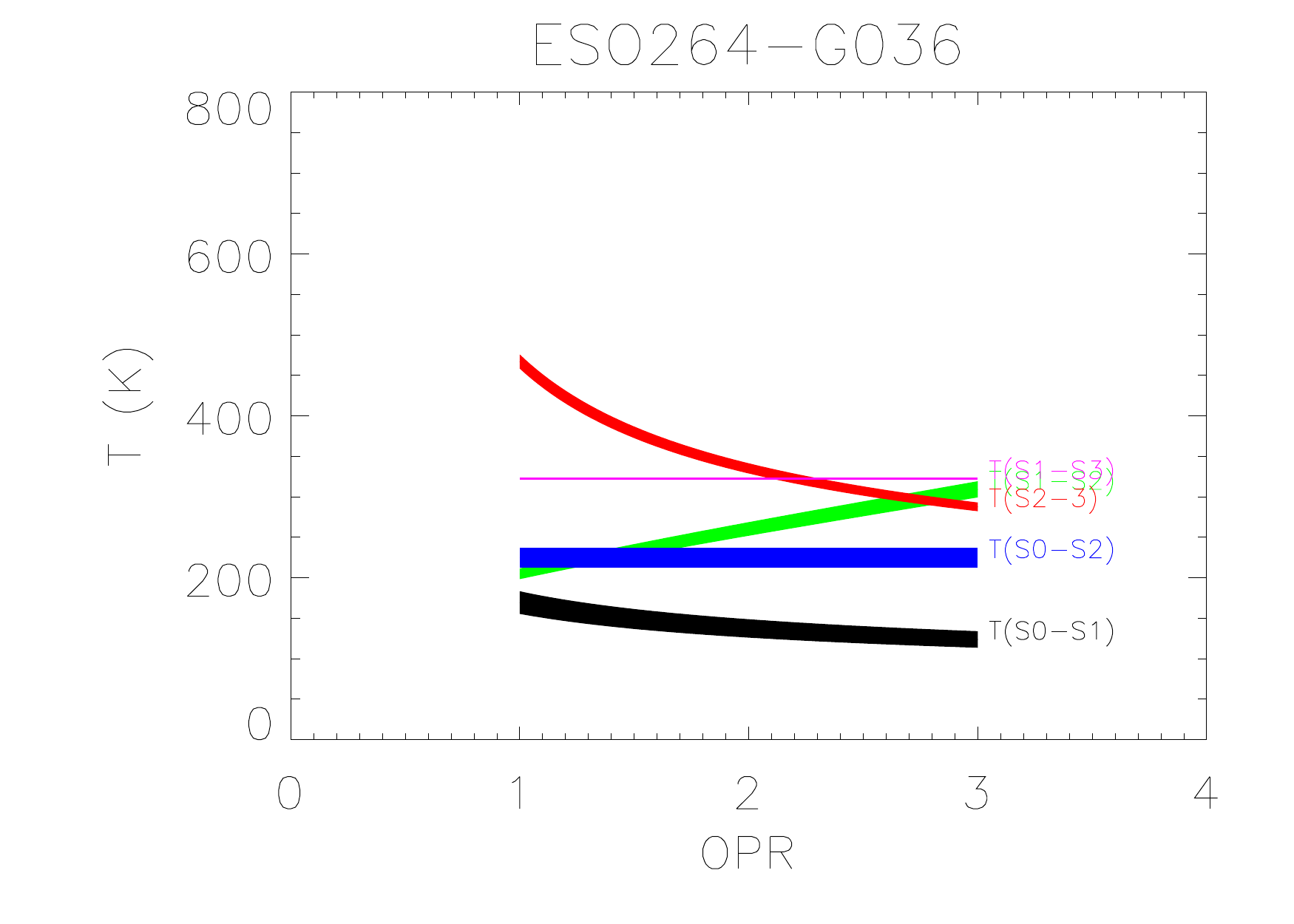}
  \includegraphics[width=0.44\linewidth]{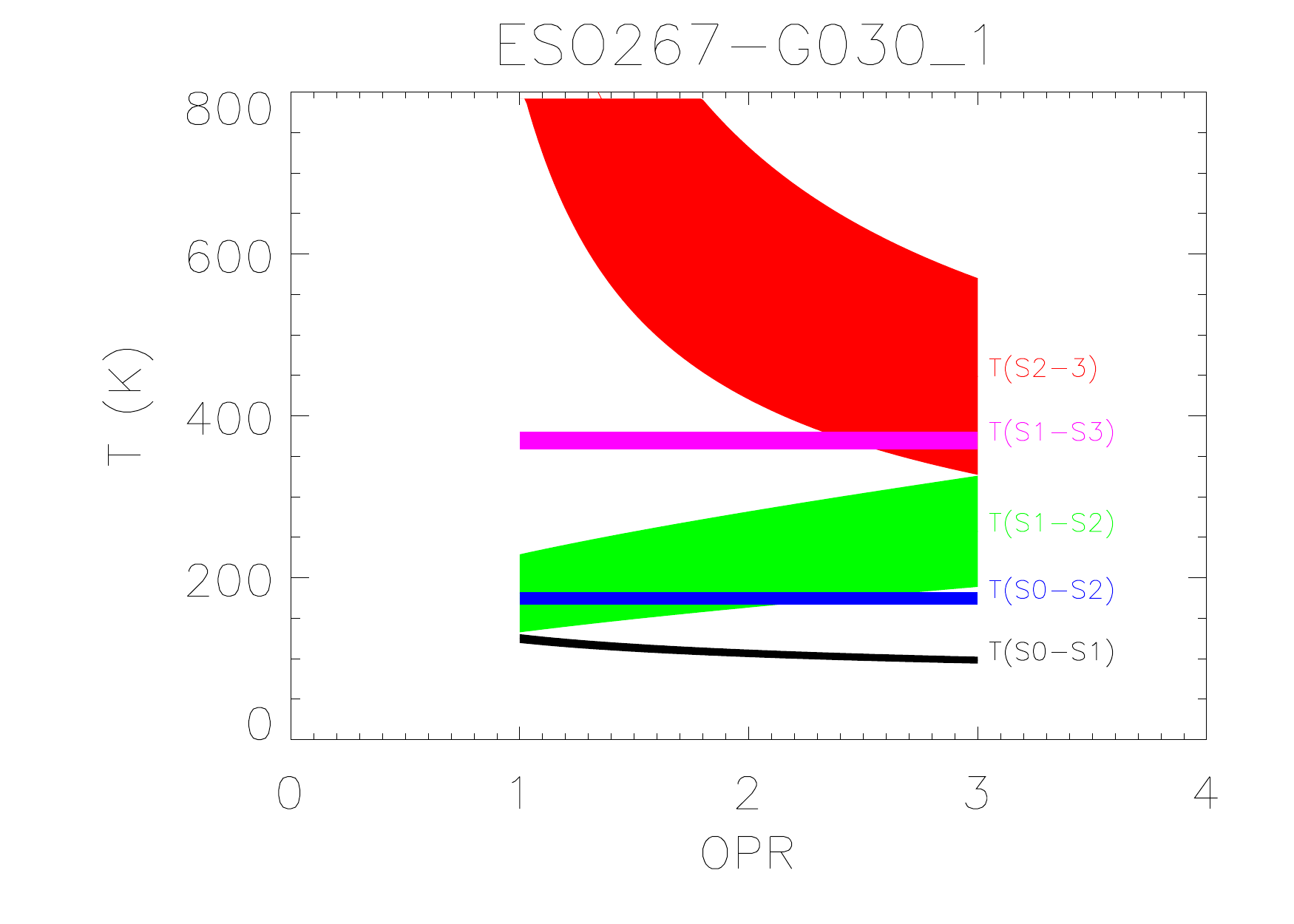}
  \caption{Temperatures as a function of OPR, the thickness of the lines represent the errors associated with the determination of temperature. Here we provide several examples representative of the observed range of properties.  \label{OPPlots}}
\end{figure*}

\subsection{H$_{2}$ Excitation Conditions} 
We seek to determine if the detected warm H$_2$ gas is emitted primarily in photon-dominated regions (PDRs). MIR rotational lines are rapidly thermalized and hence they provide few diagnostics with which to determine the excitations mechanisms. However some conjectures can be made by comparing the emission of the warm H$_2$ gas and that from other MIR coolants known to originate in PDRs such as [Si~{\sc{II}}] and PAH emission  \citep{kauf2006,cluver2010}.

About 80\% of the LIRGs nuclei in GOALS have MIR colors (using continuum fluxes measured in the IRAC 3.6, 4.5, 5.8 and 8 $\mu$m bands and 15/5.5 $\mu$m fluxes) that are consistent with a PDR origin for the observed MIR dust continuum \citep{petric2011}.  Since molecular gas and dust are closely connected we also compare the warm molecular emission properties with PDR models. We use the PDR models of \citet{kauf2006} who calculate the [Si~{\sc{ii}}] and H$_{2}$ S(0), S(1), S(2) and S(3) pure rotational line emission arising from PDRs in massive star-forming regions. 

\citet{kauf2006} computed simultaneous solutions for the chemistry, radiative transfer, and thermal balance in PDRs and assumed that in the outer layers the PDR contains singly ionized carbon, silicon and iron with a temperature greater than 100 K. The observed [Si~{\sc{ii}}] emission at 35 $\mu$m is thought to come from this outer layer of PDR. The rotational  H$_2$ transitions (seen in MIR) and the ro-vibrational H$_2$ lines (seen in NIR) are thought to come from a deeper layer where the H~{\sc I} /H$_{2}$ transition is supposed to occur. The resultant emission is a function of the PDR density $n$ and of the incident FUV ( 6.0 eV $\leq ~h \nu~\leq ~$13.6 eV) flux. The FUV incident radiation is described in terms of $G_{0} ~: ~1.6~\times 10^{-3}$ ergs cm$^{-2}$, value comparable to estimates of the local interstellar field in the Milky Way. 

\begin{figure*}[h!]
  \includegraphics[width=0.4\linewidth]{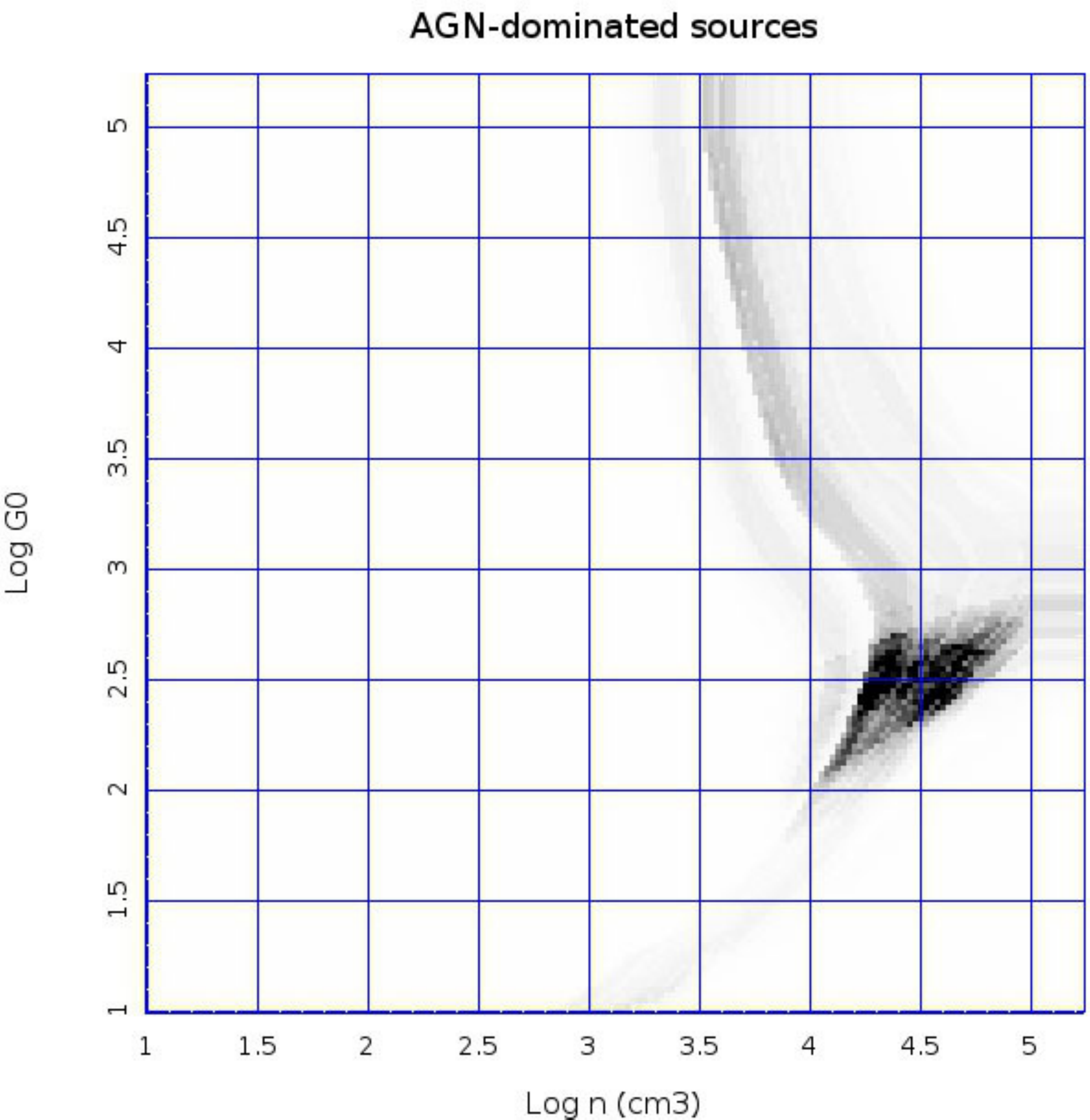}
  \includegraphics[width=0.4\linewidth]{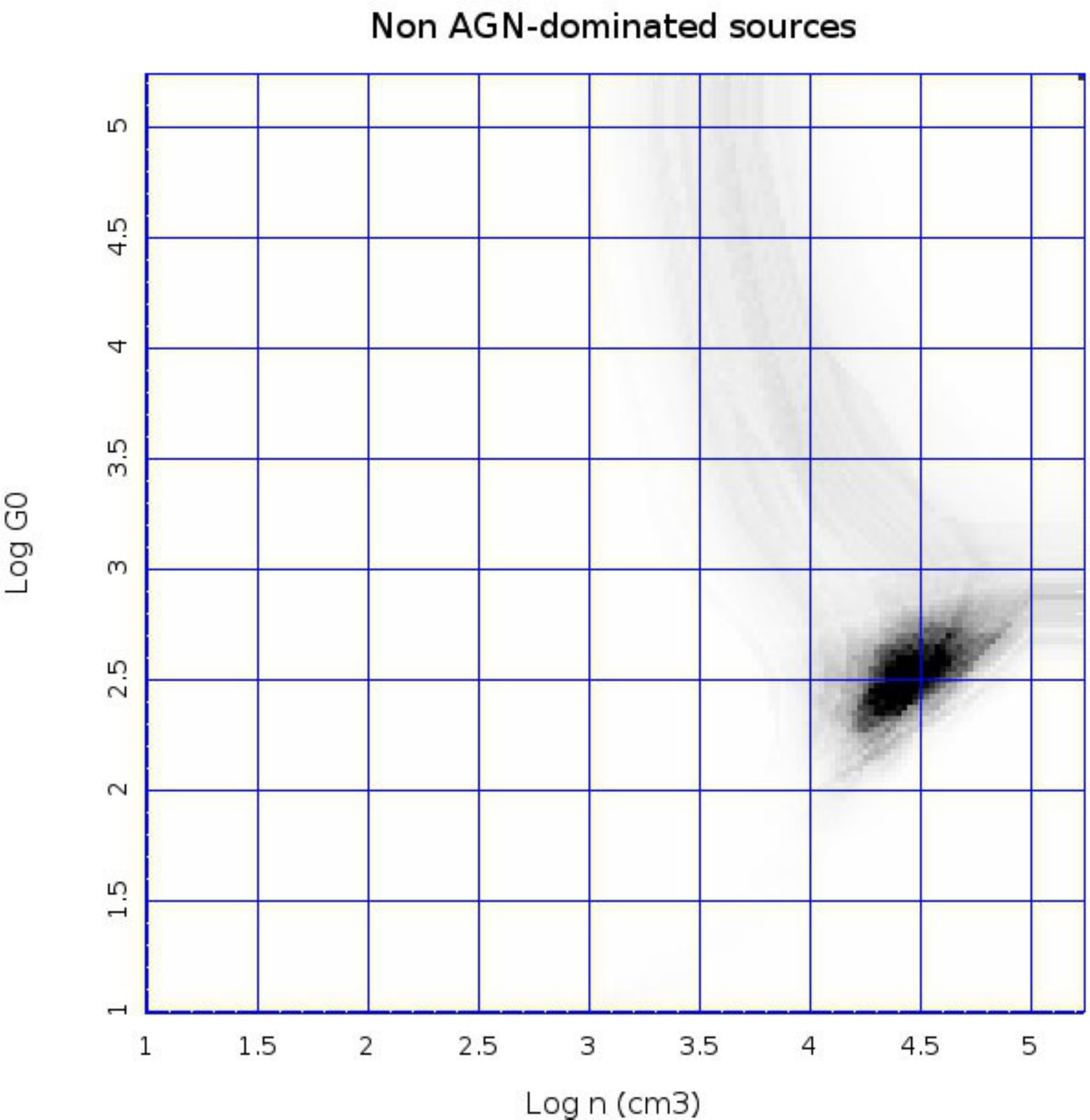}\\
  \includegraphics[width=0.4\linewidth]{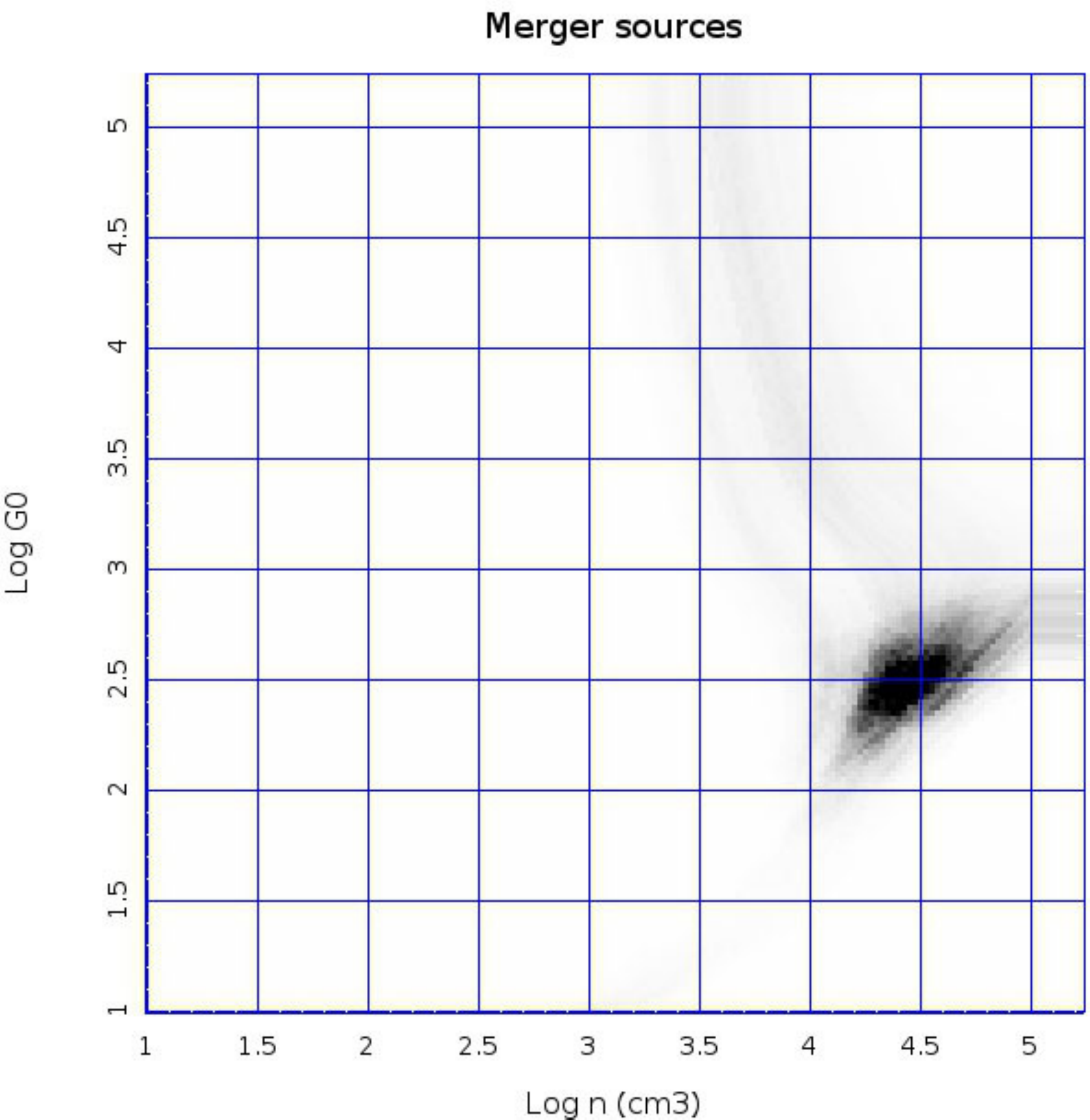}
  \includegraphics[width=0.4\linewidth]{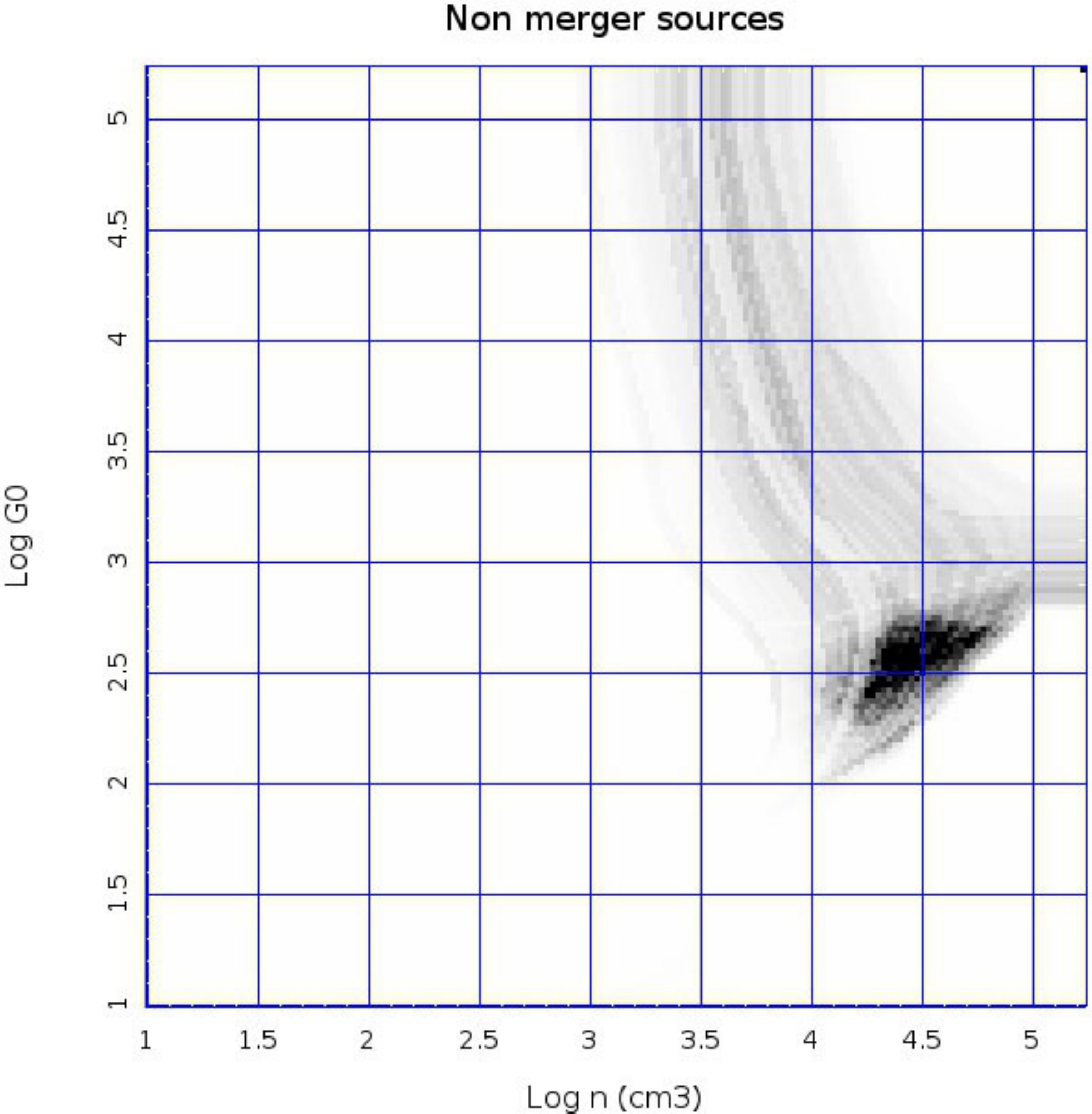}\\
  \includegraphics[width=0.4\linewidth]{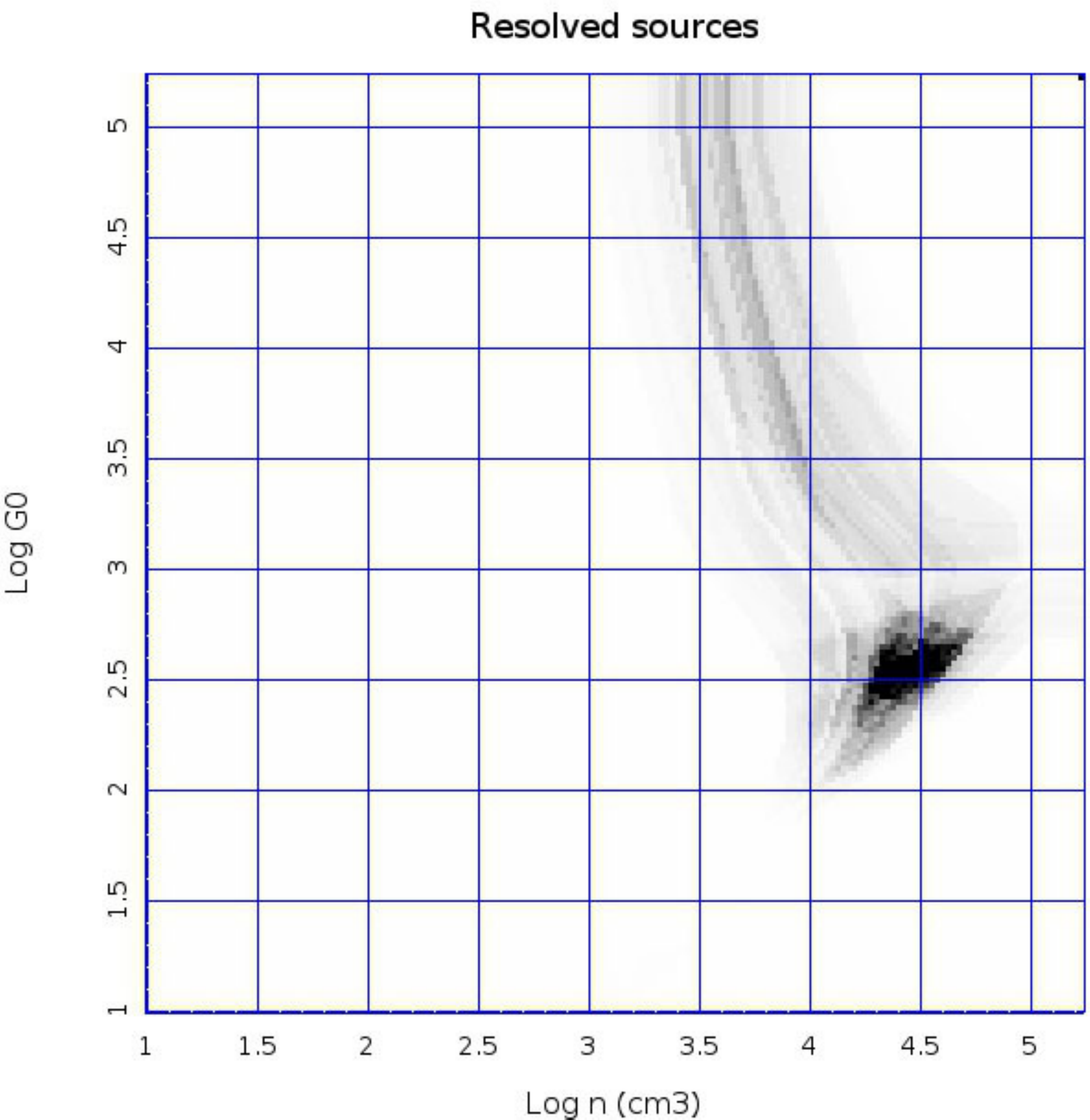}
  \includegraphics[width=0.4\linewidth]{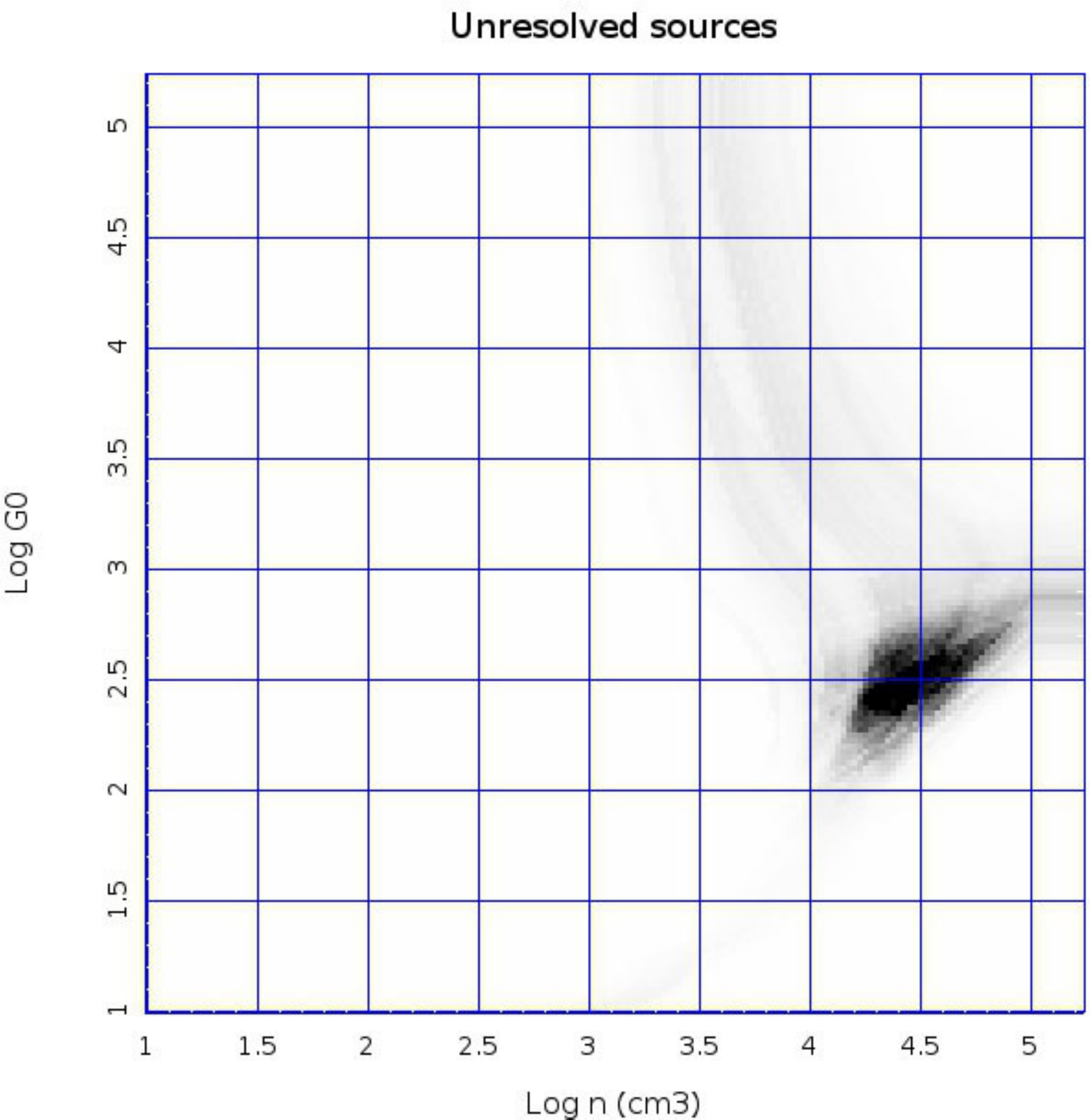}
  \caption{Grayscales showing the most probable region of the (n, G$_0$) space for {\bf{top:}} LIRGs with AGN producing more than 50\% of the MIR emission versus LIRGs in which processes associated with star formation contribute to most of the observed MIR, {\bf{middle:}} LIRGs showing signs of gravitational interactions versus those without and { \bf{bottom:}} LIRGs with relatively broad H$_2$ velocity profiles. These estimates are based on their combined S1/S0, S1/S2, and S1/SiII ratios and on the models of \citet{kauf2006} models and only include detections. See text for details. \label{NFkauf}}
\end{figure*}

Figure {\ref{NFkauf}} shows the combinations of $G_{0}$ and density $n$ possible for the observed [Si~{\sc{ii}}] and H$_2$ flux ratios.  We find that our measurements can best be modeled by a PDR/HII model with $\log(G_0)$ values between 2.3 and 2.8 and $\log(n) $ between 4.1 and 4.6.  In Figure {\ref{NFkauf}} we test for systematic differences between: AGN-dominated sources and non-AGN dominated sources, mergers and isolated galaxies,  targets with resolved H$_2$ emission and those without.  We find that AGN-dominated sources appear to have a wider range of possible $G_{0}$ and density $n$ conditions, similar to the work of Lambrides et al. (2018 submitted) who analyze a sample of 2200 active galaxies observed by Spitzer IRS and find that AGN-dominated galaxies have a wider range of dust-grain properties. 

For two sources, III Zw 35 and NGC 1961, the S(1) to [Si~{\sc{ii}}] ratios are higher than what can be expected from the PDR models. Those sources are not detected in [O~{\sc{iv}}] suggesting that if any AGN were present, its MIR emission does not contribute to the overall MIR luminosity of the galaxy because either it is obscured or the AGN is MIR-faint compared to MIR emission from star-forming regions.  III Zw 35  was  characterized as a LINER by \citet{car1999} on the basis of the \citet{vei1987} classification. Based on its detection in the 2-10 keV X-ray band with Chandra at a level of $7 \pm 2 \times 10^{-15}$ erg/s/cm$^2$, and X-ray hardness it was characterized as a Compton-thick source by \citet{gonz2009}.  Similarly NGC 1961 is part of the \citet{car1999} LINER catalogue, and was detected with the Einstein X-ray observatory  in the 0.2-4.0 keV band at  a level of $7 \pm 3 \times 10^{-13}$ erg/s/cm$^2$ \citep{fab1992}. The X-ray detections of these LINER galaxies suggest that an XDR model might be more suitable to the data here \citet{meij2007}, and that the sources may be heavily obscured.  Both objects are among a small percentage (4\%) of LIRGs where we detect the H$_2$ S(5) lines and their excitation diagrams suggest that the warm molecular gas is not well described by a single temperature.

\section{Discussion}
Nearby LIRGs appear to have higher H$_2/{\rm{PAH}}$ flux ratios than normal star-forming galaxies \citep{stir2014}.  \citet{stir2014} use simultaneous fits to the dust and gas emission and continuum features in the low resolution IRS data to determine that around 10\% of LIRGs have H$_2$ emission that is not consistent with PDR models and so could be instead excited by shocks from powerful starbursts or AGN. They also find that L$_{H_2}/\rm{L}_{PAHs}$ ratios are positively correlated to L$_{H_2}$ and are not correlated with the silicate optical strengths, unlike in ULIRGs \citep{zak2010}. \citet{stir2014} explain this difference between LIRGs and ULIRGs by suggesting that the excitation mechanisms of warm H$_2$ outside photo-dissociation regions, i.e. shocks and AGN are less common in LIRGs than they are in ULIRGs. 
Here we use data at higher spectral resolution to (1) investigate the kinematics of the warm H$_2$, (2) estimate H$_2$ masses and temperatures, and (3) use [Si~{\sc{ii}}] and OPR analysis to look at the gas excitation conditions. We discuss theoretical predictions that (1) mergers lead to inflows of gas toward the center and (2) AGN impact the distribution of central molecular gas, through outflows, shocks, and an abundance of cosmic rays.  

To compare our measurements of LIRGs to those of normal galaxies and ULIRGs, mergers to non-mergers, and pure starbursts to AGN dominated sources, we perform statistical tests that rely on both detections and upper limits to determine the probability that two samples are drawn from the same population. The interaction stages we use for this analysis are derived from those used in
\citet{petric2011}, and re-analyzed with help from the work of \citet{stir2013} and \citet{lar2016} (but see also \citet{pet2014, haan2011}). We use three broad categories: non-mergers, mergers, and early-mergers. We call a LIRG AGN dominated if its 6.2 PAH equivalent width is less than 0.27 and the high ionization [Ne~{\sc{v}}] is detected or the ratio of [O~{\sc{iv}}] /[Ne~{\sc{ii}}]  is larger than 1.75 \citep{armus2007,vei2009b,petric2011,stir2013}. 

We used the Astronomy SURVival analysis (ASURV) statistical package \citep{fn1985,isobe1986}. When we compare flux ratios or fluxes with upper-limits we present the average probability that the two samples are drawn from the same population using the following 5 non-parametric tests: (1) the Gehan Generalized Wilcoxon Test using permutation variance, (2) Gehan's generalized Wilcoxon test using hypergeometric variance, (3) Logrank test, (4) Peto \& Peto Generalized Wilcoxon Test and (5) Peto \& Prentice Generalized Wilcoxon Test. When we compare kinematics or dust gas masses we use the Kolmogorov-Smirnov (KS) statistical test. The numbers we provide are: how different two samples are based on their cumulative distributions and the statistical significance of their difference based on the probability that they are drawn from the same population. 

We compare our fluxes and luminosities with those of normal galaxies and ULIRGs. Comparisons using the ratios of S(0) to IR and S (1) to IR are shown in table \ref{AsurvTest}. We find that the most significant differences are between normal \citep[nearby, non-merging, galaxies discussed in][]{rous2007} and LIRGs, as well as between LIRGs and ULIRGs \citep{hig2006}. The sources investigated by \citet{rous2007} are at closer distances and contain fewer sources where the AGN dominates  the IR emission than the LIRGs in our sample. Because the normal galaxies presented in \citet{rous2007} are closer than the GOALS LIRGs, and because fewer sources in the \citet{rous2007}  sample are mergers, the IRS spectra probe regions of different sizes and that may experience different processes. 

\subsection{Kinematics as a function of interaction stage, and the AGN contribution to the L$_{IR}$ }

One of the goals of this work is to estimate if we can identify the kinematic signatures that gravitational interactions and AGN leave on the ISM on kpc scales.  To do this we separated the sources in two groups: 194 LIRGs with unresolved H$_2$ lines and 27 sources with resolved and marginally resolved H$_2$ lines (See section 3.1.2) to determine if they are resolved simply because of geometry, or if they have different H$_2$ properties. While LIRGs with resolved H$_2$ lines appear to have more and hotter gas the results are not highly statistically significant. Table \ref{KStab} summarizes the statistics presented here.

To test if the H$_2$ line broadening may come from rotation of a highly inclined massive galaxy, we visually inspect the 27 sources with resolved and marginally resolved H$_2$ line profiles to see how many appear highly inclined. However, because most sources with broader profiles are mergers and we lack higher resolution spatial and kinematic information, we cannot properly determine the inclination of the inner few kpc probed by the IRS spectra. ESO353-G020, UGC03351, ESO507-G070, and NGC 7771 could have broader profiles because of rotation: their profile widths are consistent with what we would expect for a spiral of inclination of more than 30$^{\circ}$, and a mass to light ratio of order 3.8.

Differences between the H$_2$ S(1) to [Si~{\sc{ii}}] flux ratios of resolved and not-resolved sources would indicate differences in excitation conditions either because of different radiation fields strengths, spectral shapes or overall densities. The marginally resolved sources appear to have above average H$_2$ luminosities relative to the 6.2 $\mu$m PAH emission and to the [Si~{\sc{ii}}] emission. Our small sample of resolved sources makes it difficult to extract statistically significant conclusions. However we provide this analysis for completeness.

\citet{gui2012} find resolved H$_2$ lines in radio galaxies, \citet{das2011} find them in optically selected AGN, and \citet{ogle2012} find them in z$\sim 2$ radio galaxies. Here we wish to test if the AGN at the cores of LIRGs also impact their surrounding ISM. From our sample of LIRGs with broad H$_{2}$ lines, only two sources show detectable [Ne~{\sc{v}}] at 14.3 $\mu$m emission. However, these objects show low [Ne~{\sc{v}}] /[Ne~{\sc{ii}}] ratios of 0.08 and 0.05, with an average AGN contribution to the total IR luminosity ($\sim 8 - 1000 \mu$m) of 13\% which is the same as the mean of the entire sample of LIRGs \citep{petric2011}. The S(1) line-widths we measure for LIRGs are lower than those observed in powerful radio galaxies or ULIRGs. 

Supermassive black holes at the centers of spheroidal galaxies can supply more energy to the galaxy than the binding energy of  the galaxy, even when they grow slowly and have a low feedback efficiency \citep{hop2006}. In simulations, AGN in gas rich mergers produce molecular gas outflows; when the outflows have components on the direction of our line of sight, they can be observed as kinematic features such as broad and/or asymmetric emission-line profiles \citep{nar2006}. Assuming that the viewing geometries of ULIRGs and LIRGs are not statistically different we would expect that the fraction of ULIRGs with AGN and broad H$_{2}$ lines should be similar to that of LIRGs, yet we find relatively fewer LIRGs with broad H$_2$ profile. Fewer LIRGs than ULIRGs have AGN that contribute significantly to their host IR, so our observation could be low number statistics or suggest that the feedback efficiency is higher in ULIRGs than it is in LIRGs.

\begin{figure*}[h!]
  \includegraphics[trim=-0.2cm 0cm 0.2cm 0cm, clip=true, width=0.9\linewidth]{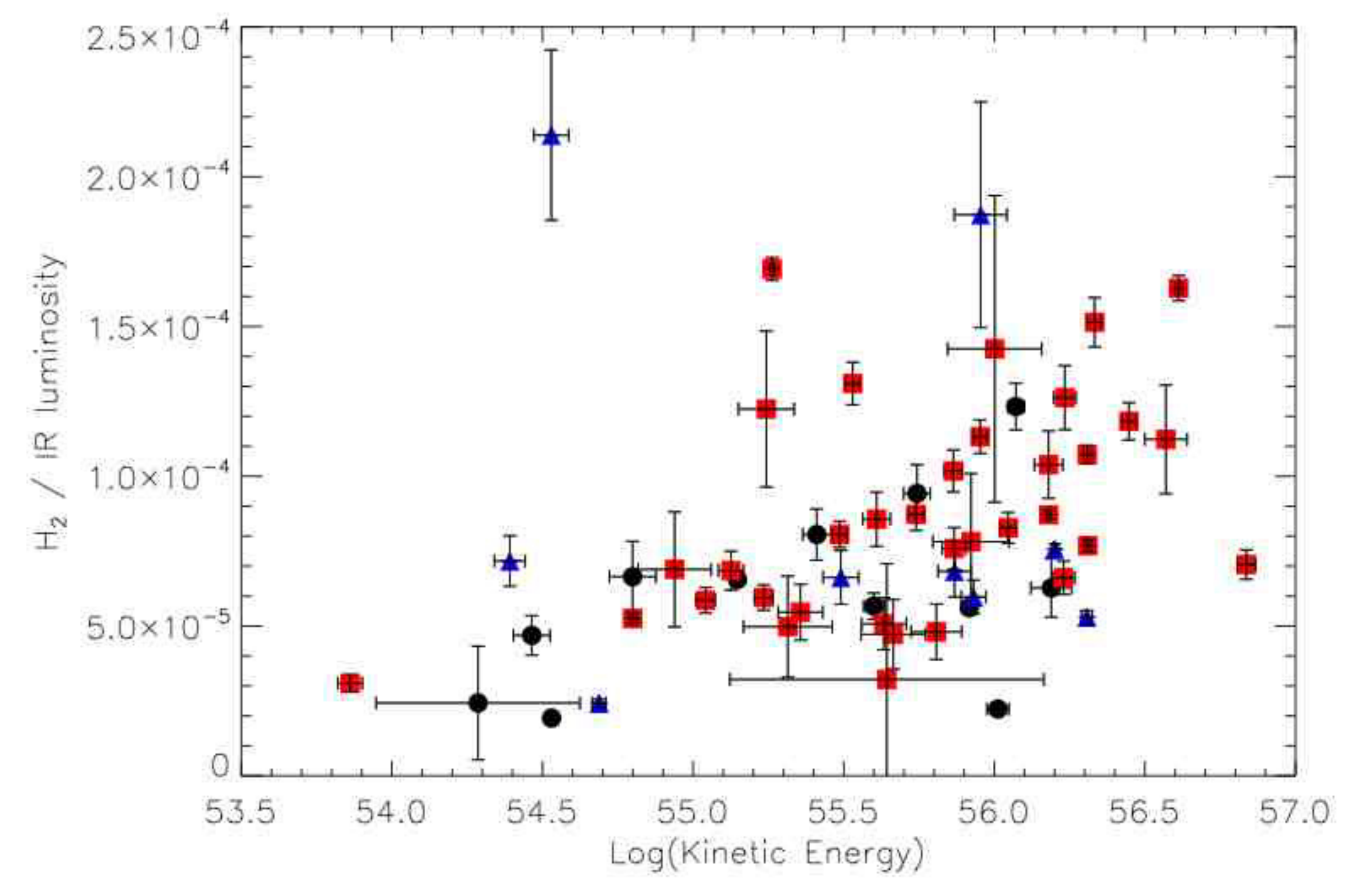}
  \caption{The estimated kinetic energy in the warm H$_2$ gas correlates well with the ratio of L$_{H_2 S(1)}$ to L$_{IR}$ for non-mergers (black circles), early stage mergers (blue triangles) and advanced mergers (red squares). } 
  \label{Kenplot}
\end{figure*}

Mergers are also known to enhance the H$_2$ emission \citep[e.g.][]{pet2012, gui2009, gui2012}, and about 40\% of LIRGs are mergers. However it may be that H$_2$ emission shocked by the tidal interactions, like that observed in the bridge of the early stage merger the Taffy Galaxies \citep{pet2012}, was not captured in our nuclear MIR spectroscopic observations which were focused on the LIRGs nuclei and not the diffuse extended IR emission. The eight galaxies with broad S(0) and S(1) emission profiles appear to be interacting. This may be consistent with shocks associated with tidal interactions being an energetically significant source of exciting the molecular gas. For all LIRGs with resolved and marginally resolved H$_2$ S(1) lines we approximate the kinetic energy ($E_{kin}$) of H$_2$ as $3/2 ~ \rm{M}_{\rm{H}_2} \sigma^2 _{\rm{H}_2}$ where M$_{H_2}$ is the mass of warm H$_2$ gas (section 3.2) and  $\sigma_{\rm{H}_2}$ is the velocity dispersion (section 3.1.2).  We find a statistically significant correlation (i.e. probability that they are not correlated is 0.04)  between the kinetic energy in the H$_2$ gas and the ratio of L$({H_2 S(1)})$ to the L$_{IR}$ (Figure \ref{Kenplot}).  While the observed L$({H_2 S(1)})$ to the L$_{IR}$ ratios are compatible with H$_2$ excitation by UV pumping, we may see evidence for collisional excitation at kinetic energies greater than 10$^{55}$ ergs.

The sources with the largest warm gas kinetic energies are mergers. A possible explanation for our findings is that mergers increase the production of  bulk in-flows leading to observable broad H$_2$ profiles and possibly denser environments and larger masses of warm molecular gas in the central regions of advanced stage mergers than in non-mergers. Outflows associated with the central region can also explain some of the broader profiles. Using the continuum source sizes from \citet{ds2011} and stellar masses of 10$^{11} ~\rm{M}_{\odot}$ we estimate escape velocities for the gas that range from $\sim$ 300 to $\sim$ 1000 km/s.
The combination of merger tidal interactions, star formation and AGN activity in those sources may dramatically affect the overall state of the molecular gas by pushing a fraction of it out of the LIRG.  We speculate that while outflows may be important, bulk molecular gas inflows may also be present \citep[e.g.][]{yam2017}; those inflows move sufficient gas to the center to fuel star formation and AGN accretion which lead to turbulence and heating of the molecular gas to 100-1000K, i.e. the warm gas observed in the MIR.  Both inflows and outflows generate turbulence and heat of the molecular gas \citep{bei2015}.

One way to put our investigation of H$_2$ in LIRGs in the context of the study of how AGN and gravitational interactions impact the ISM is to compare it with studies of neutral hydrogen. Gravitational interactions do not appear to significantly affect the neutral gas to stellar mass ratios in galaxies throughout the merger sequence and post-mergers may on average be more gas rich than isolated galaxies \citep[e.g.][]{elis2015}. However, interactions may be responsible for the cooling of halo-gas and lead to central build up of molecular gas \citep[e.g.][]{bc1993}. Our finding that in some advanced mergers the kinetic energy of the warm molecular gas in the central few kpcs is correlated to the ratio of  H$_2$ to IR luminosities is consistent with this scenario. Advanced mergers tend to host more AGN \citep[e.g][]{petric2011,vei2009b} and the ratio of  H$_2$ to PAH luminosity increases with H$_2$ luminosity in LIRGs \citep{stir2014} and in ULIRGs \citep{hill2014}. \citet{ina2013} found five nearby LIRGs with asymmetric [Ne III] and [Ne V] emission lines. We do not find any asymmetric H$_2$ resolved profiles, or with significant central velocity shifts but this may be due to the comparatively lower signal to noise of the H$_2$ lines.
Spatially resolved NIR spectroscopic studies \citep[e.g][]{med2015,rupke2010} are required to confirm that the the broader MIR profiles are associated with inflow of cooled halo gas.

\subsection{Masses and Temperatures as a function of interaction stage, and the AGN contribution to the L$_{IR}$ } 
\citet{petric2011} found there are relatively more AGN dominated sources among late-stage mergers than star formation dominated sources, compared to non-interacting LIRGs. 
While we did not find any statistically significant differences between the H$_2$ masses of mergers and non-mergers, late-stage mergers have the highest warm molecular H$_2$ masses and temperatures. However the difference decreases when we normalize the H$_2$ masses by the IR luminosities, making it difficult to extract strong conclusions from this finding. 

When we only use the 128 fits with S(0) detections we (1) no longer find that late mergers have the highest masses, but we still find that the advanced mergers show higher mass-averaged temperatures and (2) find that AGN-dominated sources have about 100K higher mass-averaged temperatures than star formation dominated LIRGs. KS tests do not indicate a statistically significant difference between the $L_{IR}/M _{H_2}$ ratios of mergers and those of non-mergers.

The median warm molecular gas mass for sources with an AGN contribution greater than 50\%  of the total IR luminosity is  2.3 $\times 10^7 M_{\odot}$, and for sources where an AGN contributes less than 10\% to the total IR luminosity it is a few times higher at 1.2 $\times 10^8 M_{\odot}$. The KS statistical difference between them is 0.3 and the probability that they are drawn from the same population is 2\%.  The mean and median ratios of L$_{\rm{IR}}$ to $M _{H_{2}}$ for AGN dominated sources are 1.2 and 0.6 $\times 10^4 ~ \rm{L}_{\odot} /\rm{M}_{\odot}$, and those for starburst dominated sources are  1.8 and 0.3 $\times 10^4  \rm{L}_{\odot} /\rm{M}_{\odot}$, with a KS statistic difference between them of 0.17 and a probability that they are the same population of 38\%. The average and median temperatures for AGN dominated sources are 313 and 296 K while those of SB dominated galaxies are  203 and 177 K respectively. The KS statistic differences between the two distributions of temperatures is the largest for any comparison done in the sample, 0.446, with a probability of effectively 0 that they are drawn from the same populations. 

Statistical differences between the warm gas properties of AGN dominated LIRGs and star formation dominated LIRGs are more significant than the differences between mergers and non-mergers and as significant as those between LIRGs and ULIRGs, and between LIRGs and normal-galaxies (table \ref{KStab}).

\subsection{H$_2$ excitation conditions as a function of interaction stage, and the AGN contribution to the L$_{IR}$ }

\begin{figure*}[h!]
  \includegraphics[width=0.95\linewidth]{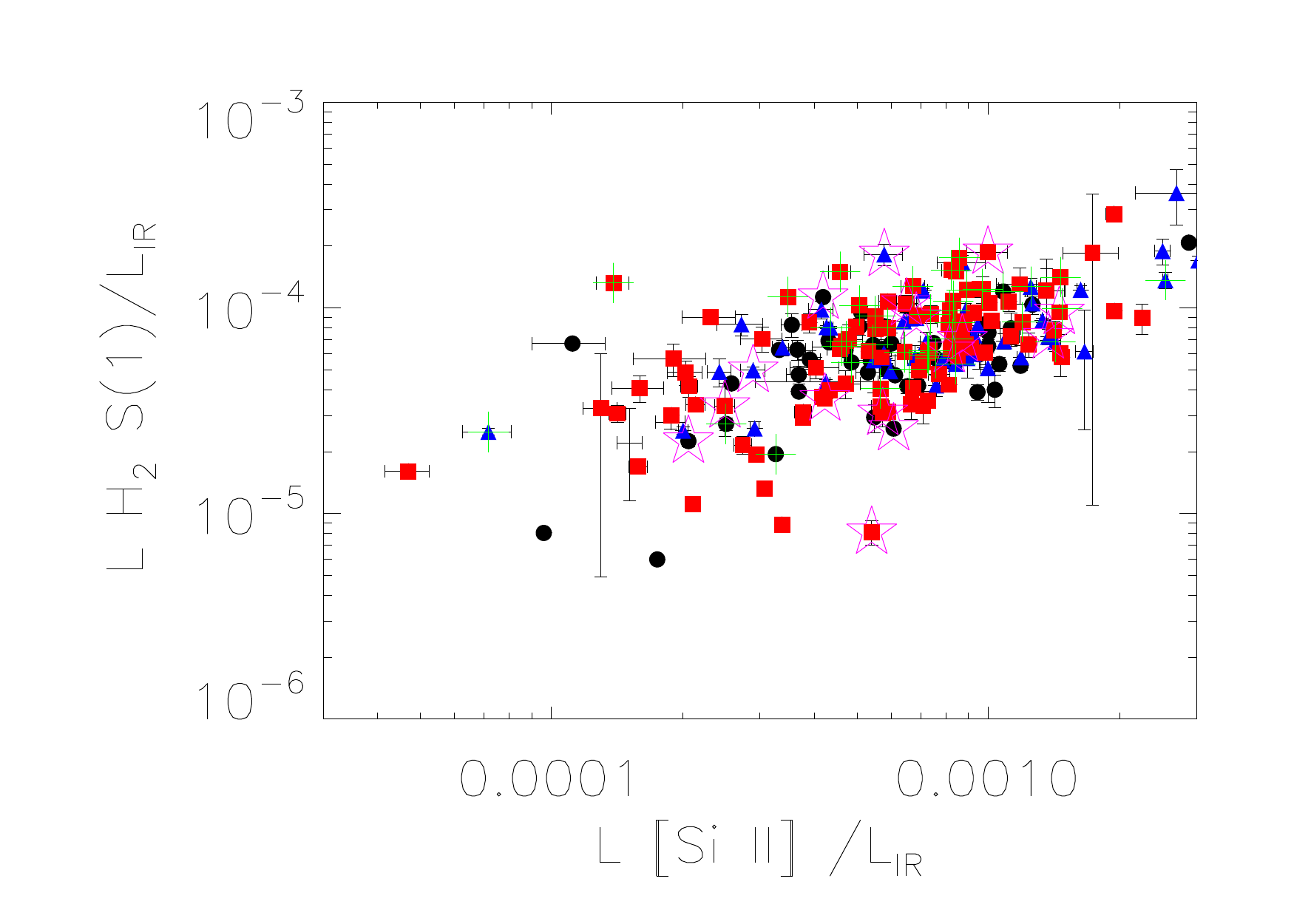} 
  \caption{Comparison between the ratios of H$_{2}$ S(1) emission to the IR luminosity versus the ratios of [Si~{\sc II}]  to the IR luminosity for non-mergers (black), early stage mergers (blue) and advanced mergers (red). Sources with resolved and marginally resolved S(1) lines are shown as (green) crosses. Sources with a significant AGN contribution to the total IR luminosity are shown as (magenta) stars. \label{ComLumsIR}}
\end{figure*}

In Figure \ref{ComLumsIR} we compare the H$_2$ S(1) emission versus the [Si~{\sc{ii}}] emission, both normalized by the L$_{IR}$ luminosity. This comparison is important because H$_2$ and [Si~{\sc{ii}}] are both tracers of the warm interstellar medium but their relative intensities vary as a function of the radiation field intensity and the metallicity \citep{kauf2006}. 

\begin{figure*}[h!]
  \includegraphics[width=0.33\linewidth]{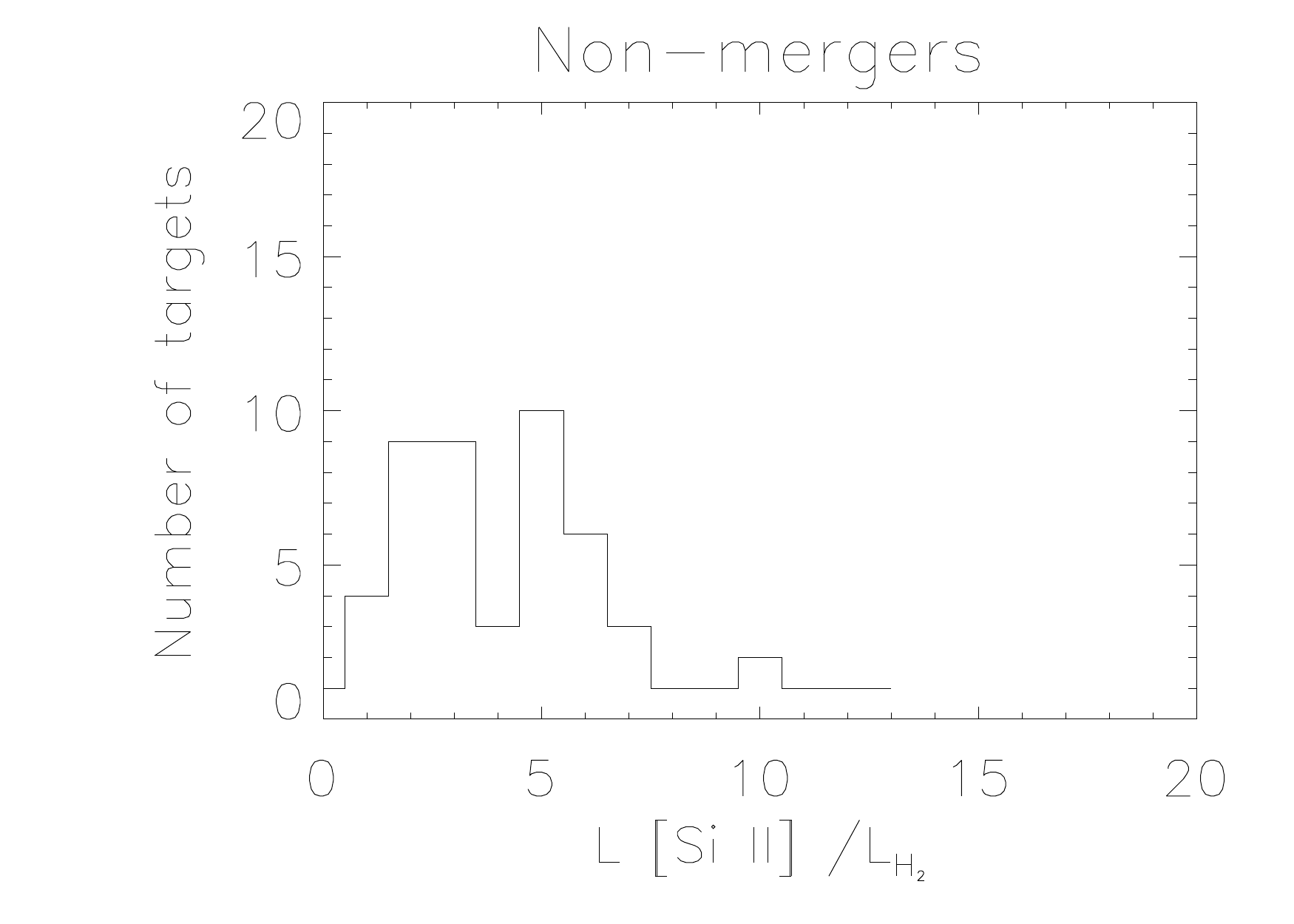}
  \includegraphics[width=0.33\linewidth]{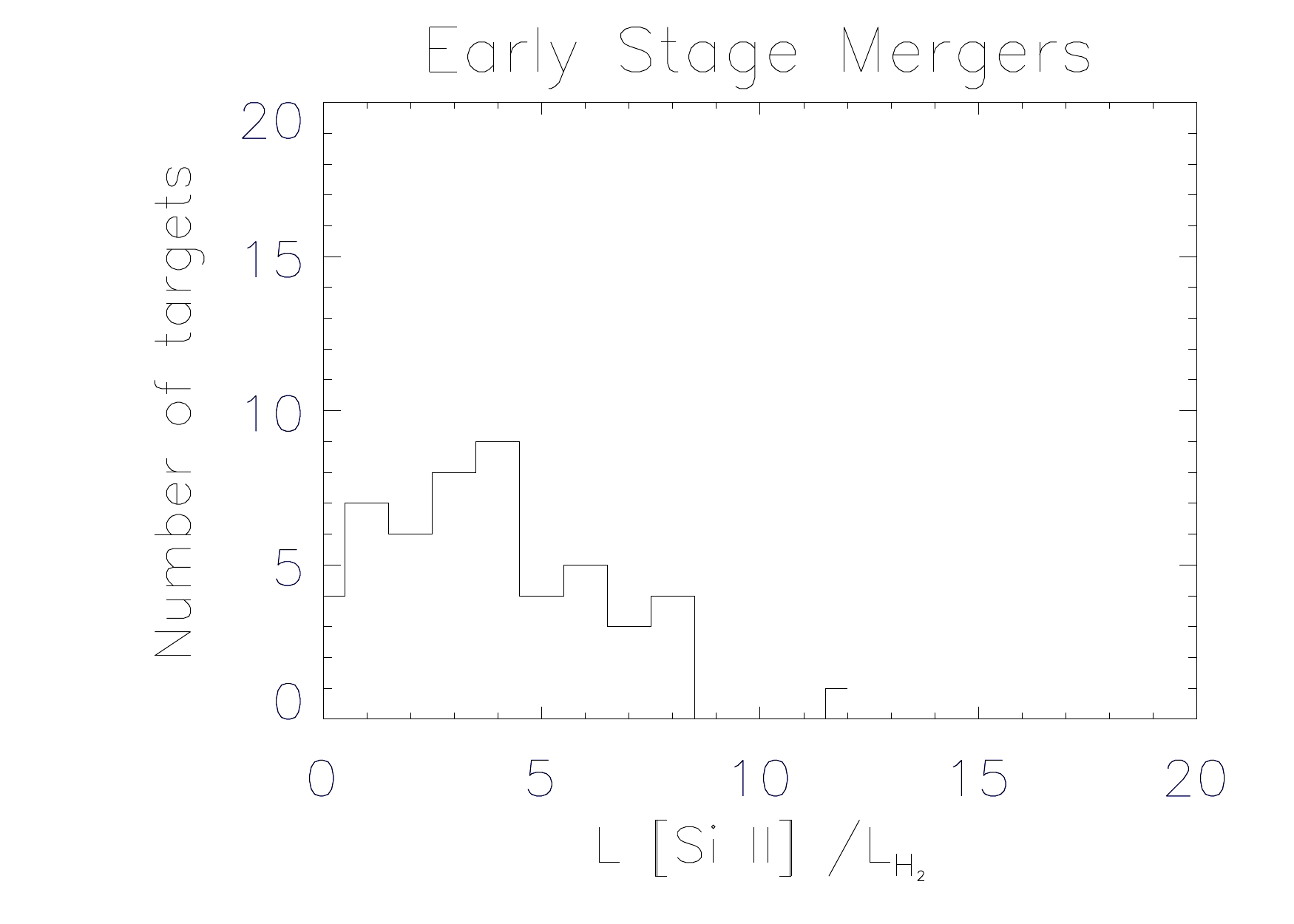}
  \includegraphics[width=0.33\linewidth]{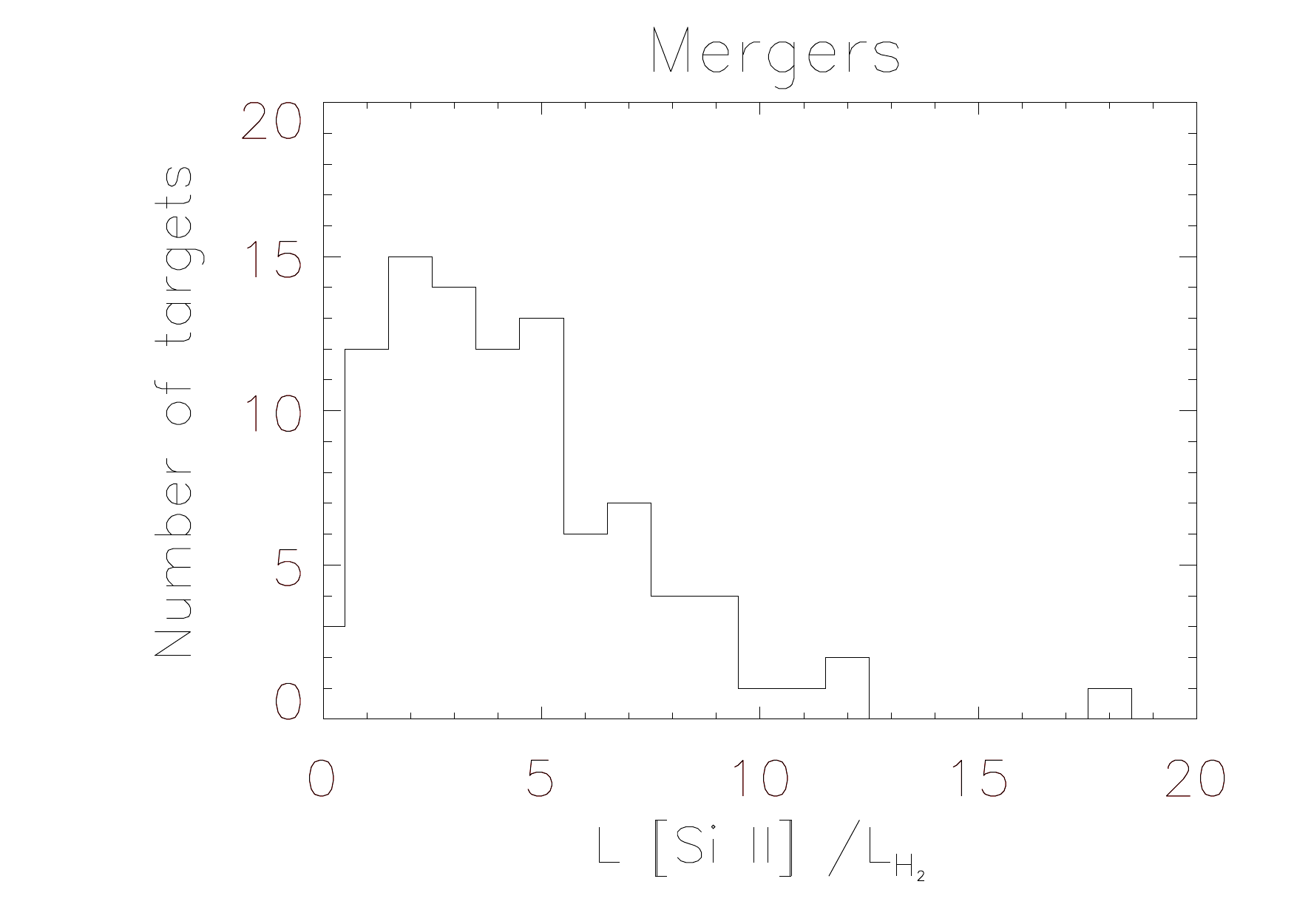}
  \caption{Histograms of [Si~{\sc II}]  to H$_{2}$ S(1) emission line ratios for: (left) non-mergers, (center) early-stage mergers, and (right) mergers.  The standard deviation of  the distribution of  [Si~{\sc II}]  to H$_2$ S(1) line emission ratios is two times larger for mergers than it is for early-stage mergers and non-mergers. The median [Si~{\sc II}]  to H$_{2}$ S(1) emission line ratios are $\sim$ 12 for non-mergers and $\sim$9 for mergers.\label{ComLumsIRmerg}}
\end{figure*}

The correlation between the H$_2$/IR and [Si~{\sc{ii}}]/IR luminosity ratios (Figure \ref{ComLumsIR}) suggests that those cooling lines have a common origin for the majority of the sources but that advanced stage mergers may have more diverse H$_2$ heating mechanisms.
While we do not find statistically significant differences between the [Si~{\sc{ii}}] to H$_2$ S(1) emission line ratios in mergers from those in non-mergers, mergers have a lower median [Si~{\sc{ii}}] to H$_2$ S(1) emission line ratios (9 versus 12) though the dispersion is large (5 for non-mergers and early stage mergers and 8 for mergers) (Figure \ref{ComLumsIRmerg}). 

These results confirm those of \citet{pet2012} who point out that while mergers can enhance the total H$_2$ emission, the observed H$_2$ properties are highly dependent on the collision geometry and on the initial conditions of the mergers. It is thus not surprising that we do not find strong trends with interaction stage or the IR emission. 

\section{Summary and Conclusions}
We present measurements of the rotational transitions of molecular hydrogen observed in a MIR spectroscopic survey of 202 LIRG nuclei with Spitzer IRS. We detect H$_{2}$ emission in at least one rotational transition in 91\% of the sources, S(1) being the most commonly detected transition. The ratio of H$_2$ S(0)+S(1)+S(2) luminosity to the IR luminosity ranges between 2.3 $\times 10^{-6}$ and 0.014.

We perform a statistical analysis of the S(0) and S(1) lines and their ratios to the IR luminosity, including upper-limits, to investigate if there are systematic differences between the H$_2$ properties of LIRGs in our sample, normal galaxies as measured by \citet{rous2007}, and ULIRGs as measured by \citet{hig2006}. We find that the probability that LIRGs and normal galaxies are drawn from the same population is null while the probability that LIRGs and ULIRGs are drawn from the same population is 16\%. We compare in a similar statistical fashion the H$_2$ S(0), S(1), and H$_2$ [S(1)+ S(2)]/L$_{\rm{IR}}$ for merging versus non-merging LIRGs and for AGN-dominated LIRGs versus star formation dominated LIRGs. We find probabilities of 39\% and 30\%, respectively, that those samples of galaxies are drawn from the same population. 

We compare the ratios of the H$_2$ lines to the [Si II] line in the context of the theoretical models of PDR/HII regions of \citet{kauf2006}. Our measurements can be modeled by FUV radiation field values between $10^{2.0} ~\rm{and}~ 10^{2.8}$ and hydrogen nucleus densities $n$ between $10^{3.5} ~\rm{and}~ 10^{4.5}$ cm$^{-3}$. 

For 78 sources where we detect at least 4 transitions with high SNR, we investigate if  the temperatures estimated from the S(0), S(1), S(2) and S(3) lines are consistent with an ortho to para ratio expected for thermalized gas. For half of the sources the observed S(0) to S(3) lines suggest that the gas is either non-thermalized or that we are observing emission from regions at different temperatures.

We compute excitation diagrams and use them to estimate the masses and temperatures of warm molecular gas in 214 LIRG nuclei in the GOALS sample. We find that the masses of warm gas in advanced mergers are slightly larger than those of non-interacting systems, and that the statistical differences between the warm gas properties of AGN dominated sources and non AGN dominated sources are more significant than the differences between mergers and non-mergers. AGN may power a fraction of the H$_2$ indirectly through dynamical perturbations which may be responsible for both extensive shocks in the ISM and an increased rate of accretion onto the supermassive black hole. One way to look for the connection between dynamical perturbations and the warm H$_2$ gas is to look for sign of kinematic peculiarities. In Petrus et al. (in prep.) we use data from the Gemini Near Infrared Spectrograph to show that such peculiarities are common among LIRGs with excess H$_2$ emission.

We find that between 10 to 15\% of LIRGs have resolved or marginally resolved S(0) and/or S(1) lines and that those sources tend to be mergers and have a slightly higher fraction of AGN dominated sources among them. As was pointed out in Peterson et al. (2012), while mergers can have a significant enhancing effect on the total H$_2$ luminosity emitted, the observed H$_2$ properties are highly dependent on the collision geometry and on the initial conditions of the mergers. Those sources with resolved lines also tend to have slightly higher H$_2$ /IR luminosity ratios suggesting either higher G$_0$ or higher densities. We find a correlation between the kinetic energy in the warm molecular gas and the intrinsic H$_2$  line widths. 

These data, in conjunction with the findings of \citet{elis2015} and \citet{yam2017}, suggest that the velocity broadening observed in some H$_2$ profiles may be due to inflow of halo gas that feeds central star formation and the central supermassive black hole which in turn produce outflows that impart kinetic energy to the central molecular gas.  Alternatively inflows of gas from the galaxies's disk may stimulate star formation and AGN activity which heat the warm molecular gas to sufficiently high temperatures to be detected via MIR H$_2$ lines. 

Both interpretations are consistent with the narrative of quasar formation that starts with two interacting gas rich galaxies, passes through an obscured, reddened QSO and ends with an optically luminous QSOs. Those optically luminous QSOs can still have large ISM reservoirs \citep[e.g][]{petric2015} if as proposed by \citet{bc1993}, as galaxies merge, hot gas from the halo cools and forms fresh molecular gas on dust grains. 

{\it{Acknowledgments:}} 
This work is based primarily on observations made with the Spitzer Space Telescope, which is operated by the Jet Propulsion Laboratory, California Institute of Technology under NASA contract 1407. We have made use of the NASA/IPAC Extragalactic Database (NED) which is operated by the Jet Propulsion Laboratory, California Institute of Technology, under contract with NASA. Support for this research was provided by NASA through an award issued by JPL/Caltech. VC would like to acknowledge partial support from the EU FP7 Grant PIRSES-GA-2012-316788. AP thanks T. Geballe for comments improving the clarity of the paper.  NL acknowledge support by the NSFC grant \#11673028 and National Key R\&D Program of China grant  \#2017YFA0402704.

\bibliographystyle{apj}
\bibliography{AP_bib}

\begin{thebibliography}{}
\expandafter\ifx\csname natexlab\endcsname\relax\def\natexlab#1{#1}\fi

\bibitem[{{Armus} {et~al.}(2007){Armus}, {Charmandaris}, {Bernard-Salas},
  {Spoon}, {Marshall}, {Higdon}, {Desai}, {Teplitz}, {Hao}, {Devost}, {Brandl},
  {Wu}, {Sloan}, {Soifer}, {Houck}, \& {Herter}}]{armus2007}
{Armus}, L., {Charmandaris}, V., {Bernard-Salas}, J., {et~al.} 2007, \apj, 656,
  148

\bibitem[{{Armus} {et~al.}(2009){Armus}, {Mazzarella}, {Evans}, {Surace},
  {Sanders}, {Iwasawa}, {Frayer}, {Howell}, {Chan}, {Petric}, {Vavilkin},
  {Kim}, {Haan}, {Inami}, {Murphy}, {Appleton}, {Barnes}, {Bothun}, {Bridge},
  {Charmandaris}, {Jensen}, {Kewley}, {Lord}, {Madore}, {Marshall},
  {Melbourne}, {Rich}, {Satyapal}, {Schulz}, {Spoon}, {Sturm}, {U}, {Veilleux},
  \& {Xu}}]{armus09}
{Armus}, L., {Mazzarella}, J.~M., {Evans}, A.~S., {et~al.} 2009, \pasp, 121,
  559

\bibitem[{{Beir{\~a}o} {et~al.}(2015){Beir{\~a}o}, {Armus}, {Lehnert},
  {Guillard}, {Heckman}, {Draine}, {Hollenbach}, {Walter}, {Sheth}, {Smith},
  {Shopbell}, {Boulanger}, {Surace}, {Hoopes}, \& {Engelbracht}}]{bei2015}
{Beir{\~a}o}, P., {Armus}, L., {Lehnert}, M.~D., {et~al.} 2015, \mnras, 451,
  2640

\bibitem[{{Black} \& {Dalgarno}(1976)}]{blad76}
{Black}, J.~H., \& {Dalgarno}, A. 1976, \apj, 203, 132

\bibitem[{{Braine} \& {Combes}(1993)}]{bc1993}
{Braine}, J., \& {Combes}, F. 1993, \aap, 269, 7

\bibitem[{{Carrillo} {et~al.}(1999){Carrillo}, {Masegosa}, {Dultzin-Hacyan}, \&
  {Ordo{\~n}ez}}]{car1999}
{Carrillo}, R., {Masegosa}, J., {Dultzin-Hacyan}, D., \& {Ordo{\~n}ez}, R.
  1999, \rmxaa, 35, 187

\bibitem[{{Cluver} {et~al.}(2010){Cluver}, {Appleton}, {Boulanger}, {Guillard},
  {Ogle}, {Duc}, {Lu}, {Rasmussen}, {Reach}, {Smith}, {Tuffs}, {Xu}, \&
  {Yun}}]{cluver2010}
{Cluver}, M.~E., {Appleton}, P.~N., {Boulanger}, F., {et~al.} 2010, \apj, 710,
  248

\bibitem[{{Dasyra} \& {Combes}(2011)}]{das2011}
{Dasyra}, K.~M., \& {Combes}, F. 2011, \aap, 533, L10

\bibitem[{{Dasyra} {et~al.}(2008){Dasyra}, {Ho}, {Armus}, {Ogle}, {Helou},
  {Peterson}, {Lutz}, {Netzer}, \& {Sturm}}]{das2008}
{Dasyra}, K.~M., {Ho}, L.~C., {Armus}, L., {et~al.} 2008, \apjl, 674, L9

\bibitem[{{D{\'{\i}}az-Santos} {et~al.}(2010){D{\'{\i}}az-Santos},
  {Charmandaris}, {Armus}, {Petric}, {Howell}, {Murphy}, {Mazzarella},
  {Veilleux}, {Bothun}, {Inami}, {Appleton}, {Evans}, {Haan}, {Marshall},
  {Sanders}, {Stierwalt}, \& {Surace}}]{ds2010}
{D{\'{\i}}az-Santos}, T., {Charmandaris}, V., {Armus}, L., {et~al.} 2010, \apj,
  723, 993

\bibitem[{{D{\'{\i}}az-Santos} {et~al.}(2011){D{\'{\i}}az-Santos},
  {Charmandaris}, {Armus}, {Stierwalt}, {Haan}, {Mazzarella}, {Howell},
  {Veilleux}, {Murphy}, {Petric}, {Appleton}, {Evans}, {Sanders}, \&
  {Surace}}]{ds2011}
---. 2011, \apj, 741, 32

\bibitem[{{D{\'{\i}}az-Santos} {et~al.}(2014){D{\'{\i}}az-Santos}, {Armus},
  {Charmandaris}, {Stacey}, {Murphy}, {Haan}, {Stierwalt}, {Malhotra},
  {Appleton}, {Inami}, {Magdis}, {Elbaz}, {Evans}, {Mazzarella}, {Surace}, {van
  der Werf}, {Xu}, {Lu}, {Meijerink}, {Howell}, {Petric}, {Veilleux}, \&
  {Sanders}}]{ds2014}
{D{\'{\i}}az-Santos}, T., {Armus}, L., {Charmandaris}, V., {et~al.} 2014,
  \apjl, 788, L17

\bibitem[{{Ellison} {et~al.}(2015){Ellison}, {Fertig}, {Rosenberg}, {Nair},
  {Simard}, {Torrey}, \& {Patton}}]{elis2015}
{Ellison}, S.~L., {Fertig}, D., {Rosenberg}, J.~L., {et~al.} 2015, \mnras, 448,
  221

\bibitem[{{Evans} {et~al.}(2008){Evans}, {Vavilkin}, {Pizagno}, {Modica},
  {Mazzarella}, {Iwasawa}, {Howell}, {Surace}, {Armus}, {Petric}, {Spoon},
  {Barnes}, {Suer}, {Sanders}, {Chan}, \& {Lord}}]{eva2008}
{Evans}, A.~S., {Vavilkin}, T., {Pizagno}, J., {et~al.} 2008, \apjl, 675, L69

\bibitem[{{Fabbiano} {et~al.}(1992){Fabbiano}, {Kim}, \&
  {Trinchieri}}]{fab1992}
{Fabbiano}, G., {Kim}, D.-W., \& {Trinchieri}, G. 1992, \apjs, 80, 531

\bibitem[{{Feigelson} \& {Nelson}(1985)}]{fn1985}
{Feigelson}, E.~D., \& {Nelson}, P.~I. 1985, \apj, 293, 192

\bibitem[{{Fern{\'a}ndez} {et~al.}(2014){Fern{\'a}ndez}, {Petric}, {Schweizer},
  \& {van Gorkom}}]{fern2014}
{Fern{\'a}ndez}, X., {Petric}, A.~O., {Schweizer}, F., \& {van Gorkom}, J.~H.
  2014, \aj, 147, 74

\bibitem[{{Flagey} {et~al.}(2013){Flagey}, {Goldsmith}, {Lis}, {Gerin},
  {Neufeld}, {Sonnentrucker}, {De Luca}, {Godard}, {Goicoechea}, {Monje}, \&
  {Phillips}}]{fla2013}
{Flagey}, N., {Goldsmith}, P.~F., {Lis}, D.~C., {et~al.} 2013, \apj, 762, 11

\bibitem[{{Gonz{\'a}lez-Mart{\'{\i}}n}
  {et~al.}(2009){Gonz{\'a}lez-Mart{\'{\i}}n}, {Masegosa}, {M{\'a}rquez},
  {Guainazzi}, \& {Jim{\'e}nez-Bail{\'o}n}}]{gonz2009}
{Gonz{\'a}lez-Mart{\'{\i}}n}, O., {Masegosa}, J., {M{\'a}rquez}, I.,
  {Guainazzi}, M., \& {Jim{\'e}nez-Bail{\'o}n}, E. 2009, \aap, 506, 1107

\bibitem[{{Guillard} {et~al.}(2009){Guillard}, {Boulanger}, {Pineau Des
  For{\^e}ts}, \& {Appleton}}]{gui2009}
{Guillard}, P., {Boulanger}, F., {Pineau Des For{\^e}ts}, G., \& {Appleton},
  P.~N. 2009, \aap, 502, 515

\bibitem[{{Guillard} {et~al.}(2012){Guillard}, {Boulanger}, {Pineau des
  For{\^e}ts}, {Falgarone}, {Gusdorf}, {Cluver}, {Appleton}, {Lisenfeld},
  {Duc}, {Ogle}, \& {Xu}}]{gui2012}
{Guillard}, P., {Boulanger}, F., {Pineau des For{\^e}ts}, G., {et~al.} 2012,
  \apj, 749, 158

\bibitem[{{Haan} {et~al.}(2011{\natexlab{a}}){Haan}, {Armus}, {Laine},
  {Charmandaris}, {Smith}, {Schweizer}, {Brandl}, {Evans}, {Surace},
  {Diaz-Santos}, {Beir{\~a}o}, {Murphy}, {Stierwalt}, {Hibbard}, {Yun}, \&
  {Jarrett}}]{haan2011}
{Haan}, S., {Armus}, L., {Laine}, S., {et~al.} 2011{\natexlab{a}}, \apjs, 197,
  27

\bibitem[{{Haan} {et~al.}(2011{\natexlab{b}}){Haan}, {Surace}, {Armus},
  {Evans}, {Howell}, {Mazzarella}, {Kim}, {Vavilkin}, {Inami}, {Sanders},
  {Petric}, {Bridge}, {Melbourne}, {Charmandaris}, {Diaz-Santos}, {Murphy},
  {U}, {Stierwalt}, \& {Marshall}}]{haan2011a}
{Haan}, S., {Surace}, J.~A., {Armus}, L., {et~al.} 2011{\natexlab{b}}, \aj,
  141, 100

\bibitem[{{Hayward} {et~al.}(2011){Hayward}, {Kere{\v s}}, {Jonsson},
  {Narayanan}, {Cox}, \& {Hernquist}}]{hay2011}
{Hayward}, C.~C., {Kere{\v s}}, D., {Jonsson}, P., {et~al.} 2011, \apj, 743,
  159

\bibitem[{{Higdon} {et~al.}(2006){Higdon}, {Armus}, {Higdon}, {Soifer}, \&
  {Spoon}}]{hig2006}
{Higdon}, S.~J.~U., {Armus}, L., {Higdon}, J.~L., {Soifer}, B.~T., \& {Spoon},
  H.~W.~W. 2006, \apj, 648, 323

\bibitem[{{Hill} \& {Zakamska}(2014)}]{hill2014}
{Hill}, M.~J., \& {Zakamska}, N.~L. 2014, \mnras, 439, 2701

\bibitem[{{Hopkins} {et~al.}(2006){Hopkins}, {Hernquist}, {Cox}, {Di Matteo},
  {Robertson}, \& {Springel}}]{hop2006}
{Hopkins}, P.~F., {Hernquist}, L., {Cox}, T.~J., {et~al.} 2006, \apjs, 163, 1

\bibitem[{{Huber} \& {Hertzberg}(1979)}]{huh79}
{Huber}, K., P., \& {Hertzberg}, G. 1979, {Constants of Diatomic Molecules}
  (Van Nostrand, New York)

\bibitem[{{Inami} {et~al.}(2010){Inami}, {Armus}, {Surace}, {Mazzarella},
  {Evans}, {Sanders}, {Howell}, {Petric}, {Vavilkin}, {Iwasawa}, {Haan},
  {Murphy}, {Stierwalt}, {Appleton}, {Barnes}, {Bothun}, {Bridge}, {Chan},
  {Charmandaris}, {Frayer}, {Kewley}, {Kim}, {Lord}, {Madore}, {Marshall},
  {Matsuhara}, {Melbourne}, {Rich}, {Schulz}, {Spoon}, {Sturm}, {U},
  {Veilleux}, \& {Xu}}]{inami2010}
{Inami}, H., {Armus}, L., {Surace}, J.~A., {et~al.} 2010, \aj, 140, 63

\bibitem[{{Inami} {et~al.}(2013){Inami}, {Armus}, {Charmandaris}, {Groves},
  {Kewley}, {Petric}, {Stierwalt}, {D{\'{\i}}az-Santos}, {Surace}, {Rich},
  {Haan}, {Howell}, {Evans}, {Mazzarella}, {Marshall}, {Appleton}, {Lord},
  {Spoon}, {Frayer}, {Matsuhara}, \& {Veilleux}}]{ina2013}
{Inami}, H., {Armus}, L., {Charmandaris}, V., {et~al.} 2013, \apj, 777, 156

\bibitem[{{Isobe} {et~al.}(1986){Isobe}, {Feigelson}, \& {Nelson}}]{isobe1986}
{Isobe}, T., {Feigelson}, E.~D., \& {Nelson}, P.~I. 1986, \apj, 306, 490

\bibitem[{{Kaufman} {et~al.}(2006){Kaufman}, {Wolfire}, \&
  {Hollenbach}}]{kauf2006}
{Kaufman}, M.~J., {Wolfire}, M.~G., \& {Hollenbach}, D.~J. 2006, \apj, 644, 283

\bibitem[{{Kewley} {et~al.}(2010){Kewley}, {Rupke}, {Zahid}, {Geller}, \&
  {Barton}}]{kew2010}
{Kewley}, L.~J., {Rupke}, D., {Zahid}, H.~J., {Geller}, M.~J., \& {Barton},
  E.~J. 2010, \apjl, 721, L48

\bibitem[{{Larson} {et~al.}(2016){Larson}, {Sanders}, {Barnes}, {Ishida},
  {Evans}, {U}, {Mazzarella}, {Kim}, {Privon}, {Mirabel}, \&
  {Flewelling}}]{lar2016}
{Larson}, K.~L., {Sanders}, D.~B., {Barnes}, J.~E., {et~al.} 2016, \apj, 825,
  128

\bibitem[{{Mazzarella} {et~al.}(2012){Mazzarella}, {Iwasawa}, {Vavilkin},
  {Armus}, {Kim}, {Bothun}, {Evans}, {Spoon}, {Haan}, {Howell}, {Lord},
  {Marshall}, {Ishida}, {Xu}, {Petric}, {Sanders}, {Surace}, {Appleton},
  {Chan}, {Frayer}, {Inami}, {Khachikian}, {Madore}, {Privon}, {Sturm}, {U}, \&
  {Veilleux}}]{mazz2012}
{Mazzarella}, J.~M., {Iwasawa}, K., {Vavilkin}, T., {et~al.} 2012, \aj, 144,
  125

\bibitem[{{Medling} {et~al.}(2015){Medling}, {U}, {Rich}, {Kewley}, {Armus},
  {Dopita}, {Max}, {Sanders}, \& {Sutherland}}]{med2015}
{Medling}, A.~M., {U}, V., {Rich}, J.~A., {et~al.} 2015, \mnras, 448, 2301

\bibitem[{{Meijerink} {et~al.}(2007){Meijerink}, {Spaans}, \&
  {Israel}}]{meij2007}
{Meijerink}, R., {Spaans}, M., \& {Israel}, F.~P. 2007, \aap, 461, 793

\bibitem[{{Mihos} \& {Hernquist}(1996)}]{mih1996}
{Mihos}, J.~C., \& {Hernquist}, L. 1996, \apj, 464, 641

\bibitem[{{Modica} {et~al.}(2012){Modica}, {Vavilkin}, {Evans}, {Kim},
  {Mazzarella}, {Iwasawa}, {Petric}, {Howell}, {Surace}, {Armus}, {Spoon},
  {Sanders}, {Wong}, \& {Barnes}}]{modi2012}
{Modica}, F., {Vavilkin}, T., {Evans}, A.~S., {et~al.} 2012, \aj, 143, 16

\bibitem[{{Narayanan} {et~al.}(2006){Narayanan}, {Cox}, {Robertson},
  {Dav{\'e}}, {Di Matteo}, {Hernquist}, {Hopkins}, {Kulesa}, \&
  {Walker}}]{nar2006}
{Narayanan}, D., {Cox}, T.~J., {Robertson}, B., {et~al.} 2006, \apjl, 642, L107

\bibitem[{{Ogle} {et~al.}(2010){Ogle}, {Boulanger}, {Guillard}, {Evans},
  {Antonucci}, {Appleton}, {Nesvadba}, \& {Leipski}}]{ogle2010}
{Ogle}, P., {Boulanger}, F., {Guillard}, P., {et~al.} 2010, \apj, 724, 1193

\bibitem[{{Ogle} {et~al.}(2012){Ogle}, {Davies}, {Appleton}, {Bertincourt},
  {Seymour}, \& {Helou}}]{ogle2012}
{Ogle}, P., {Davies}, J.~E., {Appleton}, P.~N., {et~al.} 2012, \apj, 751, 13

\bibitem[{{Patton} {et~al.}(2013){Patton}, {Torrey}, {Ellison}, {Mendel}, \&
  {Scudder}}]{pat2013}
{Patton}, D.~R., {Torrey}, P., {Ellison}, S.~L., {Mendel}, J.~T., \& {Scudder},
  J.~M. 2013, \mnras, 433, L59

\bibitem[{{Peterson} {et~al.}(2012){Peterson}, {Appleton}, {Helou}, {Guillard},
  {Jarrett}, {Cluver}, {Ogle}, {Struck}, \& {Boulanger}}]{pet2012}
{Peterson}, B.~W., {Appleton}, P.~N., {Helou}, G., {et~al.} 2012, \apj, 751, 11

\bibitem[{{Petric} {et~al.}(2015){Petric}, {Ho}, {Flagey}, \&
  {Scoville}}]{petric2015}
{Petric}, A.~O., {Ho}, L.~C., {Flagey}, N.~J.~M., \& {Scoville}, N.~Z. 2015,
  \apjs, 219, 22

\bibitem[{{Petric} {et~al.}(2011){Petric}, {Armus}, {Howell}, {Chan},
  {Mazzarella}, {Evans}, {Surace}, {Sanders}, {Appleton}, {Charmandaris},
  {D{\'{\i}}az-Santos}, {Frayer}, {Haan}, {Inami}, {Iwasawa}, {Kim}, {Madore},
  {Marshall}, {Spoon}, {Stierwalt}, {Sturm}, {U}, {Vavilkin}, \&
  {Veilleux}}]{petric2011}
{Petric}, A.~O., {Armus}, L., {Howell}, J., {et~al.} 2011, \apj, 730, 28

\bibitem[{{Petty} {et~al.}(2014){Petty}, {Armus}, {Charmandaris}, {Evans}, {Le
  Floc'h}, {Bridge}, {D{\'{\i}}az-Santos}, {Howell}, {Inami}, {Psychogyios},
  {Stierwalt}, \& {Surace}}]{pet2014}
{Petty}, S.~M., {Armus}, L., {Charmandaris}, V., {et~al.} 2014, \aj, 148, 111

\bibitem[{{Privon} {et~al.}(2013){Privon}, {Barnes}, {Evans}, {Hibbard}, {Yun},
  {Mazzarella}, {Armus}, \& {Surace}}]{privon2013}
{Privon}, G.~C., {Barnes}, J.~E., {Evans}, A.~S., {et~al.} 2013, \apj, 771, 120

\bibitem[{{Privon} {et~al.}(2015){Privon}, {Herrero-Illana}, {Evans},
  {Iwasawa}, {Perez-Torres}, {Armus}, {D{\'{\i}}az-Santos}, {Murphy},
  {Stierwalt}, {Aalto}, {Mazzarella}, {Barcos-Mu{\~n}oz}, {Borish}, {Inami},
  {Kim}, {Treister}, {Surace}, {Lord}, {Conway}, {Frayer}, \&
  {Alberdi}}]{priv2015}
{Privon}, G.~C., {Herrero-Illana}, R., {Evans}, A.~S., {et~al.} 2015, \apj,
  814, 39

\bibitem[{{Psychogyios} {et~al.}(2016){Psychogyios}, {Charmandaris},
  {Diaz-Santos}, {Armus}, {Haan}, {Howell}, {Le Floc'h}, {Petty}, \&
  {Evans}}]{psy2016}
{Psychogyios}, A., {Charmandaris}, V., {Diaz-Santos}, T., {et~al.} 2016, \aap,
  591, A1

\bibitem[{{Roussel} {et~al.}(2007){Roussel}, {Helou}, {Hollenbach}, {Draine},
  {Smith}, {Armus}, {Schinnerer}, {Walter}, {Engelbracht}, {Thornley},
  {Kennicutt}, {Calzetti}, {Dale}, {Murphy}, \& {Bot}}]{rous2007}
{Roussel}, H., {Helou}, G., {Hollenbach}, D.~J., {et~al.} 2007, \apj, 669, 959

\bibitem[{{Rupke} {et~al.}(2010){Rupke}, {Kewley}, \& {Barnes}}]{rupke2010}
{Rupke}, D.~S.~N., {Kewley}, L.~J., \& {Barnes}, J.~E. 2010, \apjl, 710, L156

\bibitem[{{Sanders} {et~al.}(2003){Sanders}, {Mazzarella}, {Kim}, {Surace}, \&
  {Soifer}}]{sanders03}
{Sanders}, D.~B., {Mazzarella}, J.~M., {Kim}, D.-C., {Surace}, J.~A., \&
  {Soifer}, B.~T. 2003, \aj, 126, 1607

\bibitem[{{Sanders} \& {Mirabel}(1996)}]{sam1996}
{Sanders}, D.~B., \& {Mirabel}, I.~F. 1996, \araa, 34, 749

\bibitem[{{Smith} {et~al.}(2007{\natexlab{a}}){Smith}, {Armus}, {Dale},
  {Roussel}, {Sheth}, {Buckalew}, {Jarrett}, {Helou}, \& {Kennicutt}}]{smith07}
{Smith}, J.~D.~T., {Armus}, L., {Dale}, D.~A., {et~al.} 2007{\natexlab{a}},
  \pasp, 119, 1133

\bibitem[{{Smith} {et~al.}(2007{\natexlab{b}}){Smith}, {Draine}, {Dale},
  {Moustakas}, {Kennicutt}, {Helou}, {Armus}, {Roussel}, {Sheth}, {Bendo},
  {Buckalew}, {Calzetti}, {Engelbracht}, {Gordon}, {Hollenbach}, {Li},
  {Malhotra}, {Murphy}, \& {Walter}}]{smith2007}
{Smith}, J.~D.~T., {Draine}, B.~T., {Dale}, D.~A., {et~al.} 2007{\natexlab{b}},
  \apj, 656, 770

\bibitem[{{Spoon} {et~al.}(2007){Spoon}, {Marshall}, {Houck}, {Elitzur}, {Hao},
  {Armus}, {Brandl}, \& {Charmandaris}}]{spoon2007}
{Spoon}, H.~W.~W., {Marshall}, J.~A., {Houck}, J.~R., {et~al.} 2007, \apjl,
  654, L49

\bibitem[{{Springel}(2000)}]{spring2000}
{Springel}, V. 2000, \mnras, 312, 859

\bibitem[{{Stierwalt} {et~al.}(2013){Stierwalt}, {Armus}, {Surace}, {Inami},
  {Petric}, {Diaz-Santos}, {Haan}, {Charmandaris}, {Howell}, {Kim}, {Marshall},
  {Mazzarella}, {Spoon}, {Veilleux}, {Evans}, {Sanders}, {Appleton}, {Bothun},
  {Bridge}, {Chan}, {Frayer}, {Iwasawa}, {Kewley}, {Lord}, {Madore},
  {Melbourne}, {Murphy}, {Rich}, {Schulz}, {Sturm}, {Vavilkin}, \&
  {Xu}}]{stir2013}
{Stierwalt}, S., {Armus}, L., {Surace}, J.~A., {et~al.} 2013, \apjs, 206, 1

\bibitem[{{Stierwalt} {et~al.}(2014){Stierwalt}, {Armus}, {Charmandaris},
  {Diaz-Santos}, {Marshall}, {Evans}, {Haan}, {Howell}, {Iwasawa}, {Kim},
  {Murphy}, {Rich}, {Spoon}, {Inami}, {Petric}, \& {U}}]{stir2014}
{Stierwalt}, S., {Armus}, L., {Charmandaris}, V., {et~al.} 2014, \apj, 790, 124

\bibitem[{{Veilleux} \& {Osterbrock}(1987)}]{vei1987}
{Veilleux}, S., \& {Osterbrock}, D.~E. 1987, \apjs, 63, 295

\bibitem[{{Veilleux} {et~al.}(2009){Veilleux}, {Rupke}, {Kim}, {Genzel},
  {Sturm}, {Lutz}, {Contursi}, {Schweitzer}, {Tacconi}, {Netzer}, {Sternberg},
  {Mihos}, {Baker}, {Mazzarella}, {Lord}, {Sanders}, {Stockton}, {Joseph}, \&
  {Barnes}}]{vei2009b}
{Veilleux}, S., {Rupke}, D.~S.~N., {Kim}, D.-C., {et~al.} 2009, \apjs, 182, 628

\bibitem[{{Yamashita} {et~al.}(2017){Yamashita}, {Komugi}, {Matsuhara},
  {Armus}, {Inami}, {Ueda}, {Iono}, {Kohno}, {Evans}, \& {Arimatsu}}]{yam2017}
{Yamashita}, T., {Komugi}, S., {Matsuhara}, H., {et~al.} 2017, \apj, 844, 96

\bibitem[{{Zakamska}(2010)}]{zak2010}
{Zakamska}, N.~L. 2010, \nat, 465, 60

\end{thebibliography}

\clearpage

\begin{deluxetable}{cccccc}
  \tableheadfrac{0.0}
  \tablecolumns{6}  
  \tabletypesize{\scriptsize}    
  \tablewidth{0pt}
  \tablecaption{H$_{2}$ Rotational Transitions\tablenotemark{a}}
  \tablehead{
    \colhead{Notation \tablenotemark{b}} 
    &\colhead{Transition \tablenotemark{c}} 
    &\colhead{Rest $\lambda \mu$m \tablenotemark{d}} 
    &\colhead{E$_{u}/k$ [K] \tablenotemark{e}} 
    &\colhead{A [10$^{-7}$ s$^{-1}$] \tablenotemark{f}} 
    &\colhead{g (J) \tablenotemark{g} } }
  \startdata
  S(0) &J~=~2~-~0 & 28.219 & 510 & 0.00029 & 5 \\
  S(1) &J~=~3~-~1 & 17.035 & 1015 &  0.00476 & 21 \\
  S(2) &J~=~4~-~2 & 12.279 & 1682&  0.0276 &9\\
  S(3) &J~=~5~-~3 & 9.665 & 2504& 0.0984 & 33 \\
  S(4) &J~=~6~-~4 & 8.025 & 3474 &  0.264  & 13 \\
  S(5) &J~=~7~-~5 & 6.910 & 4586 &  0.588 & 45  \\
  S(6) & J~=~8~-~6 &6.109 &  5829 & 1.14 & 17  \\
  S(7) & J~=~9~-~7 & 5.511 & 7197 & 2.00  &  57 \\
  \enddata
  \tablenotetext{a}{\citet{rous2007}}
  \tablenotetext{b}{ Short notation of H$_2$ transition}
  \tablenotetext{c}{The quantum numbers associated with the upper and lower energy levels, the transition from the upper to lower level results in the emission of the observed line listed in Column 1}
  \tablenotetext{d}{Rest wavelength for transition given in column 1}
  \tablenotetext{e}{The rotational upper energies from \citep{rous2007,huh79,blad76}}
  \tablenotetext{f}{Transition probabilities from \citep{rous2007,blad76}}
  \tablenotetext{g}{The statistical weight for the transition given in column 1.}
  \label{HlinesWave}
\end{deluxetable}

\clearpage

\begin{deluxetable}{ l c c c c c c c}
  \tableheadfrac{0.0}
  \tablecolumns{8}  
  \tabletypesize{\scriptsize}    
  \tablewidth{0pt}
  \tablecaption{H$_{2}$ Fluxes}
  \tablehead{\colhead{Name} &\colhead{RA} &\colhead{Dec} &\colhead{S(0) Flux} &\colhead{S(1) Flux} &\colhead{S(2) Flux} &\colhead{S(3) Flux}& \colhead{Merger} \\
    \colhead{} &\colhead{deg} &\colhead{deg} &\colhead{$10^{-18} \rm{W} \rm{m}^{-2}$} &\colhead{$10^{-18} \rm{W} \rm{m}^{-2}$} &\colhead{$10^{-18} \rm{W} \rm{m}^{-2}$} & \colhead{$10^{-18} \rm{W} \rm{m}^{-2}$} & \colhead{stage$^{a}$}  }
  \startdata
  NGC0023       &     2.4723       &    25.924       &          14     $\pm $   3       &          96     $\pm $   4       &          46     $\pm $   8       &          99     $\pm $   3   &$em$   \\
  NGC0034       &     2.7778       &   -12.107       &          40     $\pm $   9       &         139     $\pm $   7       &          41     $\pm $   5       &          58     $\pm $   1     & $m$ \\
  Arp256       &     4.7121       &   -10.377       & $<$       8     $\pm $ 1     &          48     $\pm $   6       &          18     $\pm $   5       &          25     $\pm $   2    &$m$ \\
  ESO350-IG038       &     9.2189       &   -33.555       & $<$       9     $\pm $ 1     &           8     $\pm $   1       & $<$       6     $\pm $ 1     & $<$      30     $\pm $ 1  &$m$   \\
  NGC0232-W       &    10.6910       &   -23.561       &          18     $\pm $   3       &         108     $\pm $   6       &          46     $\pm $  11       &         105     $\pm $   2   &$em$     \\
  NGC0232-E       &    10.7201       &   -23.541       & $<$       7     $\pm $ 1     &          52     $\pm $   6       &          13     $\pm $   6       &          61     $\pm $   6   &$em$  \\
  MCG+12-02-001       &    13.5162       &    73.085       &          70     $\pm $   3       &          92     $\pm $   5       &          32     $\pm $   8       &          83     $\pm $   3 & $m$     \\
  NGC0317B       &    14.4184       &    43.792       &          53     $\pm $   4       &         107     $\pm $   5       &          40     $\pm $   8       &          57     $\pm $  11   & $m$   \\
\enddata
  \tablecomments{{\bf{(a)--}} \citet{stir2014} give $H_2$ S(3) fluxes obtained by simultaneously by fitting the dust and gas absorption, emission and continuum gas and dust features. To achieve this \citet{stir2014} scale the SL spectra to match the LL spectra.   }
  
  \tablecomments{{\bf{(a)--}}The GOALS sources were classified in 5 stages, \citet{petric2011,stir2013},  e.g. Fig. 10 in \citet{petric2011}.  The merger classifications for each LIRG is given in \citet{stir2013} and a subset was re-evaluated by \citet{lar2016}. Here we compress the merger stages in three categories, non mergers ($nm$): targets without obvious signs of morphological disturbances, early-mergers ($em$): galaxies are within 1 arcmin of each other but show little or no morphological disturbance, mergers ($m$): this includes all other stages of gravitational interactions.}
  
\end{deluxetable}

\clearpage

\begin{deluxetable}{ccccccc}
  \tableheadfrac{0.0}
  \tablecolumns{7}  
  \tabletypesize{\scriptsize}    
  \tablewidth{0pt}
  \tablecaption{H$_{2}$ Detection Statistics from IRS SH and LH Spectra}
  \tablehead{
    \colhead{Line and Units} 
    &\colhead{Detection rate}
    &\colhead{Minimum} 
    &\colhead{Maximum}
    &\colhead{Mean} 
    &\colhead{Median} 
    &\colhead{Standard Dev.}
  }
  \startdata
  $\log$(S(0)[W/m$^2$]) &  53\%&    -17.7   & -15.6    & -16.5 & -16.6 & 0.5\\
  $\log$(L$_{\rm{S(0)}}$)[ L$_{\odot}$])&   ... & 5.8 & 7.9 & 6.8& 6.7& 0.8\\
  $\log$(S(1)[W/m$^2$]) &  91\%& -17.2& -15.4& -16.2&  -16.3&  0.4\\
  $\log$(L$_{\rm{S(1)}}$ [L$_{\odot}$])& ... &   6.0& 8.3 & 7.3 & 7.1&  0.9  \\
  $\log$(S(2)[W/m$^2$])&   68\% &   -16.4   &  -15.5 &     -16.5&      -16.6& 0.3\\
  $\log$(L$_{\rm{S(2)}}$[L$_{\odot}$])&  ... &  5.7 & 8.7 & 7.1 & 6.8 & 1.2 \\
  $\log$(S(3)[W/m$^2$] ) &   72\% &    -16.1&     -15.2&      -16.4&      -16.5& 0.8 \\
  $\log$(L$_{\rm{S(3)}}$[L$_{\odot}$]  &  ... &    5.2& 8.1& 7.2 & 6.9 & 0.6\\
  \enddata
\end{deluxetable}

\clearpage

\begin{deluxetable}{lcccc}
  \tableheadfrac{0.0}
  \tablecolumns{5}
  \tabletypesize{\small}
  \tablewidth{0pt}
  \tablecaption{General Properties of Galaxies where the H$_2$ S(1) is Resolved or Marginally Resolved}
  \tablehead{\colhead{Name} &\colhead{S(1) FWHM} &\colhead{S(1) FWHM intrinsic} &\colhead{[Ne V] detection flag} &\colhead{[O IV] detection flag} \\
    \colhead{} &\colhead{[km/s]} &\colhead{[km/s]} &\colhead{[1 detected]} &\colhead{[0 not detected]}}
  \startdata
  ESO353-G020 & 608 & 329 $\pm$ 46 & 0 & 1 \\
  IRASF01364-1042I  & 1318& 1215 $\pm$ 173 & 0 & 0 \\
  IIZw35 &  658 & 413 $\pm$ 59 & 0 & 0 \\
  NGC0828 & 661 & 420  $\pm$ 59 & 0 & 1 \\
  MCG+08-11-002  & 641 & 386 $\pm$ 55 & 0 & 1 \\
  NGC1961  & 624 & 358 $\pm$ 48 & 0 & 0 \\
  UGC03351 & 661& 420$\pm$  52 & 0 & 1 \\
  ESO255-IG007  & 690 & 464 $\pm$ 92 & 0 & 0 \\
  NGC2369 & 583 & 281 $\pm$ 35 & 0 & 1 \\
  ESO320-G030  & 583 & 281 $\pm$  35 & 0 & 0 \\
  ESO507-G070  & 629 & 365 $\pm$ 45 & 0 & 0 \\
  IRAS13052-5711 & 603 & 321 $\pm$ 40 & 0 & 0 \\
  VV250a  & 814 & 634 $\pm$ 126 & 0 & 0 \\
  IC4280 &  620 & 352 $\pm$ 43 & 0 & 1 \\
  NGC5256 &  649 & 400 $\pm$ 44 & 1 & 1 \\
  NGC5257$_8$  & 782 & 592 $\pm$ 73 & 0 & 0 \\
  NGC5331  & 616 & 343 $\pm$ 43 & 0 & 1 \\
  NGC5331 & 661 & 420 $\pm$ 47 & 0 & 0 \\
  IRASF16164-0746  & 612 & 336 $\pm$  37 & 0 & 1 \\
  ESO069-IG006  & 752 & 552 $\pm$ 61 & 0 & 0 \\
  IRASF16399-0937 & 661 & 420 $\pm$ 60 & 0 & 1 \\
  NGC6240 & 756 & 558 $\pm$ 56 & 1 & 1 \\
  CGCG142-034  & 653 & 407 $\pm$ 51  & 0 & 0 \\
  CGCG142-034  & 608 & 329 $\pm$ 41 & 1 & 1 \\
  ESO593-IG008 & 690 & 464 $\pm$ 52 & 0 & 1 \\
  MCG+04-48-002  & 653 & 407$\pm$  68 & 1 & 0 \\
  NGC7771& 690 & 464 $\pm$ 58 & 0 & 1 \\
  \enddata
  \label{RTAB}
\end{deluxetable}

\clearpage

\begin{deluxetable}{ccccccccc}
  \tableheadfrac{0.0} 
  \tablecolumns{9}
  \tabletypesize{\tiny}
  \tablewidth{0pt} 
  \tablecaption{Masses and Temperatures of Warm Molecular Gas as estimated from MIR spectroscopy} 
  \tablehead{ 
    \colhead{Source}
    &\colhead{RA} 
    &\colhead{Dec} 
    &\colhead{T 1} 
    &\colhead{log(M1)}
    &\colhead{T 2} 
    &\colhead{log(M2)} 
    &\colhead{$\langle{T}\rangle$}
    &\colhead{$\sum{M}$}\\
    \colhead{} 
    &\colhead{deg} 
    &\colhead{deg} 
    &\colhead{[K]}
    &\colhead{log(M$_\odot$)} 
    &\colhead{[K]} 
    &\colhead{log(M$_\odot$)}
    &\colhead{[K]} 
    &\colhead{log(M$_\odot$)}
  }
  \startdata
  NGC0023   &       2.470   &      25.924   &       113   &       7.973   &       386   &       6.548   &       123   &       7.989  \\
  NGC0034   &       2.780   &     -12.108   &       124   &       8.396   &       318   &       7.025   &       132   &       8.415  \\
  NGC0034   &       2.790   &     -12.021   &       319   &       7.013   &       ...   &         ...   &       319   &       7.013  \\
  MCG-02-01-051   &       4.710   &     -10.359   &       121   &       8.318   &       411   &       6.729   &       129   &       8.328  \\
  ESO350-IG038   &       9.220   &     -33.555   &       391   &       6.580   &       ...   &         ...   &       391   &       6.580  \\
  NGC0232   &      10.690   &     -23.561   &       116   &       8.766   &       517   &       5.962   &       116   &       8.767  \\
  NGC0232   &      10.720   &     -23.541   &       175   &       8.111   &        11   &       7.398   &       148   &       8.188  \\
  MCG+12-02-001   &      13.520   &      73.085   &       193   &       8.474   &        15   &       7.699   &       167   &       8.543  \\
\enddata
  \label{FitHMT}
\end{deluxetable}

\clearpage

\begin{deluxetable}{ccc}
  \label{AsurvTest}
  \tablecolumns{3}
  \tabletypesize{\small}
  \tablewidth{0pt}
  \tablecaption{General H$_{2}$ comparisons using censored data statistical methods }
  \tablehead{ \colhead{Samples compared} &\colhead{log(L$_{S(0)}$/L$_{IR}$)} &\colhead{log(L$_{S(1)}$/L$_{IR}$)} }
  \startdata
  ULIRGs vs LIRGs & 0\% & 8\% \\
  Normal Galaxies vs LIRGs  & 0\% & 0\%\\
  Merging vs Non-merging GOALS galaxies  \% & 90\% & 68 \%\\
  LIRGs with [NeV] detections vs SB  & 47\% & 65\%\\
  LIRGs where the AGN dominates the IR  vs SB & 25\% & 42\%\\
  \enddata
  \tablecomments{We give the probability that the two samples are
    drawn from the same population. This is the average probability we
    obtain from running each of the following statistical tests
    especially well suited for data containing upper limits, also
    known as censored data : (1) the Gehan Generalized Wilcoxon Test
    using permutation variance, (2) Gehan's generalized Wilcoxon test
    using hypergeometric variance, (3) Logrank test, (4) Peto \& Peto
    Generalized Wilcoxon Test and (5) Peto \& Prentice Generalized
    Wilcoxon Test. Those statistical tests are implemented in the
    ASURV program by E. Feigelson. Most recently this program can be
    also be found at $http://www.r-project.org.$ For the mergers vs
    non-mergers comparisons we also used the 24 $\mu$m flux determined
    by multiplying the LL IRS spectra with the MIPS 24 $\mu$m filter
    as explained in Petric et al. (2011). This was done to test for
    systematic differences associated with IRAS photometry of mergers
    and non-mergers. Using the 24 micron versus the IR micron
    luminosity gave us similar results. }
\end{deluxetable}

\clearpage

\begin{deluxetable}{lcc}
  % \tableheadfrac{0.0}
  \tablecolumns{3}
  \tabletypesize{\small}
  \tablewidth{0pt}
  \tablecaption{General H$_{2}$ comparisons using the Kolmogorov-Smirnov test }
  \tablehead{ 
    \colhead{Samples compared} &
    \colhead{$DKS^1$} 
    &\colhead{$PKS^1$}}
  \startdata
  {\bf{Sources with resolved H$_2$ lines versus sources with unresolved H$_2$ lines}} &  & \\
  Masses  & 0.3 & 3\% \\
  Temperature & 0.2 & 11\% \\
  Merger Stage &  0.2 & 13\% \\
  \cline{1-3}\\
  {\bf{All Mergers versus all non-Mergers}} & &\\
  Warm Gas Masses &  0.1 & 44\%\\
  Temperature of Warm Gas &  0.2 & 7\% \\
  \cline{1-3}\\
  {\bf{Advanced Mergers versus all non-Mergers}} & & \\
  Warm Gas Masses &  0.3 & 3\%\\
  Temperature of Warm Gas &  0.3 & 3\% \\ 
  \cline{1-3}\\
  {\bf{AGN dominated versus starburst dominated}} & & \\
  Warm Gas Masses &  0.3 & 2\%\\
  Temperature of Warm Gas &  0.4 & 0.1\% \\ 
  \enddata
  \tablecomments{(1) The Kolmogorov Ð Smirnov statistic $DKS$ gives an estimate of the difference between two distributions of measurements. It does this by comparing the cumulative cumulative distribution functions between the two sets of values being compared.  $DKS$ ~= ~Maximum (cumulative distribution of sample 1 - cumulative distribution of sample 2).  A cumulative distribution (plotted on the vertical axis) gives probability that the variable takes a value less than a certain value (shown in the x-axis).  A small $DKS$ (e.g. lower than 0.1) means that the sample properties are similar while a large $DKS$ means that they may be different.  To determine if the $DKS$ is statistically significant we calculate the $PKS$ which gives the probability that the two sets of values compared are drawn from the same population.}
  \label{KStab}
\end{deluxetable}

\clearpage
\appendix
\section{A1: Matching Apertures}
For this investigation of warm molecular H$_2$ properties we combined fluxes determined from LH, SH and SL spectra \citep{petric2011, ina2013, stir2013, stir2014}. The widths of the SL, SH, and, LH slits (3.7\arcsec , 4.7\arcsec, and 11.1\arcsec) corresponding to 1.5, 2.0, and 4.6 kpc respectively at the median galaxy distance of our sample (88 Mpc).

We scaled each of the LH H$_2$ S(0) fluxes by comparing the SH and the LH spectra. We averaged two scaling factors for simplicity called $LH2SHv1$ and $LH2SHv2$. $LH2SHv1$ is the ratio of the median flux in the SH spectra, in the wavelength region overlapping with the LH spectra, to the median flux in the LH spectra, in the wavelength region overlapping with the SH spectra. $LH2SHv2$ is the ratio of the extrapolated flux at 18 $\mu$m, from a fit to the continuum of the SH spectra at wavelengths greater than 17.5 $\mu$m, to the flux at 18 $\mu$m from the LH spectra, derived the same way from data at wavelengths shorter than 20 $\mu$m. The linear fit is done by minimizing the absolute differences and not the usual least squares fit. This is important because doing so minimizes the impact of bad pixels that are often found at  the edge on the fit. The median value of $LH2SHv1$ is 0.78 and that of $LH2SHv2$ is 0.66, and the median of the final scaling factor is 0.72. 

\begin{figure*}[h!] 
  \includegraphics[width=0.49\linewidth]{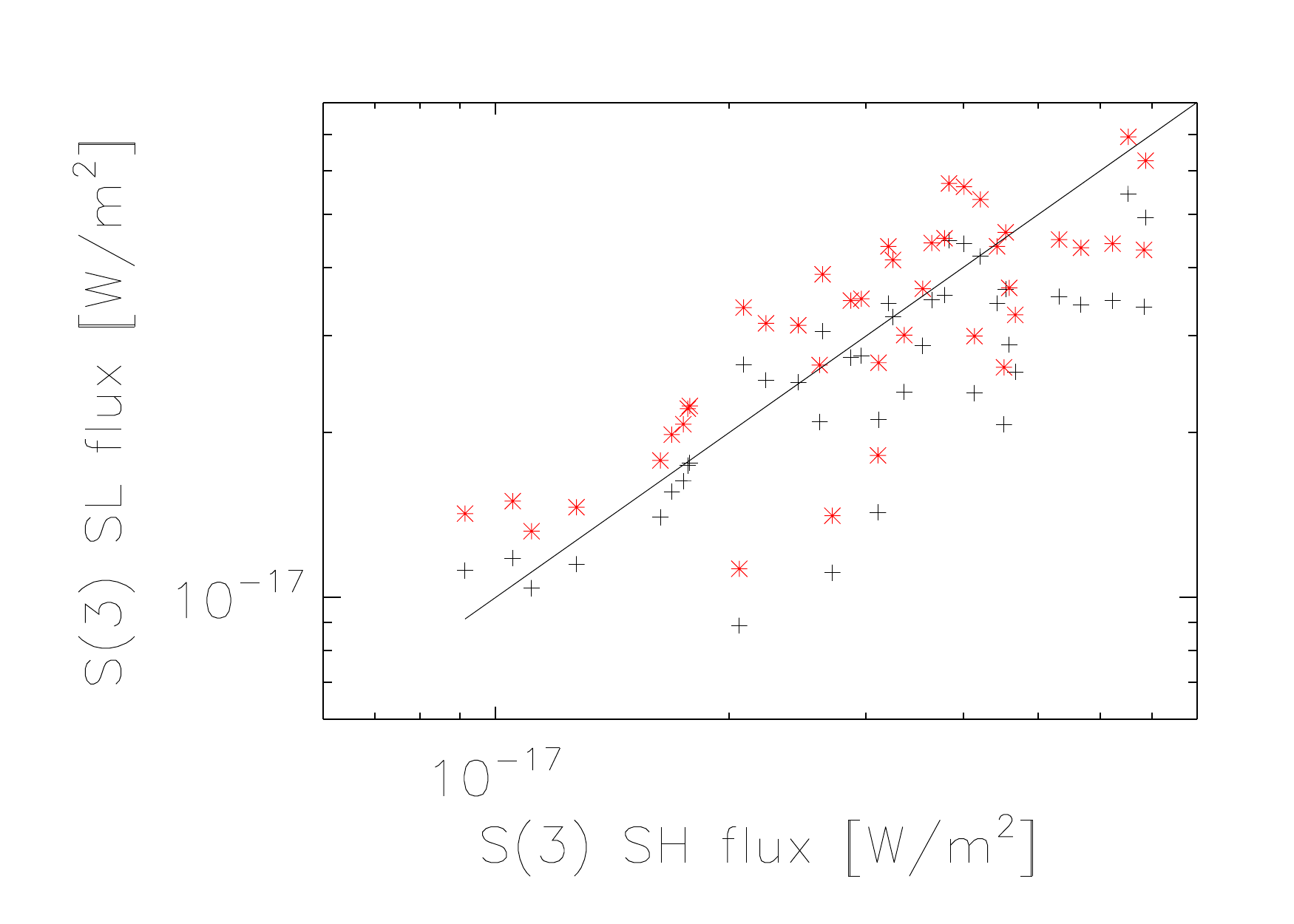}
  \includegraphics[width=0.49\linewidth]{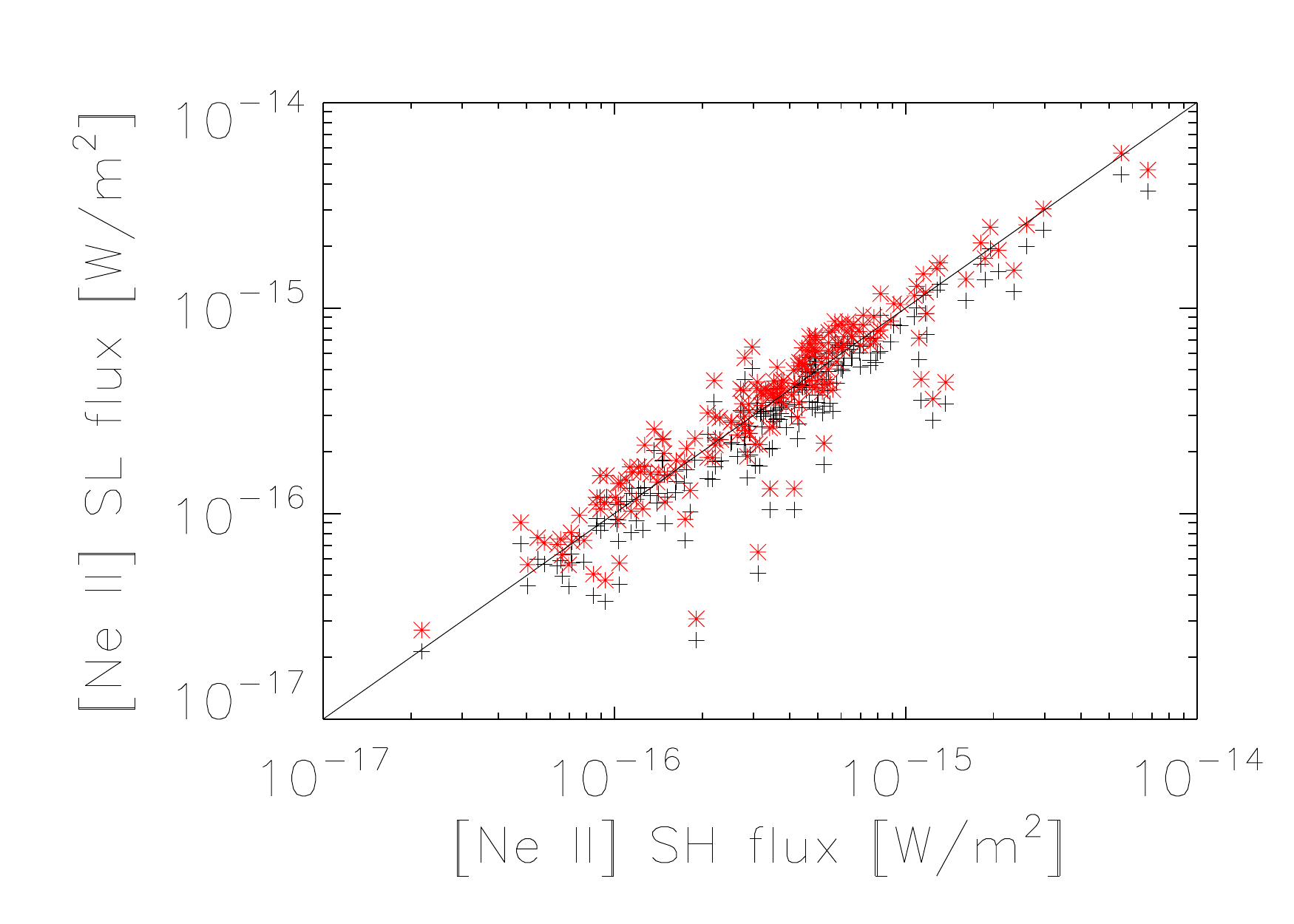}\\
  \includegraphics[width=0.49\linewidth]{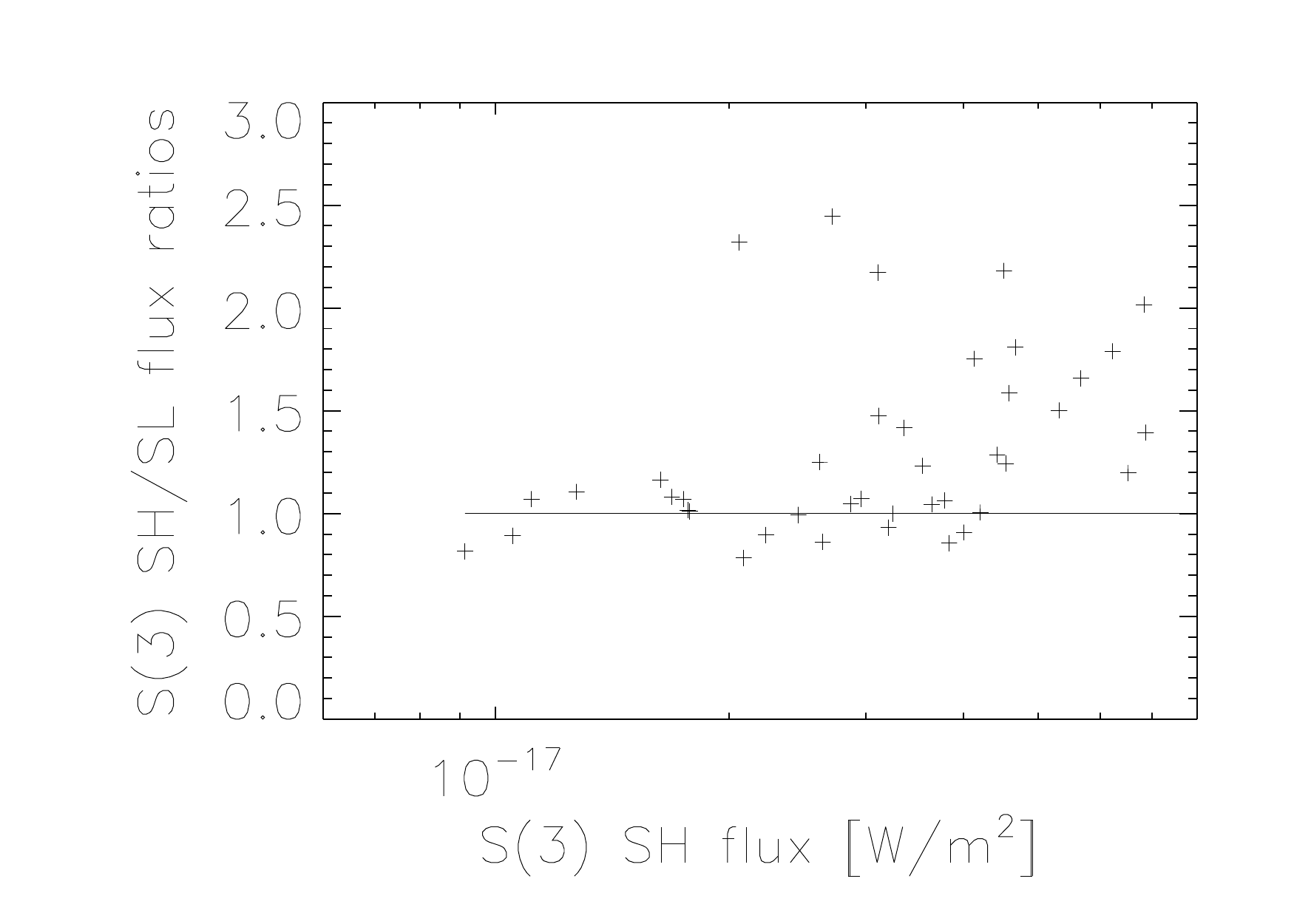}
  \includegraphics[width=0.49\linewidth]{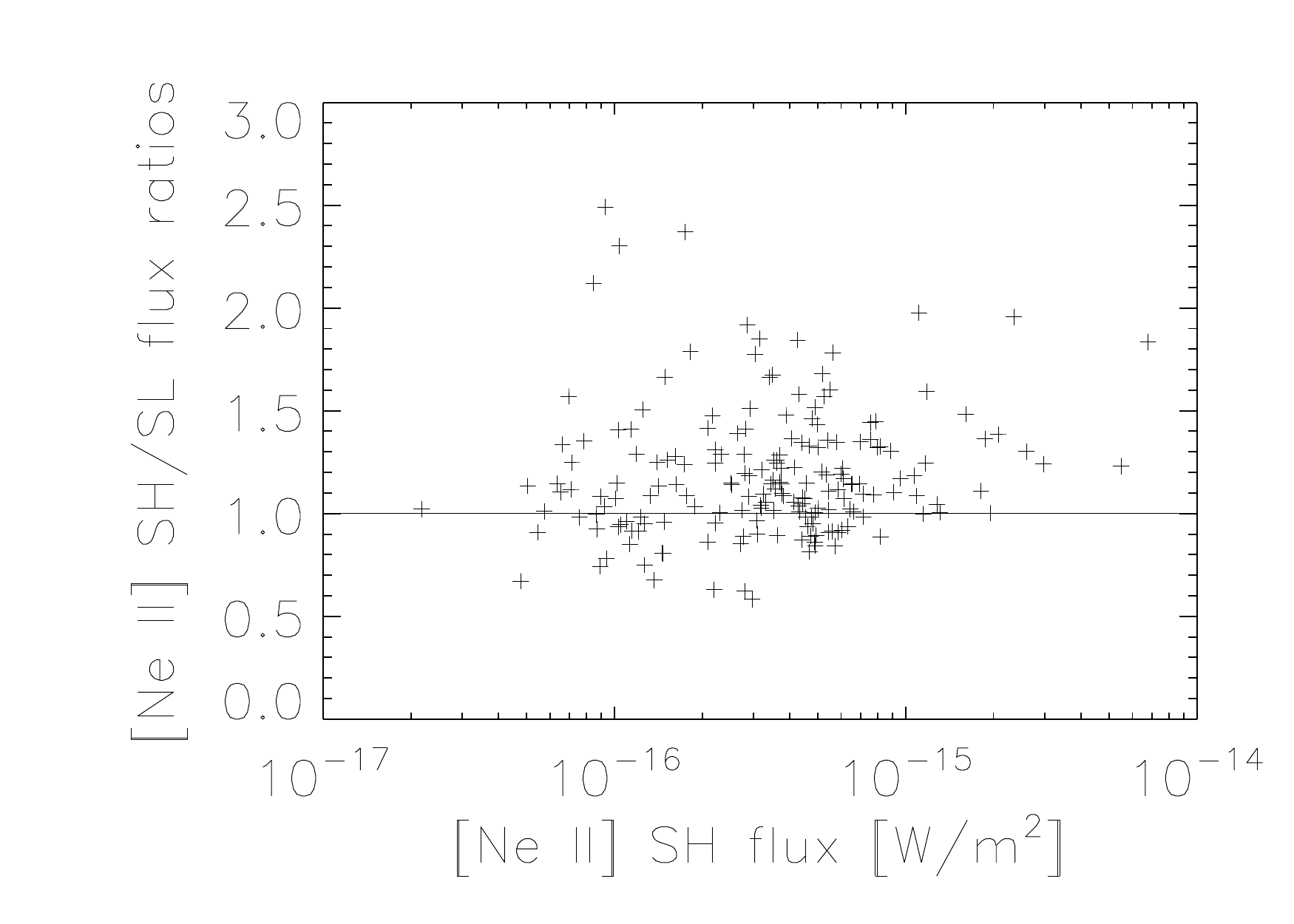}
  \caption{ {\bf Top Left:} measured H$_2$ S(3) line fluxes in SH versus those measured in SL before scaling. {\bf Top Right:} measured [Ne II] line fluxes in SH versus those measured in SL before scaling. Red points in both plots show SH measurements versus the scaled SL fluxes. Black points in both plots show SH measurements versus measured SL fluxes. {\bf Bottom Left:} ratio of SH to SL flux estimates of the H$_2$ S(3) line versus the flux measured in SH. {\bf Bottom Right:} ratio of SH to SL flux estimates of the [Ne II]  line versus the flux measured in SH.\label{ScaleSL} }
\end{figure*}

For S(3)-S(7) fluxes we rely on measurements from the lower resolution SL- spectra. We scale the SL fluxes by the ratio of the spectral extraction widths (4.7/3.7). We compute the error associated with this scaling by looking at the median ratio between the scaled S(3) SL flux and the measured S(3) SH flux. That median relative difference is 25\% for S(3). We add this to the error obtained from the gaussian fits in quadrature.  As a test of our method we apply the same scaling factor to the [Ne II]  fluxes derived in SL and compare them with those found in SH. For the [Ne II] lines we find a scatter associated with these scalings of 20\%. Figure (\ref{ScaleSL}) shows how the different scaled up values for SL H$_2$ S(3) and [Ne II] compared to the values measured in SH and suggests that our method is reliable at the 20-25\% level.  We applied those scaling factors to all SL estimates.

\section{A2: Ortho-to-Para ratios}

The OPR value in the high temperature limit (OPR$_{high~T}$), defined as:

\begin{equation}
  OPR ~=~{OPR_{high T}} ~ \frac{{\sum _{O}} (2J_{O} ~+~1)~exp[-E_{O}/(kT)] }{ \sum_{P} (2J_{P}~+~1)~exp[-E_{P}/(kT)] }
\end{equation}

\noindent where $o$ and $p$ refers to ortho (odd) and para (even) states, respectively, $I$ is the spin number, and $J$ is the rotational quantum number, is equal to 3 in LTE, and allows one to estimate departure from LTE otherwise. If all the observed H$_2$ is in LTE then T (S0-S1) $\leq$ T (S0-S2) $\leq$ T(S1-S2) $\leq$ T(S1-S3) $\leq$ T(S2-S3), as is expected from a Boltzman distribution.

Therefore, the variations of these apparent excitation temperatures as a function of OPR$_{high~T}$, combined with the condition of monotony, will define a range of OPR$_{high~T}$. If the value of 3 is allowed, then the source is compatible with LTE.

The expression for the excitation temperatures is as follows. T(S0-S2) and T(S1-S3) are independent of $OPR_{high T}$ because they only involve para or ortho levels, and determined directly from the observed fluxes:

\begin{equation}
  kT~=~\frac{E_{u2}~-~E_{u1}}{ln(N_{u1}/N_{u2} \times g_{u2}/g_{u1})}
\end{equation}

T(S0-S1), T(S1-S2), and T(S2-S3) however depend on $OPR_{high T}$ as follows:

\begin{equation}
  k~T(S_{P}~-~S_{O})~=~(E_{u,O}~-~E_{u,P})/ln(OPR_{high T} R)
\end{equation}

\noindent with 

\begin{equation}
  R~=~F_{P}/F_{O} ~\times~ A_{O}/A_{P} ~\times~ \lambda _{P}/\lambda_{O} ~\times ~ (2J_{0} ~+~1)/(2J_{P} +1).
\end{equation}

For each pair (p, o) = (0, 1), (2, 1), and (2, 3), we have computed possible temperatures as a function of $OPR_{high T}$ between 1 and 3. We are able to investigate 78 sources for which the S(0) to S(3) lines were detected and for which the uncertainties of the line flux do not lead to a significantly large uncertainty in the excitation temperature (see Figure \ref{OPPlots}).

\end{document}